\documentclass[twocolumn,longauth,traditabstract]{aa} 

\usepackage{graphicx,amsmath}
\usepackage{epsf}
\usepackage{txfonts}
\usepackage[breaklinks, colorlinks, citecolor=blue]{hyperref}  
\usepackage{natbib}
\usepackage{upgreek}
\usepackage{url}

\bibpunct{(}{)}{;}{a}{}{,} 

\def\setsymbol#1#2{\expandafter\def\csname #1\endcsname{#2}}
\def\getsymbol#1{\csname #1\endcsname}

\def\Planck{{\it Planck\/}}

\newbox\tablebox    \newdimen\tablewidth
\def\leaderfil{\leaders\hbox to 5pt{\hss.\hss}\hfil}
\def\endPlancktable{\tablewidth=\columnwidth 
    $$\hss\copy\tablebox\hss$$
    \vskip-\lastskip\vskip -2pt}

\def\tablenote#1 #2\par{\begingroup \parindent=0.8em
    \abovedisplayshortskip=0pt\belowdisplayshortskip=0pt
    \noindent
    $$\hss\vbox{\hsize\tablewidth \hangindent=\parindent \hangafter=1 \noindent
    \hbox to \parindent{\sup{\rm #1}\hss}\strut#2\strut\par}\hss$$
    \endgroup}

\def\L2{\ifmmode L_2\else $L_2$\fi}

\def\DeltaT{\ifmmode \Delta T\else $\Delta T$\fi}
\def\deltat{\ifmmode \Delta t\else $\Delta t$\fi}
\def\fknee{\ifmmode f_{\rm knee}\else $f_{\rm knee}$\fi}
\def\Fmax{\ifmmode F_{\rm max}\else $F_{\rm max}$\fi}
\def\solar{\ifmmode{\rm M}_{\mathord\odot}\else${\rm M}_{\mathord\odot}$\fi}

\def\mag{\sup{m}}
\def\inv{\ifmmode^{-1}\else$^{-1}$\fi}
\def\mo{\ifmmode^{-1}\else$^{-1}$\fi}
\def\sup#1{\ifmmode ^{\rm #1}\else $^{\rm #1}$\fi}
\def\expo#1{\ifmmode \times 10^{#1}\else $\times 10^{#1}$\fi}
\def\,{\thinspace}
\def\lsim{\mathrel{\raise .4ex\hbox{\rlap{$<$}\lower 1.2ex\hbox{$\sim$}}}}
\def\gsim{\mathrel{\raise .4ex\hbox{\rlap{$>$}\lower 1.2ex\hbox{$\sim$}}}}

\def\simprop{\mathrel{\raise .4ex\hbox{\rlap{$\propto$}\lower 1.2ex\hbox{$\sim$}}}}
\def\deg{\ifmmode^\circ\else$^\circ$\fi}
\def\pdeg{\ifmmode $\setbox0=\hbox{$^{\circ}$}\rlap{\hskip.11\wd0 .}$^{\circ}
          \else \setbox0=\hbox{$^{\circ}$}\rlap{\hskip.11\wd0 .}$^{\circ}$\fi}
\def\arcs{\ifmmode {^{\scriptstyle\prime\prime}}
          \else $^{\scriptstyle\prime\prime}$\fi}
\def\arcm{\ifmmode {^{\scriptstyle\prime}}
          \else $^{\scriptstyle\prime}$\fi}
\newdimen\sa  \newdimen\sb
\def\parcs{\sa=.07em \sb=.03em
     \ifmmode \hbox{\rlap{.}}^{\scriptstyle\prime\kern -\sb\prime}\hbox{\kern -\sa}
     \else \rlap{.}$^{\scriptstyle\prime\kern -\sb\prime}$\kern -\sa\fi}
\def\parcm{\sa=.08em \sb=.03em
     \ifmmode \hbox{\rlap{.}\kern\sa}^{\scriptstyle\prime}\hbox{\kern-\sb}
     \else \rlap{.}\kern\sa$^{\scriptstyle\prime}$\kern-\sb\fi}
\def\ra[#1 #2 #3.#4]{#1\sup{h}#2\sup{m}#3\sup{s}\llap.#4}
\def\dec[#1 #2 #3.#4]{#1\deg#2\arcm#3\arcs\llap.#4}
\def\deco[#1 #2 #3]{#1\deg#2\arcm#3\arcs}
\def\rra[#1 #2]{#1\sup{h}#2\sup{m}}

\def\dots{\relax\ifmmode \ldots\else $\ldots$\fi}
\def\WHzsr{\ifmmode $W\,Hz\mo\,sr\mo$\else W\,Hz\mo\,sr\mo\fi}
\def\mHz{\ifmmode $\,mHz$\else \,mHz\fi}
\def\GHz{\ifmmode $\,GHz$\else \,GHz\fi}
\def\mKs{\ifmmode $\,mK\,s$^{1/2}\else \,mK\,s$^{1/2}$\fi}
\def\muKs{\ifmmode \,\mu$K\,s$^{1/2}\else \,$\mu$K\,s$^{1/2}$\fi}
\def\muKRJs{\ifmmode \,\mu$K$_{\rm RJ}$\,s$^{1/2}\else \,$\mu$K$_{\rm RJ}$\,s$^{1/2}$\fi}
\def\muKHz{\ifmmode \,\mu$K\,Hz$^{-1/2}\else \,$\mu$K\,Hz$^{-1/2}$\fi}
\def\MJysr{\ifmmode \,$MJy\,sr\mo$\else \,MJy\,sr\mo\fi}
\def\MJysrmK{\ifmmode \,$MJy\,sr\mo$\,mK$_{\rm CMB}\mo\else \,MJy\,sr\mo\,mK$_{\rm CMB}\mo$\fi}
\def\microns{\ifmmode \,\mu$m$\else \,$\mu$m\fi}

\def\muK{\ifmmode \,\mu$K$\else \,$\mu$\hbox{K}\fi}
\def\microK{\ifmmode \,\mu$K$\else \,$\mu$\hbox{K}\fi}
\def\muW{\ifmmode \,\mu$W$\else \,$\mu$\hbox{W}\fi}
\def\kms{\ifmmode $\,km\,s$^{-1}\else \,km\,s$^{-1}$\fi}
\def\kmsMpc{\ifmmode $\,\kms\,Mpc\mo$\else \,\kms\,Mpc\mo\fi}
\setsymbol{LFI:center:frequency:70GHz:units}{70.3\,GHz}
\setsymbol{LFI:center:frequency:44GHz:units}{44.1\,GHz}
\setsymbol{LFI:center:frequency:30GHz:units}{28.5\,GHz}

\setsymbol{LFI:center:frequency:70GHz}{70.3}
\setsymbol{LFI:center:frequency:44GHz}{44.1}
\setsymbol{LFI:center:frequency:30GHz}{28.5}

\setsymbol{LFI:center:frequency:LFI18:Rad:M:units}{71.7\GHz}
\setsymbol{LFI:center:frequency:LFI19:Rad:M:units}{67.5\GHz}
\setsymbol{LFI:center:frequency:LFI20:Rad:M:units}{69.2\GHz}
\setsymbol{LFI:center:frequency:LFI21:Rad:M:units}{70.4\GHz}
\setsymbol{LFI:center:frequency:LFI22:Rad:M:units}{71.5\GHz}
\setsymbol{LFI:center:frequency:LFI23:Rad:M:units}{70.8\GHz}
\setsymbol{LFI:center:frequency:LFI24:Rad:M:units}{44.4\GHz}
\setsymbol{LFI:center:frequency:LFI25:Rad:M:units}{44.0\GHz}
\setsymbol{LFI:center:frequency:LFI26:Rad:M:units}{43.9\GHz}
\setsymbol{LFI:center:frequency:LFI27:Rad:M:units}{28.3\GHz}
\setsymbol{LFI:center:frequency:LFI28:Rad:M:units}{28.8\GHz}
\setsymbol{LFI:center:frequency:LFI18:Rad:S:units}{70.1\GHz}
\setsymbol{LFI:center:frequency:LFI19:Rad:S:units}{69.6\GHz}
\setsymbol{LFI:center:frequency:LFI20:Rad:S:units}{69.5\GHz}
\setsymbol{LFI:center:frequency:LFI21:Rad:S:units}{69.5\GHz}
\setsymbol{LFI:center:frequency:LFI22:Rad:S:units}{72.8\GHz}
\setsymbol{LFI:center:frequency:LFI23:Rad:S:units}{71.3\GHz}
\setsymbol{LFI:center:frequency:LFI24:Rad:S:units}{44.1\GHz}
\setsymbol{LFI:center:frequency:LFI25:Rad:S:units}{44.1\GHz}
\setsymbol{LFI:center:frequency:LFI26:Rad:S:units}{44.1\GHz}
\setsymbol{LFI:center:frequency:LFI27:Rad:S:units}{28.5\GHz}
\setsymbol{LFI:center:frequency:LFI28:Rad:S:units}{28.2\GHz}

\setsymbol{LFI:center:frequency:LFI18:Rad:M}{71.7}
\setsymbol{LFI:center:frequency:LFI19:Rad:M}{67.5}
\setsymbol{LFI:center:frequency:LFI20:Rad:M}{69.2}
\setsymbol{LFI:center:frequency:LFI21:Rad:M}{70.4}
\setsymbol{LFI:center:frequency:LFI22:Rad:M}{71.5}
\setsymbol{LFI:center:frequency:LFI23:Rad:M}{70.8}
\setsymbol{LFI:center:frequency:LFI24:Rad:M}{44.4}
\setsymbol{LFI:center:frequency:LFI25:Rad:M}{44.0}
\setsymbol{LFI:center:frequency:LFI26:Rad:M}{43.9}
\setsymbol{LFI:center:frequency:LFI27:Rad:M}{28.3}
\setsymbol{LFI:center:frequency:LFI28:Rad:M}{28.8}
\setsymbol{LFI:center:frequency:LFI18:Rad:S}{70.1}
\setsymbol{LFI:center:frequency:LFI19:Rad:S}{69.6}
\setsymbol{LFI:center:frequency:LFI20:Rad:S}{69.5}
\setsymbol{LFI:center:frequency:LFI21:Rad:S}{69.5}
\setsymbol{LFI:center:frequency:LFI22:Rad:S}{72.8}
\setsymbol{LFI:center:frequency:LFI23:Rad:S}{71.3}
\setsymbol{LFI:center:frequency:LFI24:Rad:S}{44.1}
\setsymbol{LFI:center:frequency:LFI25:Rad:S}{44.1}
\setsymbol{LFI:center:frequency:LFI26:Rad:S}{44.1}
\setsymbol{LFI:center:frequency:LFI27:Rad:S}{28.5}
\setsymbol{LFI:center:frequency:LFI28:Rad:S}{28.2}

\setsymbol{LFI:white:noise:sensitivity:70GHz:units}{152.6\muKs}
\setsymbol{LFI:white:noise:sensitivity:44GHz:units}{173.1\muKs}
\setsymbol{LFI:white:noise:sensitivity:30GHz:units}{146.8\muKs}

\setsymbol{LFI:white:noise:sensitivity:70GHz}{152.6}
\setsymbol{LFI:white:noise:sensitivity:44GHz}{173.1}
\setsymbol{LFI:white:noise:sensitivity:30GHz}{146.8}

\setsymbol{LFI:white:noise:sensitivity:LFI18:Rad:M:units}{512.0\muKs}
\setsymbol{LFI:white:noise:sensitivity:LFI19:Rad:M:units}{581.4\muKs}
\setsymbol{LFI:white:noise:sensitivity:LFI20:Rad:M:units}{590.8\muKs}
\setsymbol{LFI:white:noise:sensitivity:LFI21:Rad:M:units}{455.2\muKs}
\setsymbol{LFI:white:noise:sensitivity:LFI22:Rad:M:units}{492.0\muKs}
\setsymbol{LFI:white:noise:sensitivity:LFI23:Rad:M:units}{507.7\muKs}
\setsymbol{LFI:white:noise:sensitivity:LFI24:Rad:M:units}{462.2\muKs}
\setsymbol{LFI:white:noise:sensitivity:LFI25:Rad:M:units}{413.6\muKs}
\setsymbol{LFI:white:noise:sensitivity:LFI26:Rad:M:units}{478.6\muKs}
\setsymbol{LFI:white:noise:sensitivity:LFI27:Rad:M:units}{277.7\muKs}
\setsymbol{LFI:white:noise:sensitivity:LFI28:Rad:M:units}{312.3\muKs}
\setsymbol{LFI:white:noise:sensitivity:LFI18:Rad:S:units}{465.7\muKs}
\setsymbol{LFI:white:noise:sensitivity:LFI19:Rad:S:units}{555.6\muKs}
\setsymbol{LFI:white:noise:sensitivity:LFI20:Rad:S:units}{623.2\muKs}
\setsymbol{LFI:white:noise:sensitivity:LFI21:Rad:S:units}{564.1\muKs}
\setsymbol{LFI:white:noise:sensitivity:LFI22:Rad:S:units}{534.4\muKs}
\setsymbol{LFI:white:noise:sensitivity:LFI23:Rad:S:units}{542.4\muKs}
\setsymbol{LFI:white:noise:sensitivity:LFI24:Rad:S:units}{399.2\muKs}
\setsymbol{LFI:white:noise:sensitivity:LFI25:Rad:S:units}{392.6\muKs}
\setsymbol{LFI:white:noise:sensitivity:LFI26:Rad:S:units}{418.6\muKs}
\setsymbol{LFI:white:noise:sensitivity:LFI27:Rad:S:units}{302.9\muKs}
\setsymbol{LFI:white:noise:sensitivity:LFI28:Rad:S:units}{285.3\muKs}

\setsymbol{LFI:white:noise:sensitivity:LFI18:Rad:M}{512.0}
\setsymbol{LFI:white:noise:sensitivity:LFI19:Rad:M}{581.4}
\setsymbol{LFI:white:noise:sensitivity:LFI20:Rad:M}{590.8}
\setsymbol{LFI:white:noise:sensitivity:LFI21:Rad:M}{455.2}
\setsymbol{LFI:white:noise:sensitivity:LFI22:Rad:M}{492.0}
\setsymbol{LFI:white:noise:sensitivity:LFI23:Rad:M}{507.7}
\setsymbol{LFI:white:noise:sensitivity:LFI24:Rad:M}{462.2}
\setsymbol{LFI:white:noise:sensitivity:LFI25:Rad:M}{413.6}
\setsymbol{LFI:white:noise:sensitivity:LFI26:Rad:M}{478.6}
\setsymbol{LFI:white:noise:sensitivity:LFI27:Rad:M}{277.7}
\setsymbol{LFI:white:noise:sensitivity:LFI28:Rad:M}{312.3}
\setsymbol{LFI:white:noise:sensitivity:LFI18:Rad:S}{465.7}
\setsymbol{LFI:white:noise:sensitivity:LFI19:Rad:S}{555.6}
\setsymbol{LFI:white:noise:sensitivity:LFI20:Rad:S}{623.2}
\setsymbol{LFI:white:noise:sensitivity:LFI21:Rad:S}{564.1}
\setsymbol{LFI:white:noise:sensitivity:LFI22:Rad:S}{534.4}
\setsymbol{LFI:white:noise:sensitivity:LFI23:Rad:S}{542.4}
\setsymbol{LFI:white:noise:sensitivity:LFI24:Rad:S}{399.2}
\setsymbol{LFI:white:noise:sensitivity:LFI25:Rad:S}{392.6}
\setsymbol{LFI:white:noise:sensitivity:LFI26:Rad:S}{418.6}
\setsymbol{LFI:white:noise:sensitivity:LFI27:Rad:S}{302.9}
\setsymbol{LFI:white:noise:sensitivity:LFI28:Rad:S}{285.3}

\setsymbol{LFI:knee:frequency:70GHz:units}{29.5\mHz}
\setsymbol{LFI:knee:frequency:44GHz:units}{56.2\mHz}
\setsymbol{LFI:knee:frequency:30GHz:units}{113.7\mHz}

\setsymbol{LFI:knee:frequency:70GHz}{29.5}
\setsymbol{LFI:knee:frequency:44GHz}{56.2}
\setsymbol{LFI:knee:frequency:30GHz}{113.7}

\setsymbol{LFI:knee:frequency:LFI18:Rad:M:units}{16.3\mHz}
\setsymbol{LFI:knee:frequency:LFI19:Rad:M:units}{15.1\mHz}
\setsymbol{LFI:knee:frequency:LFI20:Rad:M:units}{18.7\mHz}
\setsymbol{LFI:knee:frequency:LFI21:Rad:M:units}{37.2\mHz}
\setsymbol{LFI:knee:frequency:LFI22:Rad:M:units}{12.7\mHz}
\setsymbol{LFI:knee:frequency:LFI23:Rad:M:units}{34.6\mHz}
\setsymbol{LFI:knee:frequency:LFI24:Rad:M:units}{46.2\mHz}
\setsymbol{LFI:knee:frequency:LFI25:Rad:M:units}{24.9\mHz}
\setsymbol{LFI:knee:frequency:LFI26:Rad:M:units}{67.6\mHz}
\setsymbol{LFI:knee:frequency:LFI27:Rad:M:units}{187.4\mHz}
\setsymbol{LFI:knee:frequency:LFI28:Rad:M:units}{122.2\mHz}
\setsymbol{LFI:knee:frequency:LFI18:Rad:S:units}{17.7\mHz}
\setsymbol{LFI:knee:frequency:LFI19:Rad:S:units}{22.0\mHz}
\setsymbol{LFI:knee:frequency:LFI20:Rad:S:units}{8.7\mHz}
\setsymbol{LFI:knee:frequency:LFI21:Rad:S:units}{25.9\mHz}
\setsymbol{LFI:knee:frequency:LFI22:Rad:S:units}{15.8\mHz}
\setsymbol{LFI:knee:frequency:LFI23:Rad:S:units}{129.8\mHz}
\setsymbol{LFI:knee:frequency:LFI24:Rad:S:units}{100.9\mHz}
\setsymbol{LFI:knee:frequency:LFI25:Rad:S:units}{38.9\mHz}
\setsymbol{LFI:knee:frequency:LFI26:Rad:S:units}{58.9\mHz}
\setsymbol{LFI:knee:frequency:LFI27:Rad:S:units}{104.4\mHz}
\setsymbol{LFI:knee:frequency:LFI28:Rad:S:units}{40.7\mHz}

\setsymbol{LFI:knee:frequency:LFI18:Rad:M}{16.3}
\setsymbol{LFI:knee:frequency:LFI19:Rad:M}{15.1}
\setsymbol{LFI:knee:frequency:LFI20:Rad:M}{18.7}
\setsymbol{LFI:knee:frequency:LFI21:Rad:M}{37.2}
\setsymbol{LFI:knee:frequency:LFI22:Rad:M}{12.7}
\setsymbol{LFI:knee:frequency:LFI23:Rad:M}{34.6}
\setsymbol{LFI:knee:frequency:LFI24:Rad:M}{46.2}
\setsymbol{LFI:knee:frequency:LFI25:Rad:M}{24.9}
\setsymbol{LFI:knee:frequency:LFI26:Rad:M}{67.6}
\setsymbol{LFI:knee:frequency:LFI27:Rad:M}{187.4}
\setsymbol{LFI:knee:frequency:LFI28:Rad:M}{122.2}
\setsymbol{LFI:knee:frequency:LFI18:Rad:S}{17.7}
\setsymbol{LFI:knee:frequency:LFI19:Rad:S}{22.0}
\setsymbol{LFI:knee:frequency:LFI20:Rad:S}{8.7}
\setsymbol{LFI:knee:frequency:LFI21:Rad:S}{25.9}
\setsymbol{LFI:knee:frequency:LFI22:Rad:S}{15.8}
\setsymbol{LFI:knee:frequency:LFI23:Rad:S}{129.8}
\setsymbol{LFI:knee:frequency:LFI24:Rad:S}{100.9}
\setsymbol{LFI:knee:frequency:LFI25:Rad:S}{38.9}
\setsymbol{LFI:knee:frequency:LFI26:Rad:S}{58.9}
\setsymbol{LFI:knee:frequency:LFI27:Rad:S}{104.4}
\setsymbol{LFI:knee:frequency:LFI28:Rad:S}{40.7}

\setsymbol{LFI:slope:70GHz:units}{$-1.03$\mHz}
\setsymbol{LFI:slope:44GHz:units}{$-0.89$\mHz}
\setsymbol{LFI:slope:30GHz:units}{$-0.87$\mHz}

\setsymbol{LFI:slope:70GHz}{$-1.03$}
\setsymbol{LFI:slope:44GHz}{$-0.89$}
\setsymbol{LFI:slope:30GHz}{$-0.87$}

\setsymbol{LFI:slope:LFI18:Rad:M:units}{$-1.04$\mHz}
\setsymbol{LFI:slope:LFI19:Rad:M:units}{$-1.09$\mHz}
\setsymbol{LFI:slope:LFI20:Rad:M:units}{$-0.69$\mHz}
\setsymbol{LFI:slope:LFI21:Rad:M:units}{$-1.56$\mHz}
\setsymbol{LFI:slope:LFI22:Rad:M:units}{$-1.01$\mHz}
\setsymbol{LFI:slope:LFI23:Rad:M:units}{$-0.96$\mHz}
\setsymbol{LFI:slope:LFI24:Rad:M:units}{$-0.83$\mHz}
\setsymbol{LFI:slope:LFI25:Rad:M:units}{$-0.91$\mHz}
\setsymbol{LFI:slope:LFI26:Rad:M:units}{$-0.95$\mHz}
\setsymbol{LFI:slope:LFI27:Rad:M:units}{$-0.87$\mHz}
\setsymbol{LFI:slope:LFI28:Rad:M:units}{$-0.88$\mHz}
\setsymbol{LFI:slope:LFI18:Rad:S:units}{$-1.15$\mHz}
\setsymbol{LFI:slope:LFI19:Rad:S:units}{$-1.00$\mHz}
\setsymbol{LFI:slope:LFI20:Rad:S:units}{$-0.95$\mHz}
\setsymbol{LFI:slope:LFI21:Rad:S:units}{$-0.92$\mHz}
\setsymbol{LFI:slope:LFI22:Rad:S:units}{$-1.01$\mHz}
\setsymbol{LFI:slope:LFI23:Rad:S:units}{$-0.95$\mHz}
\setsymbol{LFI:slope:LFI24:Rad:S:units}{$-0.73$\mHz}
\setsymbol{LFI:slope:LFI25:Rad:S:units}{$-1.16$\mHz}
\setsymbol{LFI:slope:LFI26:Rad:S:units}{$-0.79$\mHz}
\setsymbol{LFI:slope:LFI27:Rad:S:units}{$-0.82$\mHz}
\setsymbol{LFI:slope:LFI28:Rad:S:units}{$-0.91$\mHz}

\setsymbol{LFI:slope:LFI18:Rad:M}{$-1.04$}
\setsymbol{LFI:slope:LFI19:Rad:M}{$-1.09$}
\setsymbol{LFI:slope:LFI20:Rad:M}{$-0.69$}
\setsymbol{LFI:slope:LFI21:Rad:M}{$-1.56$}
\setsymbol{LFI:slope:LFI22:Rad:M}{$-1.01$}
\setsymbol{LFI:slope:LFI23:Rad:M}{$-0.96$}
\setsymbol{LFI:slope:LFI24:Rad:M}{$-0.83$}
\setsymbol{LFI:slope:LFI25:Rad:M}{$-0.91$}
\setsymbol{LFI:slope:LFI26:Rad:M}{$-0.95$}
\setsymbol{LFI:slope:LFI27:Rad:M}{$-0.87$}
\setsymbol{LFI:slope:LFI28:Rad:M}{$-0.88$}
\setsymbol{LFI:slope:LFI18:Rad:S}{$-1.15$}
\setsymbol{LFI:slope:LFI19:Rad:S}{$-1.00$}
\setsymbol{LFI:slope:LFI20:Rad:S}{$-0.95$}
\setsymbol{LFI:slope:LFI21:Rad:S}{$-0.92$}
\setsymbol{LFI:slope:LFI22:Rad:S}{$-1.01$}
\setsymbol{LFI:slope:LFI23:Rad:S}{$-0.95$}
\setsymbol{LFI:slope:LFI24:Rad:S}{$-0.73$}
\setsymbol{LFI:slope:LFI25:Rad:S}{$-1.16$}
\setsymbol{LFI:slope:LFI26:Rad:S}{$-0.79$}
\setsymbol{LFI:slope:LFI27:Rad:S}{$-0.82$}
\setsymbol{LFI:slope:LFI28:Rad:S}{$-0.91$}

\setsymbol{LFI:FWHM:70GHz:units}{13\parcm01}
\setsymbol{LFI:FWHM:44GHz:units}{27\parcm92}
\setsymbol{LFI:FWHM:30GHz:units}{32\parcm65}

\setsymbol{LFI:FWHM:70GHz}{13.01}
\setsymbol{LFI:FWHM:44GHz}{27.92}
\setsymbol{LFI:FWHM:30GHz}{32.65}

\setsymbol{LFI:FWHM:LFI18:units}{13\parcm39}
\setsymbol{LFI:FWHM:LFI19:units}{13\parcm01}
\setsymbol{LFI:FWHM:LFI20:units}{12\parcm75}
\setsymbol{LFI:FWHM:LFI21:units}{12\parcm74}
\setsymbol{LFI:FWHM:LFI22:units}{12\parcm87}
\setsymbol{LFI:FWHM:LFI23:units}{13\parcm27}
\setsymbol{LFI:FWHM:LFI24:units}{22\parcm98}
\setsymbol{LFI:FWHM:LFI25:units}{30\parcm46}
\setsymbol{LFI:FWHM:LFI26:units}{30\parcm31}
\setsymbol{LFI:FWHM:LFI27:units}{32\parcm65}
\setsymbol{LFI:FWHM:LFI28:units}{32\parcm66}

\setsymbol{LFI:FWHM:LFI18}{13.39}
\setsymbol{LFI:FWHM:LFI19}{13.01}
\setsymbol{LFI:FWHM:LFI20}{12.75}
\setsymbol{LFI:FWHM:LFI21}{12.74}
\setsymbol{LFI:FWHM:LFI22}{12.87}
\setsymbol{LFI:FWHM:LFI23}{13.27}
\setsymbol{LFI:FWHM:LFI24}{22.98}
\setsymbol{LFI:FWHM:LFI25}{30.46}
\setsymbol{LFI:FWHM:LFI26}{30.31}
\setsymbol{LFI:FWHM:LFI27}{32.65}
\setsymbol{LFI:FWHM:LFI28}{32.66}

\setsymbol{LFI:FWHM:uncertainty:LFI18:units}{0.170\arcm}
\setsymbol{LFI:FWHM:uncertainty:LFI19:units}{0.174\arcm}
\setsymbol{LFI:FWHM:uncertainty:LFI20:units}{0.170\arcm}
\setsymbol{LFI:FWHM:uncertainty:LFI21:units}{0.156\arcm}
\setsymbol{LFI:FWHM:uncertainty:LFI22:units}{0.164\arcm}
\setsymbol{LFI:FWHM:uncertainty:LFI23:units}{0.171\arcm}
\setsymbol{LFI:FWHM:uncertainty:LFI24:units}{0.652\arcm}
\setsymbol{LFI:FWHM:uncertainty:LFI25:units}{1.075\arcm}
\setsymbol{LFI:FWHM:uncertainty:LFI26:units}{1.131\arcm}
\setsymbol{LFI:FWHM:uncertainty:LFI27:units}{1.266\arcm}
\setsymbol{LFI:FWHM:uncertainty:LFI28:units}{1.287\arcm}

\setsymbol{LFI:FWHM:uncertainty:LFI18}{0.170}
\setsymbol{LFI:FWHM:uncertainty:LFI19}{0.174}
\setsymbol{LFI:FWHM:uncertainty:LFI20}{0.170}
\setsymbol{LFI:FWHM:uncertainty:LFI21}{0.156}
\setsymbol{LFI:FWHM:uncertainty:LFI22}{0.164}
\setsymbol{LFI:FWHM:uncertainty:LFI23}{0.171}
\setsymbol{LFI:FWHM:uncertainty:LFI24}{0.652}
\setsymbol{LFI:FWHM:uncertainty:LFI25}{1.075}
\setsymbol{LFI:FWHM:uncertainty:LFI26}{1.131}
\setsymbol{LFI:FWHM:uncertainty:LFI27}{1.266}
\setsymbol{LFI:FWHM:uncertainty:LFI28}{1.287}

\setsymbol{HFI:center:frequency:100GHz:units}{100\,GHz}
\setsymbol{HFI:center:frequency:143GHz:units}{143\,GHz}
\setsymbol{HFI:center:frequency:217GHz:units}{217\,GHz}
\setsymbol{HFI:center:frequency:353GHz:units}{353\,GHz}
\setsymbol{HFI:center:frequency:545GHz:units}{545\,GHz}
\setsymbol{HFI:center:frequency:857GHz:units}{857\,GHz}

\setsymbol{HFI:center:frequency:100GHz}{100}
\setsymbol{HFI:center:frequency:143GHz}{143}
\setsymbol{HFI:center:frequency:217GHz}{217}
\setsymbol{HFI:center:frequency:353GHz}{353}
\setsymbol{HFI:center:frequency:545GHz}{545}
\setsymbol{HFI:center:frequency:857GHz}{857}

\setsymbol{HFI:Ndetectors:100GHz}{8}
\setsymbol{HFI:Ndetectors:143GHz}{11}
\setsymbol{HFI:Ndetectors:217GHz}{12}
\setsymbol{HFI:Ndetectors:353GHz}{12}
\setsymbol{HFI:Ndetectors:545GHz}{3}
\setsymbol{HFI:Ndetectors:857GHz}{4}

\setsymbol{HFI:FWHM:Maps:100GHz:units}{9\parcm88}
\setsymbol{HFI:FWHM:Maps:143GHz:units}{7\parcm18}
\setsymbol{HFI:FWHM:Maps:217GHz:units}{4\parcm87}
\setsymbol{HFI:FWHM:Maps:353GHz:units}{4\parcm65}
\setsymbol{HFI:FWHM:Maps:545GHz:units}{4\parcm72}
\setsymbol{HFI:FWHM:Maps:857GHz:units}{4\parcm39}
\setsymbol{HFI:FWHM:Maps:100GHz}{9.88}
\setsymbol{HFI:FWHM:Maps:143GHz}{7.18}
\setsymbol{HFI:FWHM:Maps:217GHz}{4.87}
\setsymbol{HFI:FWHM:Maps:353GHz}{4.65}
\setsymbol{HFI:FWHM:Maps:545GHz}{4.72}
\setsymbol{HFI:FWHM:Maps:857GHz}{4.39}

\setsymbol{HFI:beam:ellipticity:Maps:100GHz}{1.15}
\setsymbol{HFI:beam:ellipticity:Maps:143GHz}{1.01}
\setsymbol{HFI:beam:ellipticity:Maps:217GHz}{1.06}
\setsymbol{HFI:beam:ellipticity:Maps:353GHz}{1.05}
\setsymbol{HFI:beam:ellipticity:Maps:545GHz}{1.14}
\setsymbol{HFI:beam:ellipticity:Maps:857GHz}{1.19}

\setsymbol{HFI:FWHM:Mars:100GHz:units}{9\parcm37}
\setsymbol{HFI:FWHM:Mars:143GHz:units}{7\parcm04}
\setsymbol{HFI:FWHM:Mars:217GHz:units}{4\parcm68}
\setsymbol{HFI:FWHM:Mars:353GHz:units}{4\parcm43}
\setsymbol{HFI:FWHM:Mars:545GHz:units}{3\parcm80}
\setsymbol{HFI:FWHM:Mars:857GHz:units}{3\parcm67}

\setsymbol{HFI:FWHM:Mars:100GHz}{9.37}
\setsymbol{HFI:FWHM:Mars:143GHz}{7.04}
\setsymbol{HFI:FWHM:Mars:217GHz}{4.68}
\setsymbol{HFI:FWHM:Mars:353GHz}{4.43}
\setsymbol{HFI:FWHM:Mars:545GHz}{3.80}
\setsymbol{HFI:FWHM:Mars:857GHz}{3.67}

\setsymbol{HFI:beam:ellipticity:Mars:100GHz}{1.18}
\setsymbol{HFI:beam:ellipticity:Mars:143GHz}{1.03}
\setsymbol{HFI:beam:ellipticity:Mars:217GHz}{1.14}
\setsymbol{HFI:beam:ellipticity:Mars:353GHz}{1.09}
\setsymbol{HFI:beam:ellipticity:Mars:545GHz}{1.25}
\setsymbol{HFI:beam:ellipticity:Mars:857GHz}{1.03}

\setsymbol{HFI:CMB:relative:calibration:100GHz}{$\lsim 1\%$}
\setsymbol{HFI:CMB:relative:calibration:143GHz}{$\lsim 1\%$}
\setsymbol{HFI:CMB:relative:calibration:217GHz}{$\lsim 1\%$}
\setsymbol{HFI:CMB:relative:calibration:353GHz}{$\lsim 1\%$}
\setsymbol{HFI:CMB:relative:calibration:545GHz}{}
\setsymbol{HFI:CMB:relative:calibration:857GHz}{}

\setsymbol{HFI:CMB:absolute:calibration:100GHz}{$\lsim 2\%$}
\setsymbol{HFI:CMB:absolute:calibration:143GHz}{$\lsim 2\%$}
\setsymbol{HFI:CMB:absolute:calibration:217GHz}{$\lsim 2\%$}
\setsymbol{HFI:CMB:absolute:calibration:353GHz}{$\lsim 2\%$}
\setsymbol{HFI:CMB:absolute:calibration:545GHz}{}
\setsymbol{HFI:CMB:absolute:calibration:857GHz}{}

\setsymbol{HFI:FIRAS:gain:calibration:accuracy:statistical:100GHz}{}
\setsymbol{HFI:FIRAS:gain:calibration:accuracy:statistical:143GHz}{}
\setsymbol{HFI:FIRAS:gain:calibration:accuracy:statistical:217GHz}{}
\setsymbol{HFI:FIRAS:gain:calibration:accuracy:statistical:353GHz}{2.5\%}
\setsymbol{HFI:FIRAS:gain:calibration:accuracy:statistical:545GHz}{1\%}
\setsymbol{HFI:FIRAS:gain:calibration:accuracy:statistical:857GHz}{0.5\%}

\setsymbol{HFI:FIRAS:gain:calibration:accuracy:systematic:100GHz}{}
\setsymbol{HFI:FIRAS:gain:calibration:accuracy:systematic:143GHz}{}
\setsymbol{HFI:FIRAS:gain:calibration:accuracy:systematic:217GHz}{}
\setsymbol{HFI:FIRAS:gain:calibration:accuracy:systematic:353GHz}{}
\setsymbol{HFI:FIRAS:gain:calibration:accuracy:systematic:545GHz}{7\%}
\setsymbol{HFI:FIRAS:gain:calibration:accuracy:systematic:857GHz}{7\%}

\setsymbol{HFI:FIRAS:zero:point:accuracy:100GHz:units}{0.8\MJysr}
\setsymbol{HFI:FIRAS:zero:point:accuracy:143GHz:units}{}
\setsymbol{HFI:FIRAS:zero:point:accuracy:217GHz:units}{}
\setsymbol{HFI:FIRAS:zero:point:accuracy:353GHz:units}{1.4\MJysr}
\setsymbol{HFI:FIRAS:zero:point:accuracy:545GHz:units}{2.2\MJysr}
\setsymbol{HFI:FIRAS:zero:point:accuracy:857GHz:units}{1.7\MJysr}

\setsymbol{HFI:FIRAS:zero:point:accuracy:100GHz}{0.8}
\setsymbol{HFI:FIRAS:zero:point:accuracy:143GHz}{}
\setsymbol{HFI:FIRAS:zero:point:accuracy:217GHz}{}
\setsymbol{HFI:FIRAS:zero:point:accuracy:353GHz}{1.4}
\setsymbol{HFI:FIRAS:zero:point:accuracy:545GHz}{2.2}
\setsymbol{HFI:FIRAS:zero:point:accuracy:857GHz}{1.7}

\setsymbol{HFI:unit:conversion:100GHz:units}{0.2415\MJysrmK}
\setsymbol{HFI:unit:conversion:143GHz:units}{0.3694\MJysrmK}
\setsymbol{HFI:unit:conversion:217GHz:units}{0.4811\MJysrmK}
\setsymbol{HFI:unit:conversion:353GHz:units}{0.2883\MJysrmK}
\setsymbol{HFI:unit:conversion:545GHz:units}{0.05826\MJysrmK}
\setsymbol{HFI:unit:conversion:857GHz:units}{0.002238\MJysrmK}

\setsymbol{HFI:unit:conversion:100GHz}{0.2415}
\setsymbol{HFI:unit:conversion:143GHz}{0.3694}
\setsymbol{HFI:unit:conversion:217GHz}{0.4811}
\setsymbol{HFI:unit:conversion:353GHz}{0.2883}
\setsymbol{HFI:unit:conversion:545GHz}{0.05826}
\setsymbol{HFI:unit:conversion:857GHz}{0.002238}

\setsymbol{HFI:colour:correction:alpha=-2:V1.01:100GHz}{0.9893}
\setsymbol{HFI:colour:correction:alpha=-2:V1.01:143GHz}{0.9759}
\setsymbol{HFI:colour:correction:alpha=-2:V1.01:217GHz}{1.0007}
\setsymbol{HFI:colour:correction:alpha=-2:V1.01:353GHz}{1.0028}
\setsymbol{HFI:colour:correction:alpha=-2:V1.01:545GHz}{1.0019}
\setsymbol{HFI:colour:correction:alpha=-2:V1.01:857GHz}{0.9889}

\setsymbol{HFI:colour:correction:alpha=0:V1.01:100GHz}{1.0008}
\setsymbol{HFI:colour:correction:alpha=0:V1.01:143GHz}{1.0148}
\setsymbol{HFI:colour:correction:alpha=0:V1.01:217GHz}{0.9909}
\setsymbol{HFI:colour:correction:alpha=0:V1.01:353GHz}{0.9888}
\setsymbol{HFI:colour:correction:alpha=0:V1.01:545GHz}{0.9878}
\setsymbol{HFI:colour:correction:alpha=0:V1.01:857GHz}{1.0014}

\newcommand{\herschel}{{\it Herschel }}
\newcommand{\cobe}{{\it COBE }}
\newcommand{\dirbe}{{\it DIRBE }}

\newcommand{\wmap}{{\it WMAP }}
\newcommand{\firas}{{\it FIRAS }}
\newcommand{\iras}{{\it IRAS }}
\newcommand{\iris}{{\it IRIS }}
\newcommand{\fermi}{{\it FERMI }}
\newcommand{\hfi}{{\it HFI }}
\newcommand{\lfi}{{\it LFI }}
\newcommand{\dht}{{\it DHT }}
\newcommand{\nanten}{{\it NANTEN }}
\newcommand{\lab}{{\it LAB }}
\newcommand{\dameHL}{{\it Dame }}

\newcommand{\DG}{dark gas}

\newcommand{\XCO}{X_{\rm CO}} 
\newcommand{\NH}{N_{\rm H}}
\newcommand{\NHHI}{N_{\rm H}^{\rm HI}}
\newcommand{\HI}{{\rm H_I}}
\newcommand{\twelveCO}{{\rm ^{12}CO}}
\newcommand{\Td}{T_{\rm D}} 
\newcommand{\NHHIhole}{\rm N_H^{hole}}
\newcommand{\TAUNHREF}{\rm \left(\frac{\taudust}{N_{H}}\right)^{ref}}
\newcommand{\TAUNHCO}{\rm \left(\frac{\taudust}{N_{H}}\right)^{CO}}
\newcommand{\AVHIHtwo}{\rm A_V^{DG}}
\newcommand{\AVHtwoCO}{\rm A_v^{H_2/CO}}

\newcommand{\NHTOT}{\rm N_H^{tot}}
\newcommand{\taumodel}{\rm \tau_M}
\newcommand{\taudust}{\rm \tau_D}

\newcommand{\whi}{\rm W_{HI}}
\newcommand{\wco}{\rm W_{CO}}
\newcommand{\lII}{$\rm l_{II}$}
\newcommand{\bII}{$\rm b_{II}$}
\newcommand{\Av}{\rm A_V}
\newcommand{\Hdeux}{\rm H_2}
\newcommand{\sigmaII}{\sigma_{\rm II}}
\newcommand{\offset}{I_\nu^0}

\newcommand{\XCOUNIT}{\rm \,H_{2}cm^{-2}/(K\,km\,s^{-1})} 
\newcommand{\Kkms}{\rm \,Kkms^{-1}}
\newcommand{\NHUNIT}{\rm \,Hcm^{-2}} 
\newcommand{\mic}{\,{\rm \mu m} } 

\newcommand{\NHHIholemaxv}{\rm 2.0\times10^{19}\NHUNIT}
\newcommand{\NHHIholev}{\rm 1.75\times10^{19}\NHUNIT}
\newcommand{\NHHIrefmaxv}{\rm 1.2\times10^{21}\NHUNIT}

\newcommand{\IRASfourfreq}{2998}
\newcommand{\HFIonefreq}{\getsymbol{HFI:center:frequency:857GHz}}
\newcommand{\HFItwofreq}{\getsymbol{HFI:center:frequency:545GHz}}
\newcommand{\HFIthreefreq}{\getsymbol{HFI:center:frequency:353GHz}}
\newcommand{\HFIfourfreq}{\getsymbol{HFI:center:frequency:217GHz}}
\newcommand{\HFIfivefreq}{\getsymbol{HFI:center:frequency:143GHz}}
\newcommand{\HFIsixfreq}{\getsymbol{HFI:center:frequency:100GHz}}
\newcommand{\LFIonefreq}{\getsymbol{LFI:center:frequency:70GHz}}
\newcommand{\LFItwofreq}{\getsymbol{LFI:center:frequency:44GHz}}
\newcommand{\LFIthreefreq}{\getsymbol{LFI:center:frequency:30GHz}}
\newcommand{\DIRBEheightfreq}{2998}
\newcommand{\DIRBEninefreq}{2141}
\newcommand{\DIRBEtenfreq}{1249}
\newcommand{\WMAPonefreq}{93.7}
\newcommand{\WMAPtwofreq}{61.2}
\newcommand{\WMAPthreefreq}{41.1}
\newcommand{\WMAPfourfreq}{32.9}
\newcommand{\WMAPfivefreq}{23.1}
\newcommand{\IRASfourreso}{4.30}
\newcommand{\HFIonereso}{\getsymbol{HFI:FWHM:Mars:857GHz}}
\newcommand{\HFItworeso}{\getsymbol{HFI:FWHM:Mars:545GHz}}
\newcommand{\HFIthreereso}{\getsymbol{HFI:FWHM:Mars:353GHz}}
\newcommand{\HFIfourreso}{\getsymbol{HFI:FWHM:Mars:217GHz}}
\newcommand{\HFIfivereso}{\getsymbol{HFI:FWHM:Mars:143GHz}}
\newcommand{\HFIsixreso}{\getsymbol{HFI:FWHM:Mars:100GHz}}

\newcommand{\DHTonereso}{8.8}
\newcommand{\DAMEHLonereso}{8.4}
\newcommand{\NANTENonereso}{2.6}
\newcommand{\LABonereso}{36.0}
\newcommand{\IRASfourabserr}{13.6}
\newcommand{\HFIoneabserr}{\getsymbol{HFI:FIRAS:gain:calibration:accuracy:systematic:857GHz}}
\newcommand{\HFItwoabserr}{\getsymbol{HFI:FIRAS:gain:calibration:accuracy:systematic:545GHz}}
\newcommand{\HFIthreeabserr}{\getsymbol{HFI:CMB:absolute:calibration:353GHz}}
\newcommand{\HFIfourabserr}{\getsymbol{HFI:CMB:absolute:calibration:217GHz}}

\newcommand{\HFIsixabserr}{\getsymbol{HFI:CMB:absolute:calibration:100GHz}}

\newcommand{\DHToneabserr}{24.0}
\newcommand{\DAMEHLoneabserr}{24.0}
\newcommand{\NANTENoneabserr}{10.0}
\newcommand{\LABoneabserr}{10.0}

\newcommand{\Healpix}{\rm HEALPix}

\newcommand{\AVGXCO}{2.54}              
\newcommand{\AVGXCOERR}{0.13}           
\newcommand{\AVGAVHIHtwo}{0.4}          
\newcommand{\AVGAVHIHtwoERR}{0.029}     
\newcommand{\AVGNHHIHtwo}{8.0}          
\newcommand{\AVGNHHIHtwoERR}{0.58}      
\newcommand{\AVGMXMH}{28\%}             
\newcommand{\AVGMXMHERR}{3\%}             
\newcommand{\AVGMXMCO}{118\%}           
\newcommand{\AVGMXMCOERR}{12\%}           

\begin{document}

\author{\small
Planck Collaboration:
P.~A.~R.~Ade\inst{71}
\and
N.~Aghanim\inst{47}
\and
M.~Arnaud\inst{58}
\and
M.~Ashdown\inst{56, 77}
\and
J.~Aumont\inst{47}
\and
C.~Baccigalupi\inst{69}
\and
A.~Balbi\inst{29}
\and
A.~J.~Banday\inst{75, 6, 63}
\and
R.~B.~Barreiro\inst{53}
\and
J.~G.~Bartlett\inst{3, 54}
% ...
}
\institute{\small
Aalto University Mets\"{a}hovi Radio Observatory, Mets\"{a}hovintie 114, FIN-02540 Kylm\"{a}l\"{a}, Finland\\
\and
Agenzia Spaziale Italiana Science Data Center, c/o ESRIN, via Galileo Galilei, Frascati, Italy\\
\and
Astroparticule et Cosmologie, CNRS (UMR7164), Universit\'{e} Denis Diderot Paris 7, B\^{a}timent Condorcet, 10 rue A. Domon et L\'{e}onie Duquet, Paris, France\\
\and
Atacama Large Millimeter/submillimeter Array, ALMA Santiago Central Offices Alonso de Cordova 3107, Vitacura, Casilla 763 0355, Santiago, Chile\\
\and
CITA, University of Toronto, 60 St. George St., Toronto, ON M5S 3H8, Canada\\
\and
CNRS, IRAP, 9 Av. colonel Roche, BP 44346, F-31028 Toulouse cedex 4, France\\
\and
California Institute of Technology, Pasadena, California, U.S.A.\\
\and
Centre of Mathematics for Applications, University of Oslo, Blindern, Oslo, Norway\\
\and
DAMTP, Centre for Mathematical Sciences, Wilberforce Road, Cambridge CB3 0WA, U.K.\\
\and
DSM/Irfu/SPP, CEA-Saclay, F-91191 Gif-sur-Yvette Cedex, France\\
\and
DTU Space, National Space Institute, Juliane Mariesvej 30, Copenhagen, Denmark\\
\and
Departamento de F\'{\i}sica, Universidad de Oviedo, Avda. Calvo Sotelo s/n, Oviedo, Spain\\
\and
Department of Astronomy and Astrophysics, University of Toronto, 50 Saint George Street, Toronto, Ontario, Canada\\
\and
Department of Astronomy and Earth Sciences, Tokyo Gakugei University, Koganei, Tokyo 184-8501, Japan\\
\and
Department of Physical Science, Graduate School of Science, Osaka Prefecture University, 1-1 Gakuen-cho, Naka-ku, Sakai, Osaka 599-8531, Japan\\
\and
Department of Physics \& Astronomy, University of British Columbia, 6224 Agricultural Road, Vancouver, British Columbia, Canada\\
\and
Department of Physics, Gustaf H\"{a}llstr\"{o}min katu 2a, University of Helsinki, Helsinki, Finland\\
\and
Department of Physics, Nagoya University, Chikusa-ku, Nagoya, 464-8602, Japan\\
\and
Department of Physics, Princeton University, Princeton, New Jersey, U.S.A.\\
\and
Department of Physics, Purdue University, 525 Northwestern Avenue, West Lafayette, Indiana, U.S.A.\\
\and
Department of Physics, University of California, Berkeley, California, U.S.A.\\
\and
Department of Physics, University of California, One Shields Avenue, Davis, California, U.S.A.\\
\and
Department of Physics, University of California, Santa Barbara, California, U.S.A.\\
\and
Department of Physics, University of Illinois at Urbana-Champaign, 1110 West Green Street, Urbana, Illinois, U.S.A.\\
\and
Dipartimento di Fisica G. Galilei, Universit\`{a} degli Studi di Padova, via Marzolo 8, 35131 Padova, Italy\\
\and
Dipartimento di Fisica, Universit\`{a} La Sapienza, P. le A. Moro 2, Roma, Italy\\
\and
Dipartimento di Fisica, Universit\`{a} degli Studi di Milano, Via Celoria, 16, Milano, Italy\\
\and
Dipartimento di Fisica, Universit\`{a} degli Studi di Trieste, via A. Valerio 2, Trieste, Italy\\
\and
Dipartimento di Fisica, Universit\`{a} di Roma Tor Vergata, Via della Ricerca Scientifica, 1, Roma, Italy\\
\and
Discovery Center, Niels Bohr Institute, Blegdamsvej 17, Copenhagen, Denmark\\
\and
Dpto. Astrof\'{i}sica, Universidad de La Laguna (ULL), E-38206 La Laguna, Tenerife, Spain\\
\and
European Southern Observatory, ESO Vitacura, Alonso de Cordova 3107, Vitacura, Casilla 19001, Santiago, Chile\\
\and
European Space Agency, ESAC, Planck Science Office, Camino bajo del Castillo, s/n, Urbanizaci\'{o}n Villafranca del Castillo, Villanueva de la Ca\~{n}ada, Madrid, Spain\\
\and
European Space Agency, ESTEC, Keplerlaan 1, 2201 AZ Noordwijk, The Netherlands\\
\and
Harvard-Smithsonian Center for Astrophysics, 60 Garden Street, Cambridge, MA 02138, U.S.A.\\
\and
Helsinki Institute of Physics, Gustaf H\"{a}llstr\"{o}min katu 2, University of Helsinki, Helsinki, Finland\\
\and
INAF - Osservatorio Astrofisico di Catania, Via S. Sofia 78, Catania, Italy\\
\and
INAF - Osservatorio Astronomico di Padova, Vicolo dell'Osservatorio 5, Padova, Italy\\
\and
INAF - Osservatorio Astronomico di Roma, via di Frascati 33, Monte Porzio Catone, Italy\\
\and
INAF - Osservatorio Astronomico di Trieste, Via G.B. Tiepolo 11, Trieste, Italy\\
\and
INAF/IASF Bologna, Via Gobetti 101, Bologna, Italy\\
\and
INAF/IASF Milano, Via E. Bassini 15, Milano, Italy\\
\and
INRIA, Laboratoire de Recherche en Informatique, Universit\'{e} Paris-Sud 11, B\^{a}timent 490, 91405 Orsay Cedex, France\\
\and
IPAG: Institut de Plan\'{e}tologie et d'Astrophysique de Grenoble, Universit\'{e} Joseph Fourier, Grenoble 1 / CNRS-INSU, UMR 5274, Grenoble, F-38041, France\\
\and
Imperial College London, Astrophysics group, Blackett Laboratory, Prince Consort Road, London, SW7 2AZ, U.K.\\
\and
Infrared Processing and Analysis Center, California Institute of Technology, Pasadena, CA 91125, U.S.A.\\
\and
Institut d'Astrophysique Spatiale, CNRS (UMR8617) Universit\'{e} Paris-Sud 11, B\^{a}timent 121, Orsay, France\\
\and
Institut d'Astrophysique de Paris, CNRS UMR7095, Universit\'{e} Pierre \& Marie Curie, 98 bis boulevard Arago, Paris, France\\
\and
Institut de Ci\`{e}ncies de l'Espai, CSIC/IEEC, Facultat de Ci\`{e}ncies, Campus UAB, Torre C5 par-2, Bellaterra 08193, Spain\\
\and
Institute of Astronomy and Astrophysics, Academia Sinica, Taipei, Taiwan\\
\and
Institute of Theoretical Astrophysics, University of Oslo, Blindern, Oslo, Norway\\
\and
Instituto de Astrof\'{\i}sica de Canarias, C/V\'{\i}a L\'{a}ctea s/n, La Laguna, Tenerife, Spain\\
\and
Instituto de F\'{\i}sica de Cantabria (CSIC-Universidad de Cantabria), Avda. de los Castros s/n, Santander, Spain\\
\and
Jet Propulsion Laboratory, California Institute of Technology, 4800 Oak Grove Drive, Pasadena, California, U.S.A.\\
\and
Jodrell Bank Centre for Astrophysics, Alan Turing Building, School of Physics and Astronomy, The University of Manchester, Oxford Road, Manchester, M13 9PL, U.K.\\
\and
Kavli Institute for Cosmology Cambridge, Madingley Road, Cambridge, CB3 0HA, U.K.\\
\and
LERMA, CNRS, Observatoire de Paris, 61 Avenue de l'Observatoire, Paris, France\\
\and
Laboratoire AIM, IRFU/Service d'Astrophysique - CEA/DSM - CNRS - Universit\'{e} Paris Diderot, B\^{a}t. 709, CEA-Saclay, F-91191 Gif-sur-Yvette Cedex, France\\
\and
Laboratoire Traitement et Communication de l'Information, CNRS (UMR 5141) and T\'{e}l\'{e}com ParisTech, 46 rue Barrault F-75634 Paris Cedex 13, France\\
\and
Laboratoire de Physique Subatomique et de Cosmologie, CNRS, Universit\'{e} Joseph Fourier Grenoble I, 53 rue des Martyrs, Grenoble, France\\
\and
Laboratoire de l'Acc\'{e}l\'{e}rateur Lin\'{e}aire, Universit\'{e} Paris-Sud 11, CNRS/IN2P3, Orsay, France\\
\and
Lawrence Berkeley National Laboratory, Berkeley, California, U.S.A.\\
\and
Max-Planck-Institut f\"{u}r Astrophysik, Karl-Schwarzschild-Str. 1, 85741 Garching, Germany\\
\and
MilliLab, VTT Technical Research Centre of Finland, Tietotie 3, Espoo, Finland\\
\and
National University of Ireland, Department of Experimental Physics, Maynooth, Co. Kildare, Ireland\\
\and
Niels Bohr Institute, Blegdamsvej 17, Copenhagen, Denmark\\
\and
Observational Cosmology, Mail Stop 367-17, California Institute of Technology, Pasadena, CA, 91125, U.S.A.\\
\and
Optical Science Laboratory, University College London, Gower Street, London, U.K.\\
\and
SISSA, Astrophysics Sector, via Bonomea 265, 34136, Trieste, Italy\\
\and
SUPA, Institute for Astronomy, University of Edinburgh, Royal Observatory, Blackford Hill, Edinburgh EH9 3HJ, U.K.\\
\and
School of Physics and Astronomy, Cardiff University, Queens Buildings, The Parade, Cardiff, CF24 3AA, U.K.\\
\and
Space Sciences Laboratory, University of California, Berkeley, California, U.S.A.\\
\and
Spitzer Science Center, 1200 E. California Blvd., Pasadena, California, U.S.A.\\
\and
Stanford University, Dept of Physics, Varian Physics Bldg, 382 Via Pueblo Mall, Stanford, California, U.S.A.\\
\and
Universit\'{e} de Toulouse, UPS-OMP, IRAP, F-31028 Toulouse cedex 4, France\\
\and
Universities Space Research Association, Stratospheric Observatory for Infrared Astronomy, MS 211-3, Moffett Field, CA 94035, U.S.A.\\
\and
University of Cambridge, Cavendish Laboratory, Astrophysics group, J J Thomson Avenue, Cambridge, U.K.\\
% ...
}


\title{\textit{Planck} Early Results: All sky temperature and dust
optical depth from \textit{Planck} and \textit{IRAS}: Constraints on the ``dark gas'' in
our Galaxy}

\titlerunning{Constraints on the {\DG} in our galaxy}
\authorrunning{{\Planck} collaboration}


\abstract{
An all sky map of the apparent temperature and optical depth of
thermal dust emission is constructed using the {\Planck}-{\hfi}
($350\mic$ to 2 mm) and {\iras} ($100\mic$) data. The optical depth
maps are correlated with tracers of the atomic (\ion{H}{i}) and
molecular gas traced by CO. The correlation with the column density of
observed gas is linear in the lowest column density regions at high
Galactic latitudes. At high $\NH$, the correlation is consistent with
that of the lowest $\NH$, for a given choice of the CO-to-$\Hdeux$
conversion factor.  In the intermediate $\NH$ range, a departure from
linearity is observed, with the dust optical depth in excess of the
correlation.  This excess emission is attributed to thermal emission
by dust associated with a {\DG} phase, undetected in the available
\ion{H}{i} and CO surveys. The 2D spatial distribution of the {\DG} in
the solar neighbourhood ($\rm |$\bII$|>10\degr$) is shown to extend
around known molecular regions traced by CO.

The average dust emissivity in the \ion{H}{i} phase in the solar
neighbourhood is found to be $\rm
\taudust/\NHTOT=5.2\times10^{-26}\,cm^2$ at {\HFIonefreq}{\GHz}. It
follows roughly a power law distribution with a spectral index $\rm
\beta=1.8$ all the way down to 3 mm, although the SED flattens
slightly in the millimetre.  Taking into account the spectral shape of
the dust optical depth, the emissivity is consistent with previous
values derived from {\firas} measurements at high latitudes within
10\%.  The threshold for the existence of the {\DG} is found at $\rm
\NHTOT=(\AVGNHHIHtwo\pm\AVGNHHIHtwoERR)\times10^{20}\,\NHUNIT$ ($\rm
\Av=\AVGAVHIHtwo\,mag$). Assuming the same high frequency emissivity
for the dust in the atomic and the molecular phases leads to an
average $\XCO=(\AVGXCO\pm\AVGXCOERR)\times10^{20}\XCOUNIT$. The mass
of {\DG} is found to be $\rm \AVGMXMH$ of the atomic gas and $\rm
\AVGMXMCO$ of the CO emitting gas in the solar neighbourhood. The
Galactic latitude distribution shows that its mass fraction is
relatively constant down to a few degrees from the Galactic plane.

A possible explanation for the {\DG} lies in a dark molecular phase,
where $\Hdeux$ survives photodissociation but CO does not. The
observed transition for the onset of this phase in the solar
neighbourhood ($\rm \Av=\AVGAVHIHtwo\,mag$) appears consistent with
recent theoretical predictions. It is also possible that up to half of
the {\DG} could be in atomic form, due to optical depth effects in the
\ion{H}{i} measurements.
}


\maketitle

\section{Introduction}

The matter that forms stars, that is left over after star formation,
or that has never experienced star formation comprises the
interstellar medium (ISM). The life-cycle and the duration of the
various observable phases remains largely unknown, because the nature
of the diffuse interstellar medium is difficult to discern, owing to
its low temperatures and large angular scales.

The distribution of diffuse interstellar gas, by which we mean gas not
in gravitationally--bound structures and not in the immediate vicinity
of active star-formation regions, has primarily been assessed using
the 21-cm hyperfine line of atomic hydrogen. That line is easily
excited by collisions and is optically thin for gas with temperature
$\rm T_K>50\,K$ and velocity dispersion $\rm \delta V>10\kms$ as long
as the column density is less than $\rm 9\times10^{21}$ cm$^{-2}$
\citep{Kulkarni1988}.  Such conditions are typical of the diffuse ISM
pervaded by the interstellar radiation field (ISRF), because
photoelectric heating from grain surfaces keeps the gas warm ($\rm T>50\,K$),
and observed velocity dispersions (presumably due to turbulence) are
typically $\rm >10\kms$. Based on the observed dust extinction per
unit column density, $\rm N(HI)/\Av=1.9\times 10^{21}$ cm$^{-2}$ mag$^{-1}$
\citep{Bohlin1978}, the upper column density for optically thin 21-cm
lines corresponds to visible extinctions $\rm \Av<4.7$. Thus the 21-cm
line is expected to trace diffuse, warm atomic gas
accutately throughout the diffuse ISM, except for lines of sight that are
visibly opaque or are particularly cold.

Molecular gas is typically traced by the 2.6-mm $^{12}$CO($J$=1$\rightarrow$0)
rotational line in emission, which, like the 21-cm \ion{H}{i}
line, can be easily excited because it involves energy levels that can
be obtained by collisions. The CO emission line, however, is
commonly optically thick, due to its high radiative transition
rate. In the limit where the lines are optically thick, the primary
information determining the amount of molecular gas in the beam is the
line width. If the material is gravitationally bound, then the virial
mass is measured and CO can be used as a tracer of molecular mass.  It
is common astronomical practice to consider the velocity-integrated CO
line intensity as measuring the molecular column density, with the
implicit assumption that the material is virialized and the mass of
the virialized structures is being measured.  In the diffuse ISM,
these conditions typically do not apply. On a physical scale of R
(measured in parsecs), interstellar material is only virialized if its
column density $\rm N>5.2 \times 10^{21} \delta V^2 R^{-1}$ cm$^{-2}$
where $\rm \delta V$ is the velocity dispersion (measured in $\kms$).
Thus the diffuse ISM is typically gravitationally unbound,
invalidating the usage of CO as a virial tracer of the molecular gas
mass, except in very compact regions or in regions that are visibly
opaque.  Although CO can emit in gas with low
density, the critical density required for collisional equilibrium is
of order $10^3$ cm$^{-3}$, which further complicates the usage of CO
as a tracer.  This again is not typical of the diffuse ISM.

To measure the amount and distribution of the molecular ISM, as well
as the cold atomic ISM, other tracers of the interstellar gas are
required. At least three tracers have been used in the past. These are
UV absorption in Werner bands of $\Hdeux$, infrared emission from
dust, and $\gamma$-ray emission from pion productiondue to cosmic-rays
colliding with interstellar nucleons. The UV absorption is
exceptionally sensitive to even very low $\Hdeux$ column densities of
$10^{17}$ cm$^{-2}$.  Using {\it Copernicus} \citep{Savage1977} and
FUSE data, atomic and molecular gas could be measured simultaneously
on the sightlines to UV-bright stars and some galaxies. A survey at
high Galactic latitudes with FUSE showed that the molecular fraction
of the ISM, $\rm f(\Hdeux)\equiv
2N(\Hdeux)/[2N(\Hdeux)+N(HI)]<10^{-3}$ for lines of sight with total
column density less than $\rm 10^{20}\,cm^{-2}$, but there is a
tremendous dispersion from $\rm 10^{-4}$ to $\rm 10^{-1}$ for
higher-column density lines of sight \citep{Wakker2006}. Since
UV-bright sources are preferentially found towards the
lowest-extinction sightlines, an accurate average $\rm f(\Hdeux)$ is
extremely difficult to determine from the stellar absorption
measurements. Along lines of sight toward AGNs behind diffuse
interstellar clouds, \cite{Gillmon2006} found molecular hydrogen
fractions of 1--30\% indicating significant molecular content even for
low-density clouds.

The dust column density has been used as a total gas column density
tracer, with the assumption that gas and dust are well mixed. The
possibility that dust traces the column density better than \ion{H}{i}
and CO was recognized soon after the first all-sky infrared survey by
{\iras}, which for the first time revealed the distribution of dust on
angular scales down to $5\arcm$. Molecular gas without CO was inferred
from comparing {\iras} $100\mic$ surface brightness to surveys of the
21-cm and 2.6-mm lines of \ion{H}{i} and CO on $9\arcm$ or degree
scale \cite{deVries1987,Heiles1988,Blitz1990}. At $3\arcm$ scale using
Arecibo, the cloud G236+39 was found to have significant infrared
emission unaccounted for by 21-cm or 2.6-mm lines, with a large
portion of the cloud being possibly $\Hdeux$ with CO emission
below detection threshold \citep{Reach1994}. \cite{Meyerdierks1996}
also detected IR emission surrounding the Polaris flare in excess of
what was expectated from the \ion{H}{i} and CO emission, which they
attributed to diffuse molecular gas.  The all sky far-infrared
observations by {\cobe}-{\dirbe} \citep{Hauser1998} made it possible
to survey the molecular gas not traced by \ion{H}{i} or CO at the
$1^\circ$ scale \citep{Reach1998}. This revealed numerous ``infrared
excess'' clouds, many of which were confirmed as molecular after
detection of faint CO with NANTEN \citep{Onishi2001}. Finally, there
are also indications of more dust emission than seen in nearby
external galaxies such as the Large Magellanic Cloud
\citep{Bernard2008,Roman-Duval2010} and the Small Magellanic Cloud
\citep{Leroy2007}. This suggests that large fractions of the gas
masses of these galaxies are not detected using standard gas tracers.

The $\gamma$-rays from the interstellar medium provide an independent
tracer of the total nucleon density. As was the case with the
dust column density, the $\gamma$-ray inferred nucleon column density
appears to show an extra component of the ISM not associated with
detected 21-cm or 2.6-mm emission; this extra emission was referred to
as "{\DG}" \citep[e.g.][]{Grenier2005,Abdo2010}, a term we will adopt in
this paper to describe interstellar material beyond what is traced by
\ion{H}{i} and CO emission.  \cite{Grenier2005} inferred {\DG} column
densities of order 50\% of the total column density toward locations
with little or beyound detection threshold CO emission, and general
consistency between infrared and $\gamma$-ray methods of detection.
Recent observations using {\fermi} have significantly advanced this
method, allowing $\gamma$-ray emission to be traced even by the
high-latitude diffuse ISM.  In the Cepheus, Cassiopeia, and Polaris
Flare clouds, the correlated excess of dust and $\gamma$ rays yields
{\DG} masses that range from 40 \% to 60 \% of the CO-bright molecular
mass \citep{Abdo2010}.

Theoretical work predicts a dark molecular gas layer in regions where
the balance between photodissociation and molecular formation allows
$\Hdeux$ to form in significant quantity while the gas-phase C remains
atomic or ionized \citep{Wolfire2010,Glover2010}.  In this paper we
describe new observations made with {\Planck} \footnote{\Planck\
(http://www.esa.int/\Planck ) is a project of the European Space
Agency (ESA) with instruments provided by two scientific consortia
funded by ESA member states (in particular the lead countries: France
and Italy) with contributions from NASA (USA), and telescope
reflectors provided in a collaboration between ESA and a scientific
consortium led and funded by Denmark.}  \citep{planck2011-1.1} that
trace the distribution of submillimeter emission at $350\mic$ and
longer wavelengths. In combination with observations up to $100\mic$
wavelength by {\iras} and {\cobe}-{\dirbe}, we are uniquely able to
trace the distribution of interstellar dust with temperatures down to
$\rm \sim 10\,K$. The surface brightness sensitivity of {\Planck}, in
particular on angular scales of $\rm 5^\prime$ to $\rm 7^\circ$, is
unprecedented.  Because we can measure the dust optical depth more
accurately by including the {\Planck} data, we can now reassess the
relationship between dust and gas, and relate it to previous infrared
and independent UV and $\gamma$-ray results, and compare it to
theoretical explanations to determine just how important the {\DG} is
for the evolution of the interstellar medium.

\section{Observations} 
\label{sec:obs}

\subsection{Planck data}
\label{sec:planckdata}

\begin{table}[tmb]
\begingroup
\newdimen\tblskip \tblskip=5pt
\caption{\label{tab:ancillary} Characteristics of the data used in this study}
\nointerlineskip
\footnotesize
\vskip -8pt
\setbox\tablebox=\vbox{
\newdimen\digitwidth
\setbox0=\hbox{\rm 0}
\digitwidth=\wd0
\catcode`*=\active
\def*{kern\digitwidth}
\newdimen\signwidth
\setbox0=\hbox{+}
\signwidth=\wd0
\catcode`!=\active
\def!{\kern\signwidth}
\halign to \hsize{\hbox to 0.1in{#\hfil}\tabskip=2.0em&
	\hfil#\hfil&
	\hfil#\hfil&
	\hfil#\hfil&
	\hfil#\hfil&
	\hfil#\hfil\tabskip 8pt\cr
\noalign{\vskip 3pt\hrule\vskip 1.5pt\hrule\vskip 5pt}
\omit\hfil Data   & $\lambda_{\rm ref}$ & $\nu_{\rm ref}$ & $\theta$ & $\sigmaII$ & $\sigma_{\rm abs}$ \cr
\omit\hfil  & [$\mic$]   & [{\GHz}]       & [arcmin]    & [MJy/sr] & [\%] \cr
\noalign{\vskip 4pt\hrule\vskip 6pt}
 \iras &    100.0    &   \IRASfourfreq   &  \IRASfourreso &      0.06$^\dag$$^\ddag$ &   \IRASfourabserr$^\ddag$ \cr
  \hfi &    349.8 &    \HFIonefreq    &  \HFIonereso   &      0.12$^\flat$       &    \HFIoneabserr \cr
  \hfi &    550.1 &    \HFItwofreq    &  \HFItworeso   &      0.12$^\flat$       &    \HFItwoabserr \cr
  \hfi &    849.3 &    \HFIthreefreq  &  \HFIthreereso &      0.08$^\flat$       &    \HFIthreeabserr \cr
  \hfi &   1381.5 &    \HFIfourfreq   &  \HFIfourreso  &      0.08$^\flat$       &    \HFIfourabserr \cr
  \hfi &   2096.4 &    \HFIfivefreq   &  \HFIfivereso  &      0.08$^\flat$       &    \HFIfourabserr \cr
  \hfi &   2997.9 &    \HFIsixfreq    &  \HFIsixreso   &      0.07$^\flat$       &    \HFIsixabserr \cr
\noalign{\vskip 3pt\hrule\vskip 4pt}
\omit\hfil Data   & line  & $\lambda_{\rm ref}$ & $\theta$  & $\sigma$ & $\sigma_{\rm abs}$ \cr
\omit\hfil  &   & [$\mic$] & [arcmin]  &     [$\Kkms$]   &       [\%] \cr
\noalign{\vskip 3pt\hrule\vskip 4pt}
\lab     &  $\HI$         &    21 cm   & \LABonereso    &      1.70$^\sharp$       &  \LABoneabserr \cr
\dht     &  $\twelveCO$   &    2.6 mm  & \DHTonereso    &      1.20               &  \DHToneabserr \cr
\dameHL  &  $\twelveCO$   &    2.6 mm  & \DAMEHLonereso &      0.6                &  \DAMEHLoneabserr \cr
\nanten  &  $\twelveCO$   &    2.6 mm  & \NANTENonereso &      1.20               &  \NANTENoneabserr \cr
\noalign{\vskip 3pt\hrule\vskip 4pt}
}}
\endPlancktable
$^\dag$ Assumed to be for the average \iras coverage. $\sigmaII$
computed by rescaling this value to actual coverage. 
$^\ddag$ From \cite{Miville2005}. 
$^\flat$ $1\sigma$ average value in one beam scaled from
\cite{planck2011-1.7}. We actually use internal variance maps for
$\sigmaII$ 
$^\sharp$ $1\sigma$ average value. We actually use a map of the
uncertainties (see Sect.\,\ref{sec:hidata}).  
\endgroup
\end{table}

The {\Planck} first mission results are presented in
\cite{planck2011-1.1} and the in-flight performances of the two focal
plane instruments \hfi (High Frequency Instrument) and {\lfi} (Low
Frequency Instrument) are given in \cite{planck2011-1.5} and
\cite{planck2011-1.4} respectively.  The data processing and
calibration of the \hfi and \lfi data used here is described in
\cite{planck2011-1.7} and \cite{planck2011-5.1a} respectively.

Here we use only the {\hfi} (DR2 release) data, the processing and
calibration of which are described in \cite{planck2011-1.7}. In this data the
CMB component was identified and subtracted through a Needlet Internal
Linear Combination (NILC) \citep{planck2011-1.7}.

We use the internal variance on intensity ($\sigmaII^2$) estimated
during the {\Planck} data processing and provided with the
{\Planck}-\hfi data, which we assume represents the white noise on
the intensity. Note that this variance is inhomogeneous over the
sky, owing to the {\Planck} scanning strategy \citep{planck2011-1.1},
with lower values in the {\Planck} deep fields near the ecliptic
poles.  We have checked that, within a small factor ($<2$), the data
variance above is consistent with ``Jack-Knife'' maps obtained from
differencing the two halves of the {\Planck} rings.  We also use the
absolute uncertainties due to calibration uncertainties given in
\cite{planck2011-1.7} for {\hfi} and summarized in
Table\,\ref{tab:ancillary}. We note that, for a large scale analysis
such as carried out here, variances contribute to a small fraction of the
final uncertainty resulting from combining data over large sky regions,
so that most of the final uncertainty is due to absolute uncertainties.

\subsection{Ancillary data}
\label{sec:ancillarydata}

\subsubsection{HI data}
\label{sec:hidata}

In order to trace the atomic medium, we use the LAB
(Leiden/Argentine/Bonn) survey which contains the final data release
of observations of the \ion{H}{i} 21-cm emission line over the entire sky
\citep{Kalberla2005}. This survey merged the Leiden/Dwingeloo Survey
\citep{Hartmann1997} of the sky north at $\rm \delta>-30\degr$ with the
IAR (Instituto Argentino de Radioastronomia) Survey \citep{Arnal2000,
Bajaja2005} of the Southern sky at $\rm \delta<-25\degr$.  The angular
resolution and the velocity resolution of the survey are $\rm \sim
0.6\degr$ and $\rm \sim 1.3\kms$. The LSR velocity range $\rm -450 < V _{\rm
LSR} < 400\kms$ is fully covered by the survey with 891 channels with
a velocity separation of $\rm \Delta V_{\rm ch}=1.03\,\kms$.

The data were corrected for stray radiation at the Institute for
Radioastronomy of the University of Bonn. The rms
brightness-temperature noise of the merged database is slightly lower
in the southern sky than in the northern sky, ranging over 0.07-0.09
K.  Residual uncertainties in the profile wings, due to defects in the
correction for stray radiation, are for most of the data below a level
of 20 to 40 mK.  We integrated the LAB data in the velocity range
$\rm -400<V_{\rm LSR}< 400\kms$ to produce an all sky map of the \ion{H}{i}
integrated intensity ($\whi$), which was finally projected into the
{\Healpix} pixelisation scheme using the method described in
Sect.\,\ref{sec:drizzling}.

We estimate the noise level of the $\whi$ map as $\rm \Delta T_{rms}
\Delta V_{ch} \sqrt{N_{ch}}$ where $\rm N_{ch}(=777)$ is the
number of channels used for the integration, and $\rm \Delta T_{rms}$
is the rms noise of the individual spectra measured in the
emission-free velocity range mainly in $\rm -400 < V_{LSR} < 350\kms$.
The resulting noise of the $\whi$ map is mostly less than $\sim
2.5\Kkms$ all over the sky with an average value of $\rm \sim 1.7\Kkms$,
except for some limited positions showing somewhat larger noise ($\rm \sim
10\Kkms$).

\subsubsection{CO data}
\label{sec:codata}

\begin{figure*}[ht]
\begin{center}
\includegraphics[width=17cm,angle=180]{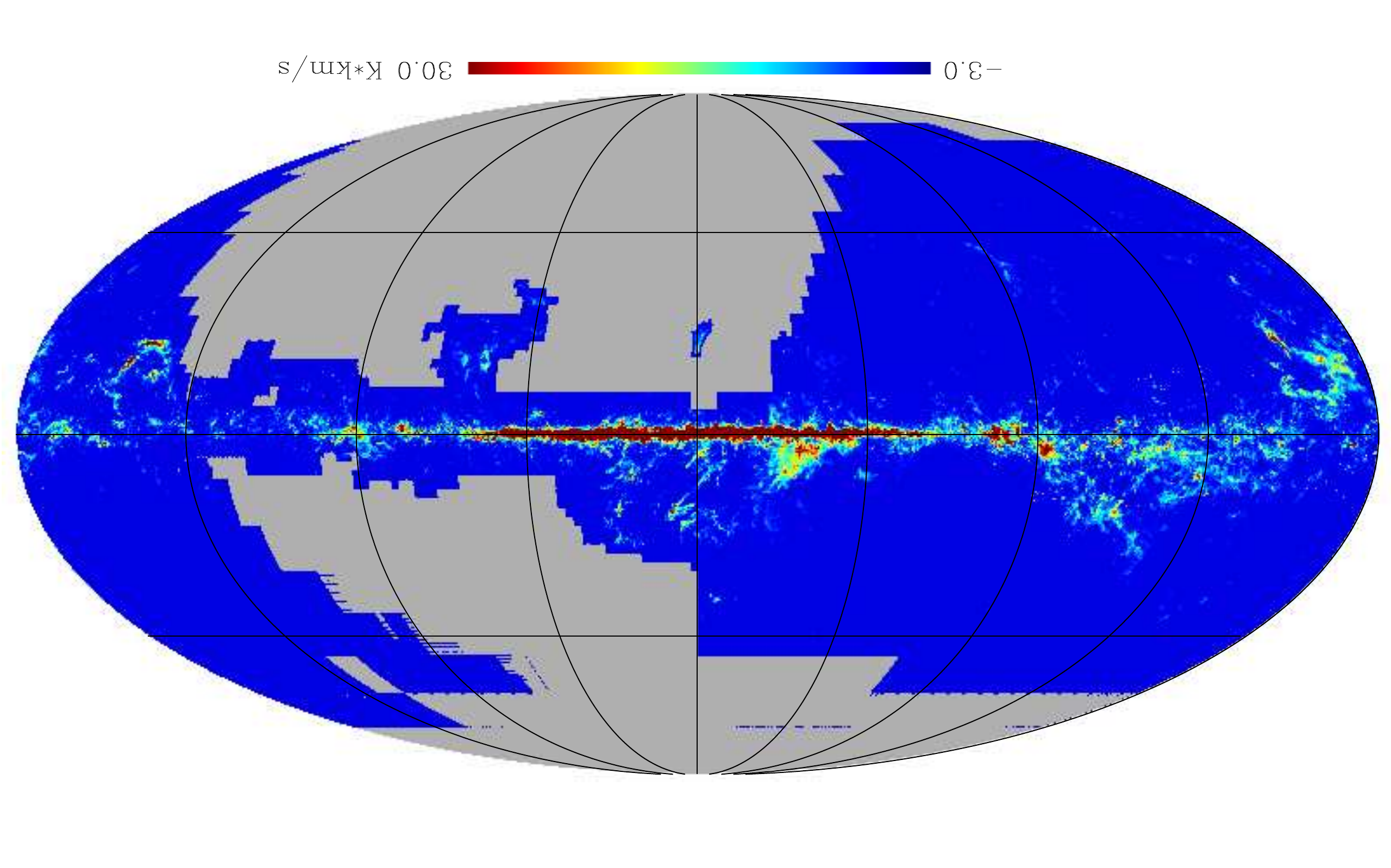}
\caption{
Map of the $^{12}$CO($J$=1$\rightarrow$0) integrated intensity used in this paper
combining the \cite{Dame2001} and high latitude survey and the NANTEN
survey. The data shown cover 62.8\% of the sky. The map is shown at a
common resolution of all the sub-surveys of 8.8'. Many small clouds at
high latitude are not visible in this rendering of the data.
\label{fig_COmap}}
\end{center}
\end{figure*}

In order to trace the spatial distribution of the CO emission, we use
a combination of 3 large scale surveys in the $^{12}$CO($J$=1$\rightarrow$0) line.

In the Galactic plane, we use the \cite{Dame2001} survey obtained with
the CfA telescope in the north and the CfA-Chile telescope in the
south, referred to here as \dht (Dame, Hartmann \& Thaddeus). These data have an angular resolution
of $\rm 8.4\arcm\pm0.2\arcm$ and $\rm 8.8\arcm\pm0.2\arcm$
respectively.  The velocity
coverage and the velocity resolution for these data vary from region
to region on the sky, depending on the individual observations
composing the survey.  The most heavily used spectrometer is the
500\,kHz filter bank providing a velocity coverage and resolution of
332\,\kms and 1.3\,\kms, respectively.  Another 250\,kHz filter bank
providing the 166\,\kms coverage and 0.65\,\kms resolution was also
frequently used . The rms noises of these data are suppressed down to
0.1--0.3 K (for details, see their Table 1).  The data cubes have been
transformed into the velocity-integrated intensity of the line
($\wco$) by integrating the velocity range where the CO emission is
significantly detected using the moment method proposed by
\cite{Dame2011}. The noise level of the $\wco$ map is typically $\rm
\sim 1.2\Kkms$, but it varies by a factor of a few
depending on the integration range used.

We also use the unpublished high latitude survey obtained using
the CfA telescope (Dame et al. 2010, private communication). This
survey is still on-going and covers the northern sky up to latitudes
as high as $\rm |$\bII$|=70\degr$ which greatly increases the overall sky
coverage. The noise level of the CO spectra are suppressed to
$\rm \sim0.18$\,K for the 0.65\,\kms velocity resolution, and the total CO
intensity was derived by integrating typically 10--20 velocity
channels, which results in a noise level of 0.4--$0.6\Kkms$.

Finally, we combined the above survey with the {\nanten} $^{12}$CO($J$=1$\rightarrow$0)
survey obtained from Chile.  This survey complements
some of the intermediate Galactic latitudes not covered by the
\cite{Dame2001} maps with an angular resolution of $2.6\arcm$.  Most of
the survey along the Galactic plane has a velocity coverage of $\rm
\sim650$\,\kms with a wide band spectrometer, but a part of the survey
has a coverage of $\rm \sim100$\,\kms with a narrow band spectrometer.
The noise level achieved was 0.4--0.5 K at a velocity resolution of
0.65\,\kms.  The CO spectra were sampled with the $2\arcm$ grid in the
Galactic centre, and with the $4\arcm$ and $8\arcm$ grid along the Galactic
plane in the latitude range $\rm |b|< 5$\,\degr and $\rm |b| >5$\,\degr,
respectively.  The integrated intensity maps were obtained by
integrating over the whole velocity range, excluding regions of the
spectra where no emission is observed. The resulting rms noise in the
velocity-integrated intensity map varies depending on the width of the
emission.  This survey along the Galactic plane is still not published
in full, but parts of the survey have been analyzed
(e.g. \cite{fukui1999,matsunaga2001,mizuno2004}). A large amount of
the sky at intermediate Galactic latitude toward the nearby clouds is
also covered with a higher velocity resolution of $\rm \sim0.1\kms$
with a narrow band spectrometer with a $\rm \lesssim100\kms$ band
(e.g. \cite{onishi1999,kawamura1999,mizuno2001}).  The velocity
coverage, the grid spacing, and the noise level for these data vary,
depending on the characteristics of the individual clouds observed,
but the quality of the data is high enough to trace the total CO
intensity of the individual clouds.

The three surveys were repixelised into the {\Healpix} pixelisation
scheme \citep{Gorski2005} with the appropriate pixel size to ensure
Shannon sampling of the beam (Nside=2048 for the NANTEN2 survey and
Nside=1024 for the CfA surveys) using the procedure described in
Sect.\,\ref{sec:drizzling}.

Each survey was smoothed to a common resolution of 8.8\arcm through
convolution with a Gaussian with kernel size adjusted to go from the
original resolution of each survey to a goal resolution of 8.8\arcm, using
the smoothing capabilities of the {\Healpix} software.
We checked the consistency of the different surveys in the common region
observed with NANTEN and CfA.

We found a reasonably good correlation between the two but a slope
indicating that the NANTEN survey yields 24\% larger intensities than
the CfA values. The origin of this discrepancy is currently unknown.
We should note that the absolute intensity scale in CO observations
is not highly accurate as noted often in the previous CO papers.
Since the CfA survey covers most of the regions used in this paper and
has been widely used for calibrating the H$_2$ mass
calibration factor $\XCO$, in particular by several generations of gamma
ray satellites, we assumed the CfA photometry when merging the data,
and therefore rescaled the {\nanten} data down by 24\% before merging.
Note that this an arbitrary choice. The implications on our results
will be discussed in Sec.\,\ref{sec:diffatomic}.

The 3 surveys were then combined into a single map. In doing so, data
from different surveys falling into the same pixel were averaged using
$\sigma^2$ as a weight. The resulting combined map was then smoothed
to the resolution appropriate to this study. The resulting CO
integrated intensity map is shown in Fig.\,\ref{fig_COmap}.

\subsubsection{IR data}
\label{sec:irdata}

We use the IRIS (Improved Reprocessing of the {\iras} Survey) {\iras}
$100\mic$ data \citep[see]{Miville2005} in order to constrain the dust
temperature. The data, provided in the original format of individual
tiles spread over the entire sky were combined into the {\Healpix}
pixelisation using the method described in Sect.\,\ref{sec:drizzling}
at a {\Healpix} resolution ($\rm Nside=2048$ corresponding to a pixel
size of $1.7^\prime$). The {\iras} coverage maps were also processed
in the same way.  We assume the noise properties given in
\cite{Miville2005} and given in Table\,\ref{tab:ancillary}. The noise
level of 0.06\,$MJy\,sr^{-1}$ at $100\mic$ was assumed to represent
the average data noise level and was appropriately multiplied by the
coverage map to lead to the pixel variance of the data.

\begin{table}[tmb]
\begin{center}
\begingroup
\newdimen\tblskip \tblskip=5pt
\caption{\label{tab:offset} Thermal Dust emissivity derived from the correlation with $\HI$ emission
in the reference region with $\rm |$\bII$|>20\degr$ and $\rm \NHHI < \NHHIrefmaxv$
($\rm (I_\nu/N_H)^{ref}$). Offsets derived from an empty region with
$\rm \NHHI < \NHHIholemaxv$, assuming the same emissivity
$\rm (I_\nu/N_H)^{ref}$. The uncertainties are given at the $\rm 1\sigma$ level.
The corresponding data are plotted in Fig.\,\ref{fig:offset}.}
\nointerlineskip
\footnotesize
\vskip -8pt
\setbox\tablebox=\vbox{
\newdimen\digitwidth
\setbox0=\hbox{\rm 0}
\digitwidth=\wd0
\catcode`*=\active
\def*{kern\digitwidth}
\newdimen\signwidth
\setbox0=\hbox{+}
\signwidth=\wd0
\catcode`!=\active
\def!{\kern\signwidth}
\halign to \hsize{\hbox to 0.9in{#\hfil}\tabskip=2.2em&
	\hfil#\hfil&
	\hfil#\hfil\tabskip 0pt\cr
\noalign{\vskip 3pt\hrule\vskip 1.5pt\hrule\vskip 5pt}
\omit $\nu$ \hfil    & $\rm (I_\nu/N_H)^{ref}$        & offset \hfil \cr
\omit ({\GHz}) \hfil & [$\rm MJy/sr/10^{20}\NHUNIT$] &  [MJy/sr] \hfil \cr
\noalign{\vskip 4pt\hrule\vskip 6pt}
\multispan3\iras\hfil\cr
\noalign{\vskip 4pt\hrule\vskip 6pt}
   \IRASfourfreq     & (6.95$\pm$0.94)$\times10^{-1}$ & (7.36$\pm$0.03)$\times10^{-1}$\cr
\noalign{\vskip 4pt\hrule\vskip 6pt}
\multispan3\dirbe:\hfil\cr
\noalign{\vskip 4pt\hrule\vskip 6pt}
\DIRBEheightfreq     & (6.60$\pm$0.01)$\times10^{-1}$ & (7.72$\pm$0.22)$\times10^{-1}$\cr
  \DIRBEninefreq     & 1.16$\pm$0.01                 & 1.41$\pm$0.16\cr
  \DIRBEtenfreq      & (8.85$\pm$0.04)$\times10^{-1}$ & (8.46$\pm$0.90)$\times10^{-1}$\cr
\noalign{\vskip 4pt\hrule\vskip 6pt}
\multispan3\Planck-\hfi:\hfil\cr
\noalign{\vskip 4pt\hrule\vskip 6pt}
\HFIonefreq          & (5.43$\pm$0.38)$\times10^{-1}$ & (2.57$\pm$0.05)$\times10^{-1}$\cr
    \HFItwofreq      & (1.82$\pm$0.13)$\times10^{-1}$ & (1.83$\pm$0.05)$\times10^{-1}$\cr 
 \HFIthreefreq       & (4.84$\pm$0.10)$\times10^{-2}$ & (9.50$\pm$0.24)$\times10^{-2}$\cr
    \HFIfourfreq     & (1.14$\pm$0.03)$\times10^{-2}$ & (3.45$\pm$0.11)$\times10^{-2}$\cr
    \HFIfivefreq     & (2.92$\pm$0.07)$\times10^{-3}$ & (1.01$\pm$0.06)$\times10^{-2}$\cr
    \HFIsixfreq      & (1.13$\pm$0.04)$\times10^{-3}$ & (3.34$\pm$0.60)$\times10^{-3}$\cr
\noalign{\vskip 4pt\hrule\vskip 6pt}
\multispan3\Planck-\lfi:\hfil\cr
\noalign{\vskip 4pt\hrule\vskip 6pt}
\LFIonefreq          & (8.74$\pm$2.66)$\times10^{-5}$ & (-6.40$\pm$0.93)$\,10^{-3}$\cr
    \LFItwofreq      & (1.14$\pm$0.16)$\times10^{-4}$ & (-7.02$\pm$0.45)$\,10^{-3}$\cr
  \LFIthreefreq      & (2.11$\pm$0.14)$\times10^{-4}$ & (-6.95$\pm$0.15)$\,10^{-3}$\cr
\noalign{\vskip 4pt\hrule\vskip 6pt}
\multispan3\wmap:\hfil\cr
\noalign{\vskip 4pt\hrule\vskip 6pt}
  \WMAPonefreq       & (-1.05$\pm$0.57)$\times10^{-4}$ & (-1.65$\pm$1.02)$\times10^{-3}$\cr
  \WMAPtwofreq       & (-1.14$\pm$0.27)$\times10^{-4}$ & (-4.90$\pm$3.63)$\times10^{-4}$\cr
  \WMAPthreefreq     & (3.52$\pm$1.13)$\times10^{-5}$  & (9.76$\pm$1.28)$\times10^{-4}$\cr
  \WMAPfourfreq      & (1.39$\pm$0.08)$\times10^{-4}$  & (4.10$\pm$0.97)$\times10^{-4}$\cr
  \WMAPfivefreq      & (2.69$\pm$0.04)$\times10^{-4}$  & (8.49$\pm$0.44)$\times10^{-4}$\cr
\noalign{\vskip 3pt\hrule\vskip 4pt}
}}
\endPlancktable
\endgroup
\end{center}
\end{table}

\subsection{Additional Data processing}

\subsubsection{Common angular resolution and pixelisation}
\label{sec:drizzling}

The individual maps are then combined into {\Healpix} using the
intersection surface as a weight. This procedure was shown to preserve
photometry accuracy.

The ancillary data described in Sect.\,\ref{sec:ancillarydata} were
brought to the {\Healpix} pixelisation, using a method where the
surface of the intersection between each {\Healpix} pixel with each
FITS pixel of the survey data is computed and used as a weight to
regrid the data.  The {\Healpix} resolution was chosen so as to match
the Shannon sampling of the original data at resolution $\theta$, with
a {\Healpix} resolution set so that the pixel size is
$<\theta/2.4$. The ancillary data and the description of their
processing will be presented in \cite{Paradis2011}.

All ancillary data were then smoothed to an appropriate resolution by
convolution with a Gaussian smoothing function with appropriate FWHM
using the smoothing {\Healpix} function, and were brought to a pixel
size matching the Shannon sampling of the final resolution.

\subsubsection{Background levels}
\label{sec:background}

\begin{figure}[ht]
\begin{center}
\includegraphics[width=8cm,angle=0]{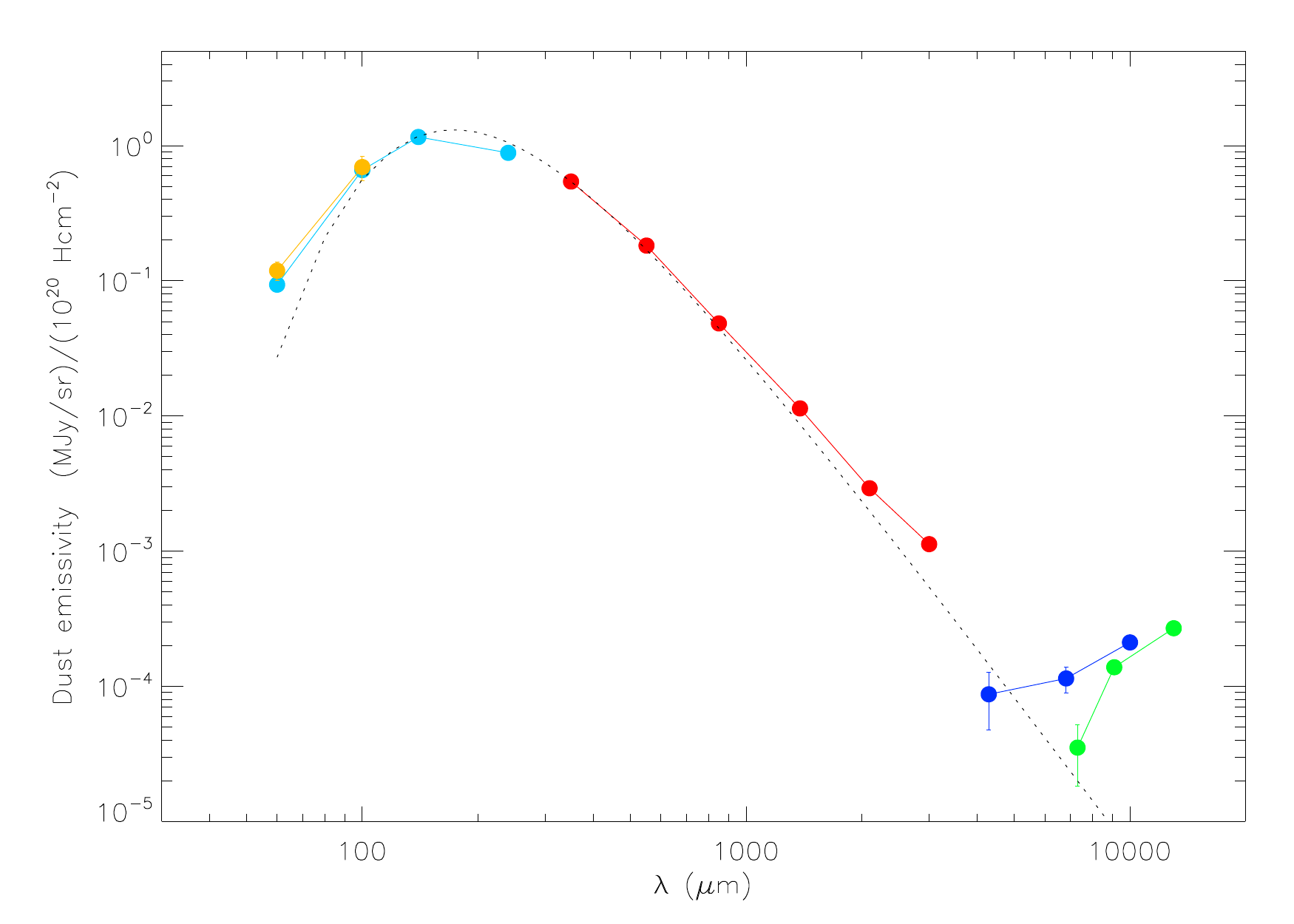}
\includegraphics[width=8cm,angle=0]{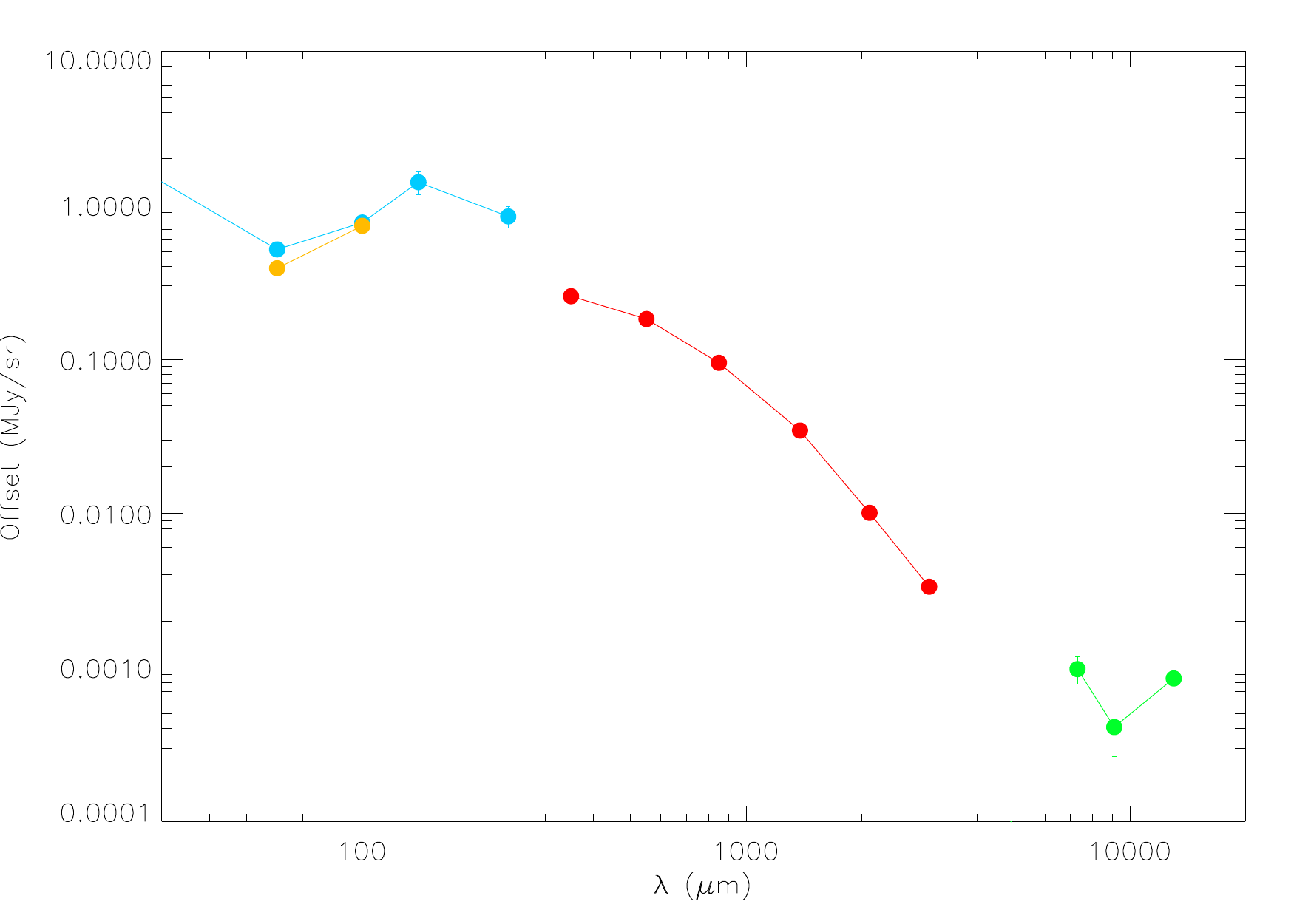}
\caption{\label{fig:offset}
Upper panel: Thermal Dust emissivity $\rm (I_\nu/N_H)^{ref}$ from
Table\,\ref{tab:offset}.  The dot curve showns a grey body at
$\Td=17.5\,K$ and $\rm \beta=1.8$
normalized at {\HFIonefreq}{\GHz}, for comparison. The various colours are for
different instruments: {\iras} (Yellow), {\dirbe} (light blue),
{\Planck}-{\hfi} (red), {\wmap} (dark blue) and {\Planck}-{\lfi} (green).
Lower panel: Offsets from Table\,\ref{tab:offset}.  The error bars are
plotted at $\pm$3$\sigma$.
}
\end{center}
\end{figure}

Computing the apparent temperature and optical depth of thermal dust
over the whole sky requires accurate subtraction of any offset ($\offset$) in the
intensity data, either of instrumental or astrophysical origin.
Although both the {\iris} and the {\Planck}-{\hfi} data used in this
study have been carefully treated with respect to residual offsets
during calibration against the {\firas} data, the data still
contains extended sources of emission unrelated to the Galactic
emission, such as the Cosmic InfraRed Background (CIB) signal
\citep{Miville2002,planck2011-6.6} or zodical light which could affect
the determination of the dust temperature and optical depth at low
surface brightness.

In order to estimate the above data offsets, we first compute the
correlation between IR and \ion{H}{i} emission in a reference region
such that $\rm |$\bII$|>20\degr$ and $\NHHI < \NHHIrefmaxv$. This
was done using the IDL {\it regress} routine and iterative removal of
outliers. The derived dust emissivities ($\rm (I_\nu/N_H)^{ref}$) are
given in Table \,\ref{tab:offset}. The uncertainties given are
those derived from the correlation using the data variance as the data
uncertainty. The derived emissivities are in agreement with the
ensemble average of the values found for the local \ion{H}{i}
velocities in \cite{planck2011-7.12} (see their Table 2) for
individual smaller regions at high Galactic latitude, within the
uncertainties quoted in Table\,\ref{tab:offset}. Note that these
emissivities are used only to derive the offsets in this study.

We then select all sky pixels with minimum \ion{H}{i}
column density defined as $\rm \NHHI < \NHHIholemaxv$ and compute the
average \ion{H}{i} column density in this region to be $\rm \NHHIhole
= \NHHIholev$. The offsets are then computed assuming that the dust
emissivity in this region is the same as in the reference region, ie,
\begin{equation}
\label{eq:offset}
\offset=I_\nu^{\rm hole}-(I_\nu/N_{\rm H})^{\rm ref} \times \NHHIhole
\end{equation}
where $I_\nu^{\rm hole}$ is the average brightness in the hole region
at frequency $\nu$.

The offset values derived from the above procedure are given in
Table\,\ref{tab:offset} and were subtracted from the maps used in the
rest of this analysis. The offset uncertainties also listed in
Table\,\ref{tab:offset} were derived from the emissivity uncertainties
propagated to the offset values through Eq.\,\ref{eq:offset}.  When
subtracting the above offsets from the {\iras} and {\Planck} intensity
maps, the data variances were combined with the offset uncertainties
in order to reflect uncertainty on the offset determination.  Note
that, for consistency and future use, Table\,\ref{tab:offset} also
lists emissivities and offset values for FIR-mm datasets not used in
this study. Note also that these offsets for {\Planck} data are
not meant to replace the official values provided with the data, since
they suppress any large scale emission not correlated with \ion{H}{i},
whatever their origin.

\section{Dust temperature and emissivity}

\subsection{Temperature determination}
\label{sec:temperature}

\begin{figure*}[ht!]
\begin{center}
\includegraphics[width=17cm,angle=180]{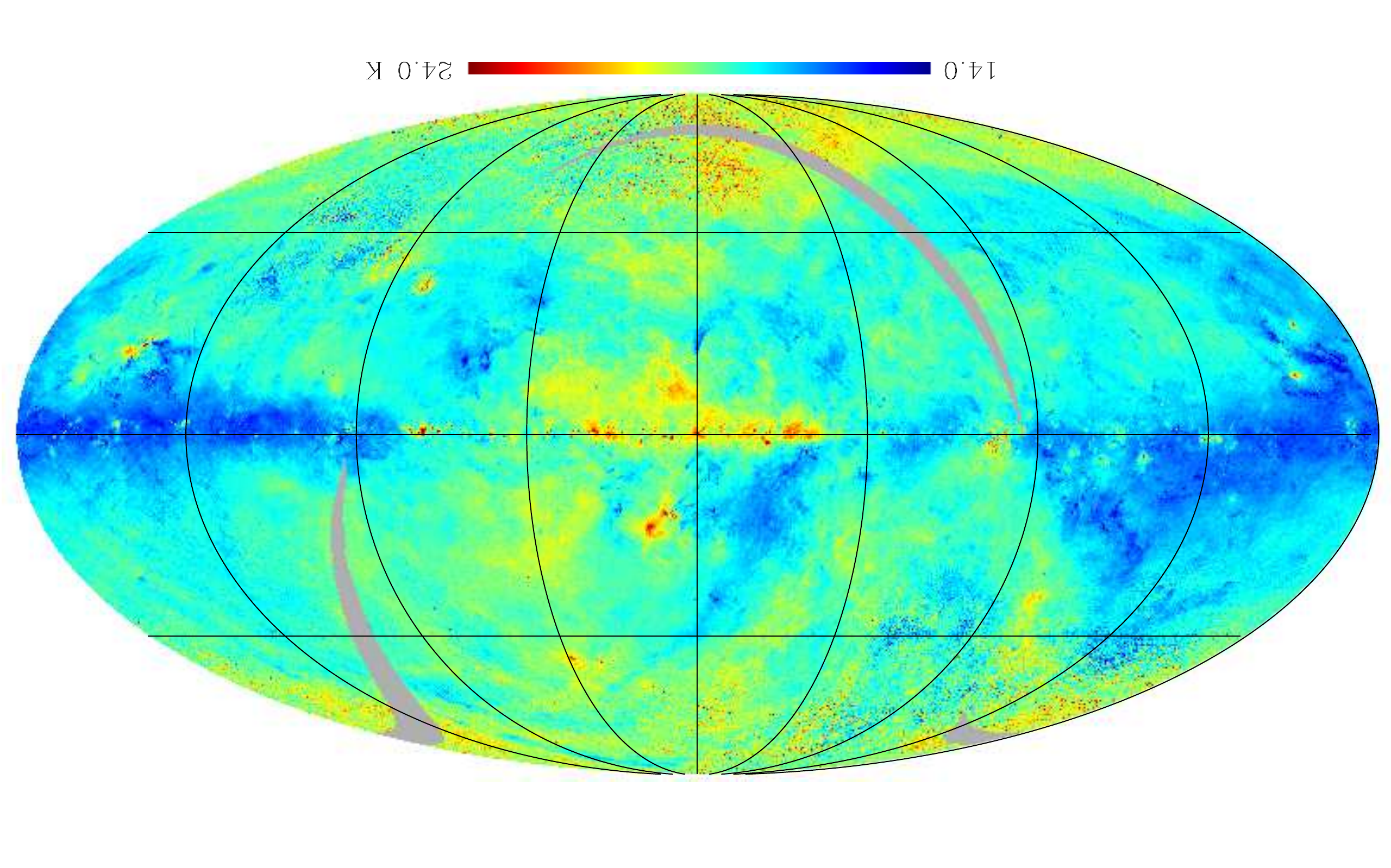}
\includegraphics[width=17cm,angle=180]{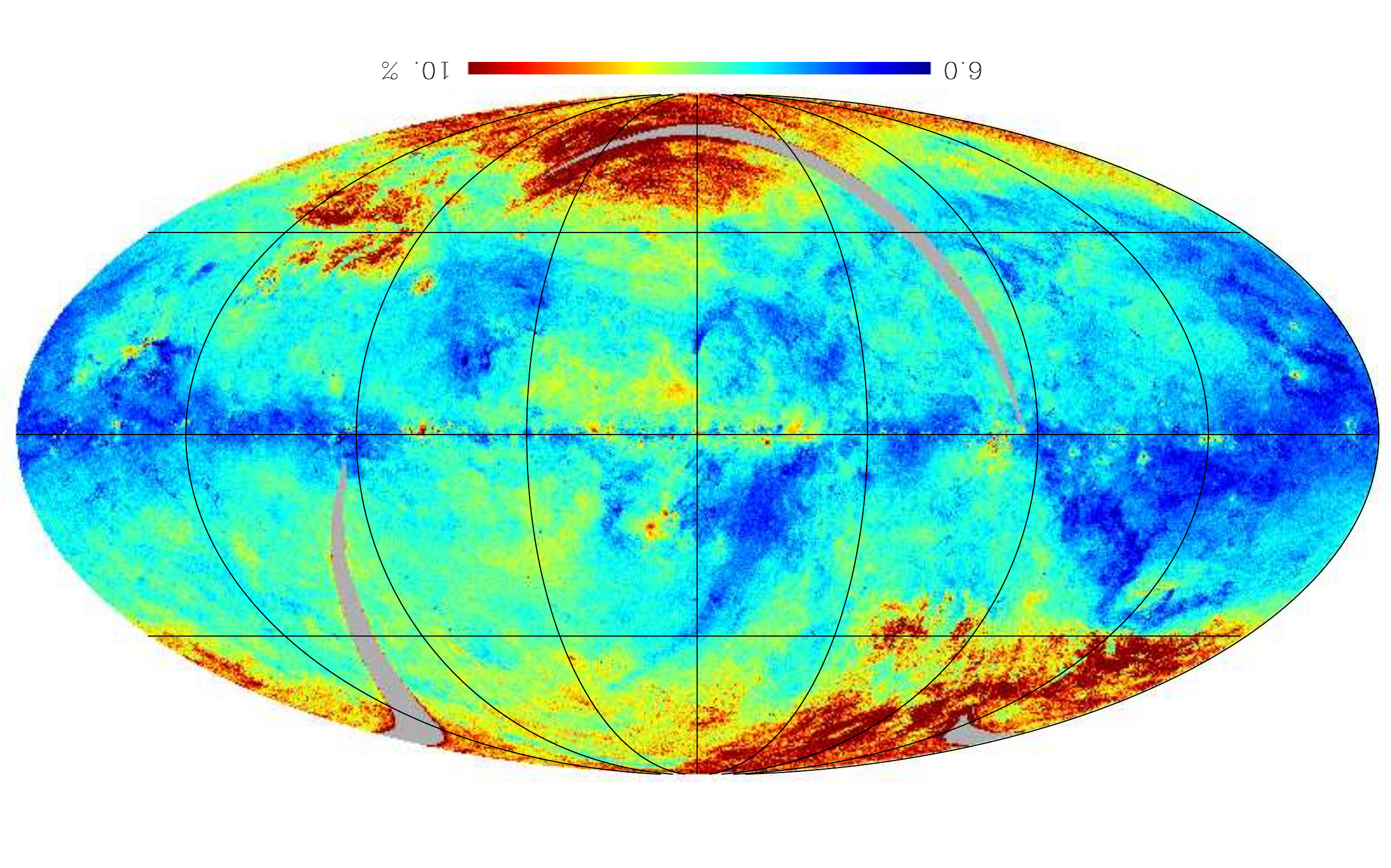}
\caption{
Upper panel: All sky map of the dust temperature in K. The temperature is derived
from modeling the {\iris} $100\mic$ and the {\Planck}-{\hfi} emission at {\HFIonefreq}
and {\HFItwofreq}{\GHz}. Lower panel: All sky map of the dust temperature uncertainty in \%.
The maps are shown in Galactic coordinates with the
Galactic centre at the centre of the image. Grey regions correspond to
missing {\iras} data.
\label{fig:Tmap}}
\end{center}
\end{figure*}

As shown in previous studies
\citep[e.g.][]{Reach1995,finkbeiner1999,Paradis2009b,planck2011-7.0,planck2011-7.12},
the dust emissivity spectrum in our Galaxy cannot be represented by a
single dust emissivity index $\beta$ over the full FIR-submm
domain. The data available indicate that $\beta$ is usually steeper
in the FIR and flatter in the
submm band, with a transition around $500\mic$.  As dust temperature is
best derived around the emission peak, we limit the range of
frequencies used in the determination to the FIR, which limits the
impact of a potential change of $\beta$ with frequency.

In addition, the dust temperature derived will depend on the
assumption made about $\beta$, since these two parameters are somewhat
degenerate in $\chi^2$ space.  In order to minimize the above effect,
we derived dust temperature maps using a fixed value of the dust
emissivity index $\beta$.  The selected $\beta$ value was derived by
fitting each pixel of the maps with a modified black body of the form
$\rm I_{\nu} \propto \nu^\beta B_\nu (\Td)$ in the above spectral range
(method referred to as ``free $\beta$''). This leads to a median value of
$\rm \Td=17.7\,K$ and $\beta=1.8$ in the region at $\rm |$\bII$|>10\degr$. Note that
the $\beta$ value is consistent with that derived from the combination
of the \firas and {\Planck}-{\hfi} data at low column density in
\cite{planck2011-7.12}.  Inspection of the corresponding $\Td$ and
$\beta$ maps indeed showed spurious values of both parameters, caused
by their correlation and the presence of noise in the
data, in particular in low brightness regions of the maps.

We then performed fits to the FIR emission using the fixed $\beta=1.8$
value derived above (method referred to as ``fixed $\beta$'').  In the
determination of $\Td$, we used the {\iris} $100\mic$ map and the two
highest {\hfi} frequencies at {\HFIonefreq} and {\HFItwofreq}{\GHz}.
Although the median reduced $\chi^2$ is slightly higher than for the
``free $\beta$'' method, the temperature maps show many fewer spurious
values, in particular in low brightness regions. This results in a
sharper distribution of the temperature histogram.  Since we later use
the temperature maps to investigate the spectral distribution of the
dust optical depth and the dust temperature is a source of
uncertainty, we adopt the ``fixed $\beta$'' method maps in the
following. The corresponding temperature and uncertainty maps are
shown in Fig.\,\ref{fig:Tmap}.

Temperature maps were derived at the common resolution of those three
channels as well as at the resolution of lower intensity data.  The
model was used to compute emission in each photometric channels of the
instruments used here ({\iras}, {\Planck}-{\hfi}), taking into account
the colour corrections using the actual transmission profiles for each
instrument and following the adopted flux convention.  In the interest
of computing efficiency, the predictions of a given model were
tabulated for a large set of parameters ($\Td$, $\beta$).  For each
map pixel, the $\chi^2$ was computed for each entry of the table and
the shape of the $\chi^2$ distribution around the minimum value was
used to derive the uncertainty on the free parameters.  This included
the effect of the data variance $\sigmaII^2$ and the absolute
uncertainties.

\subsection{Angular distribution of dust temperature}
\label{sec:temperature_dist}

\begin{figure*}[ht!]
\begin{center}
\includegraphics[width=6cm,angle=180]{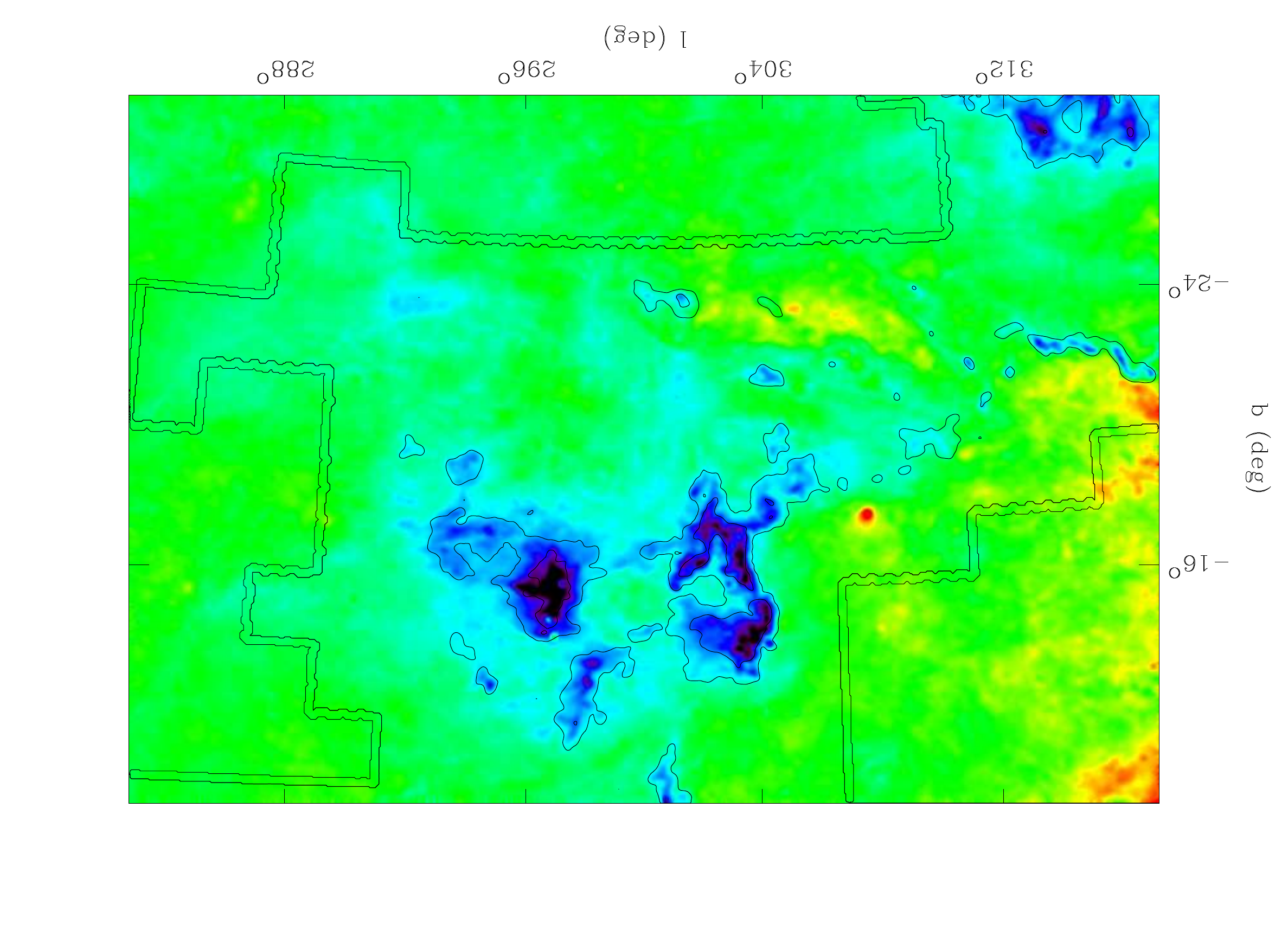}
\includegraphics[width=6cm,angle=180]{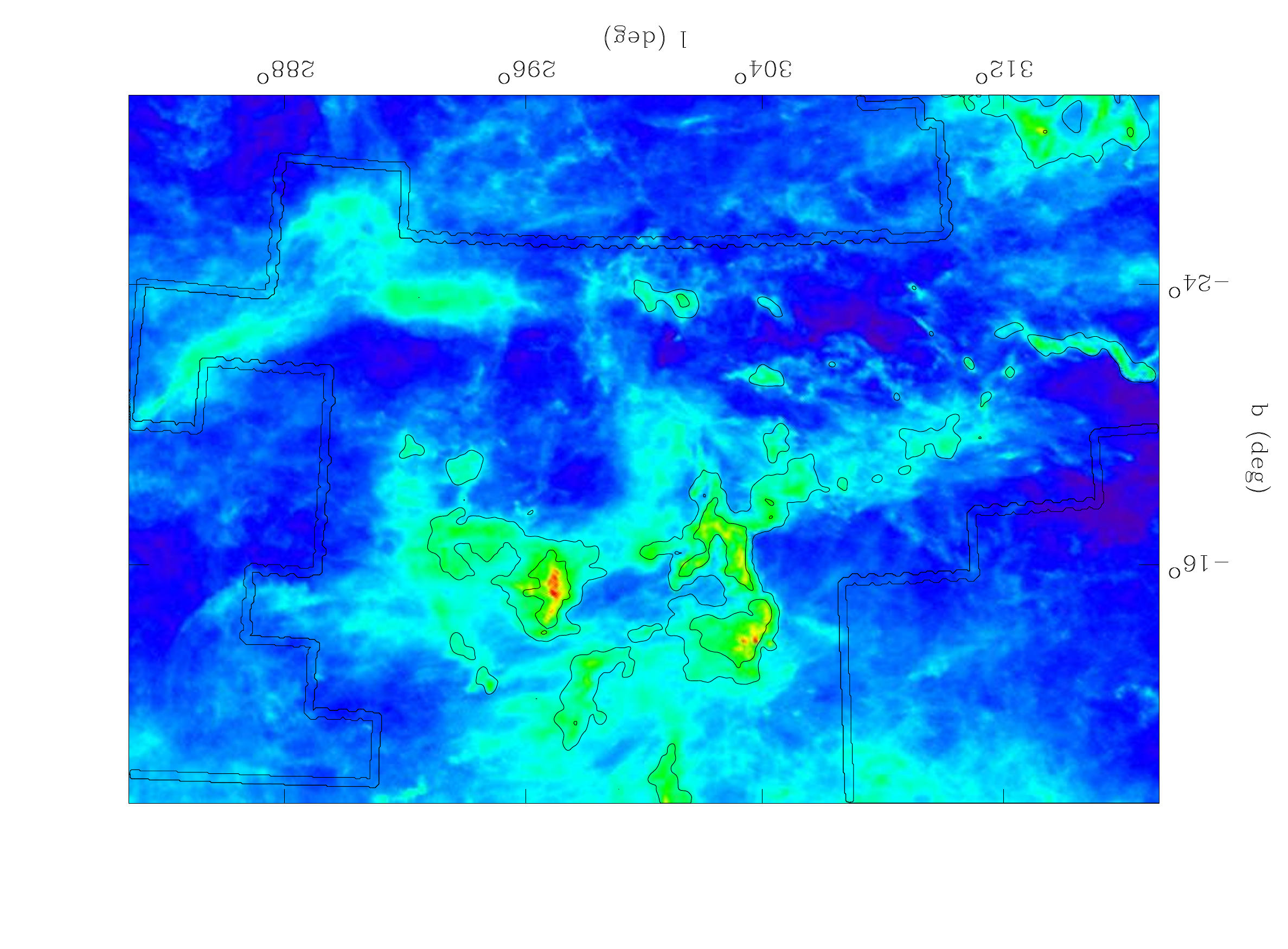}
\includegraphics[width=6cm,angle=180]{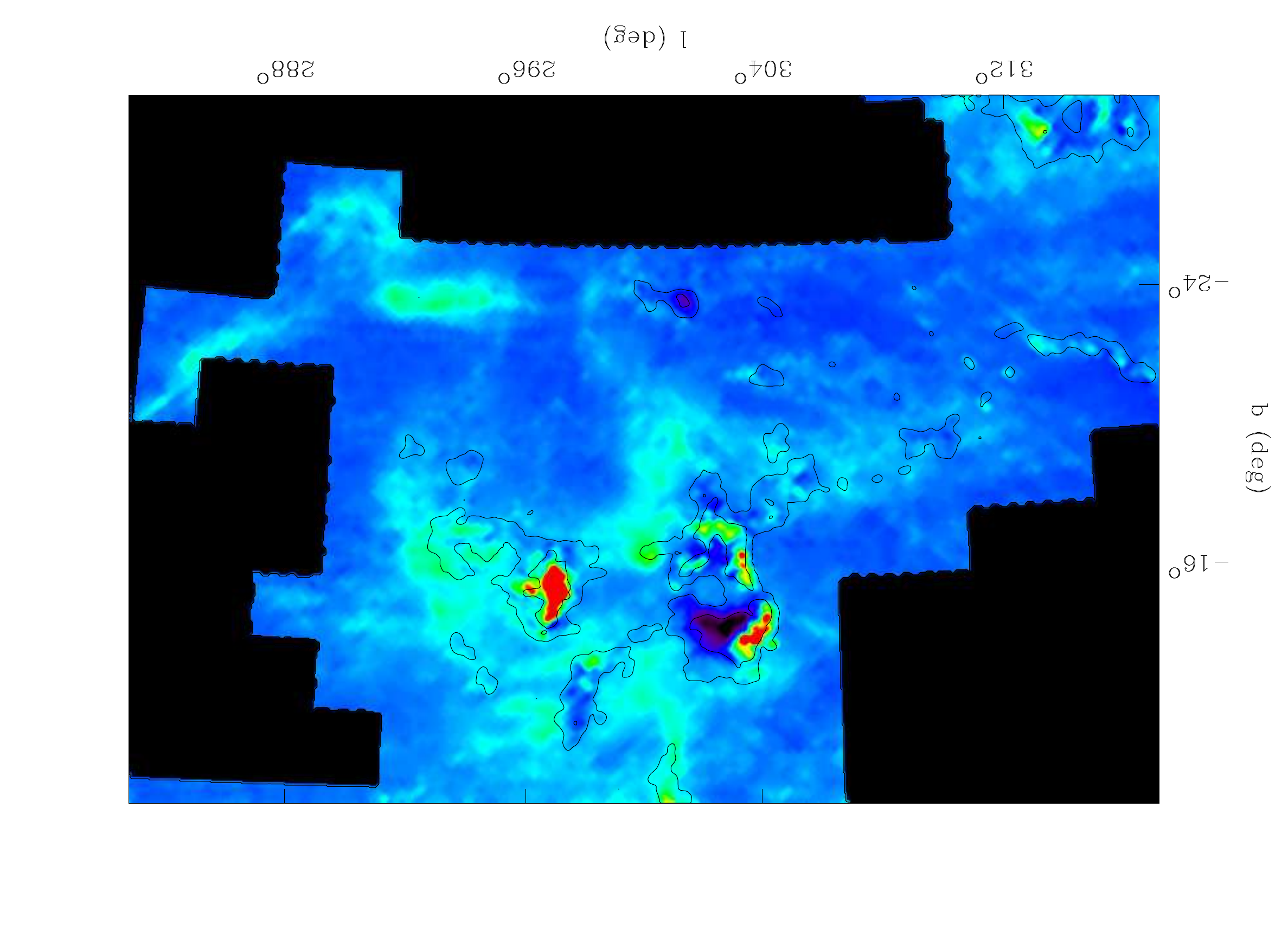}
\includegraphics[width=6cm,angle=180]{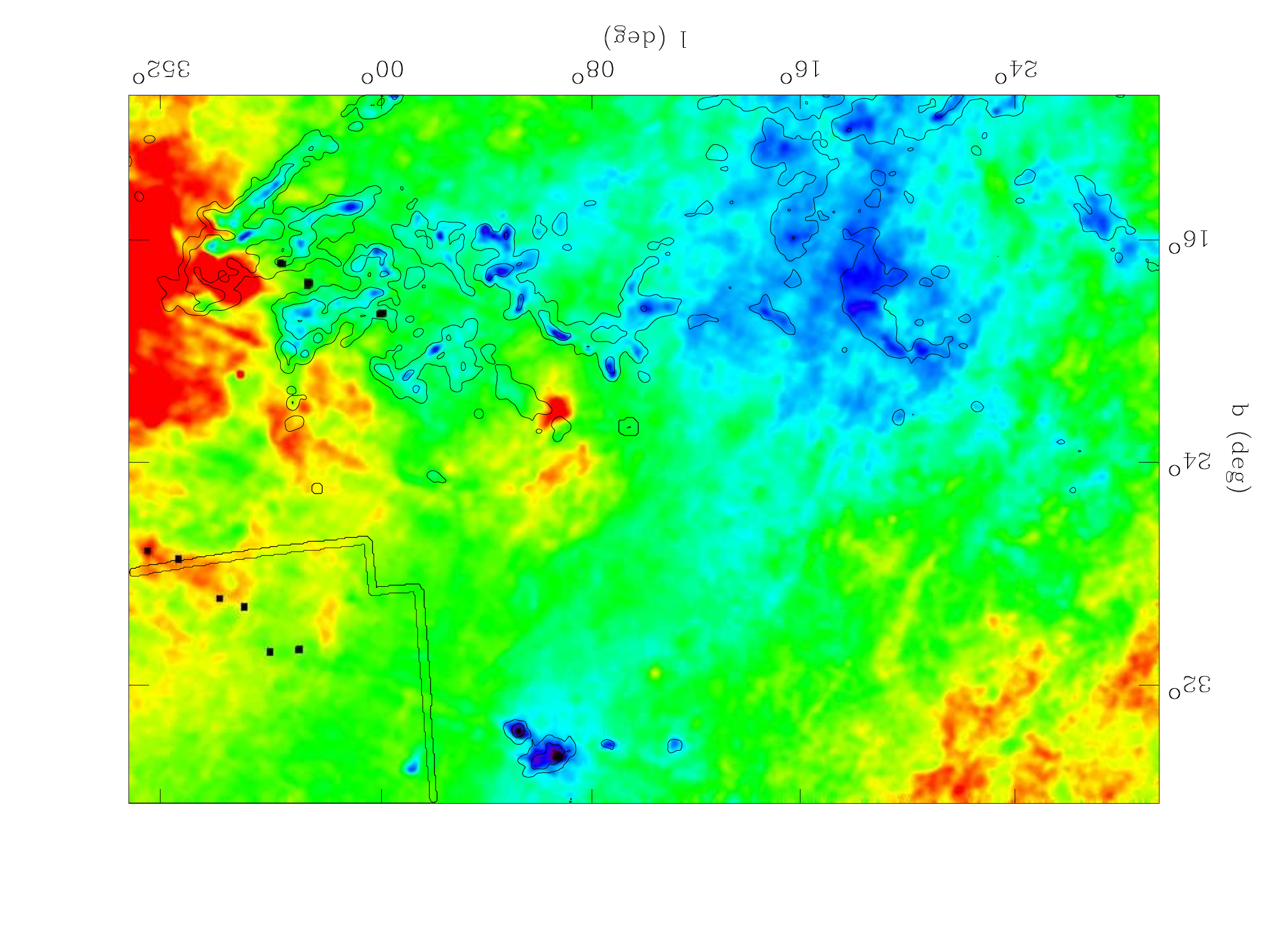}
\includegraphics[width=6cm,angle=180]{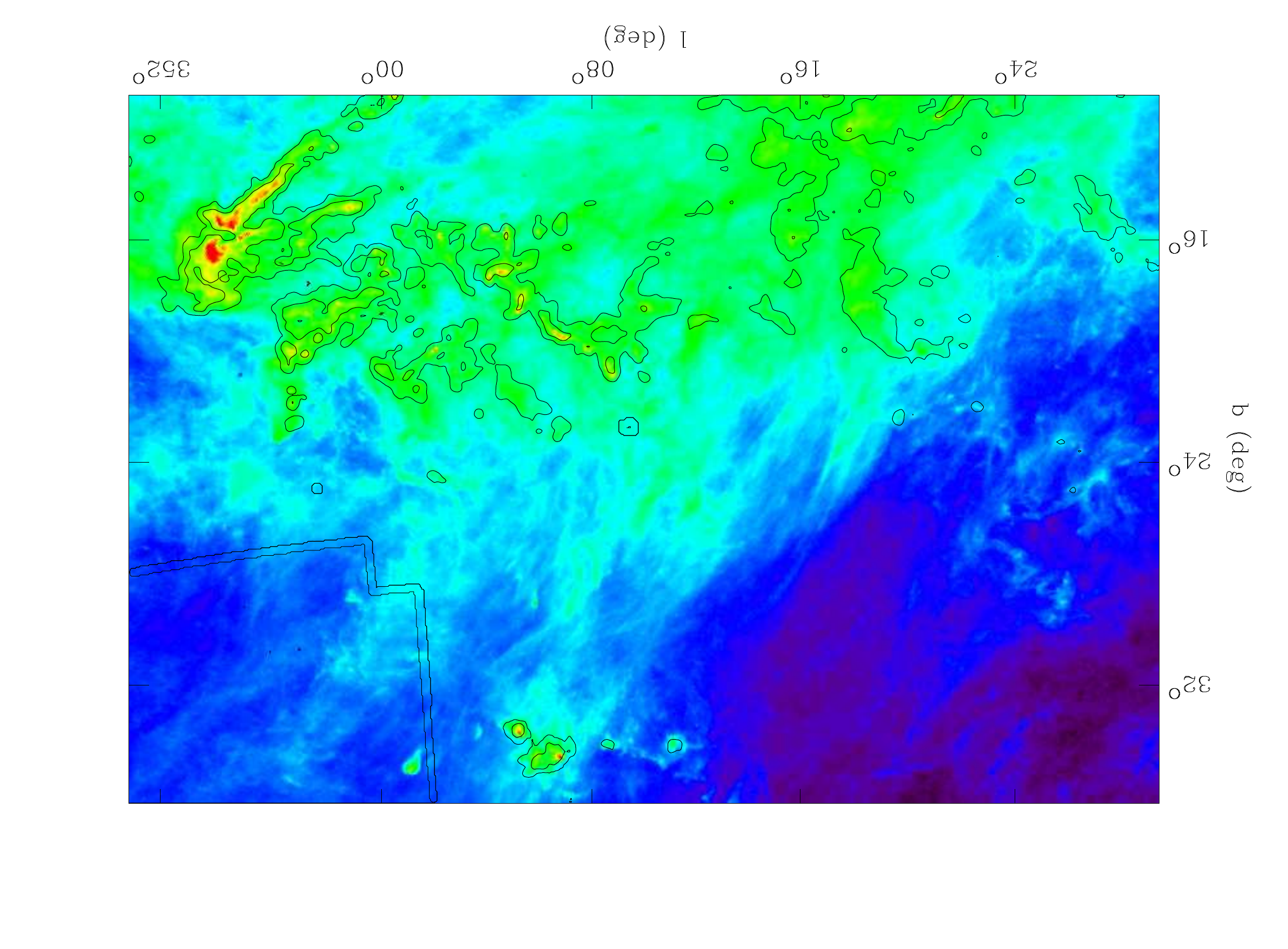}
\includegraphics[width=6cm,angle=180]{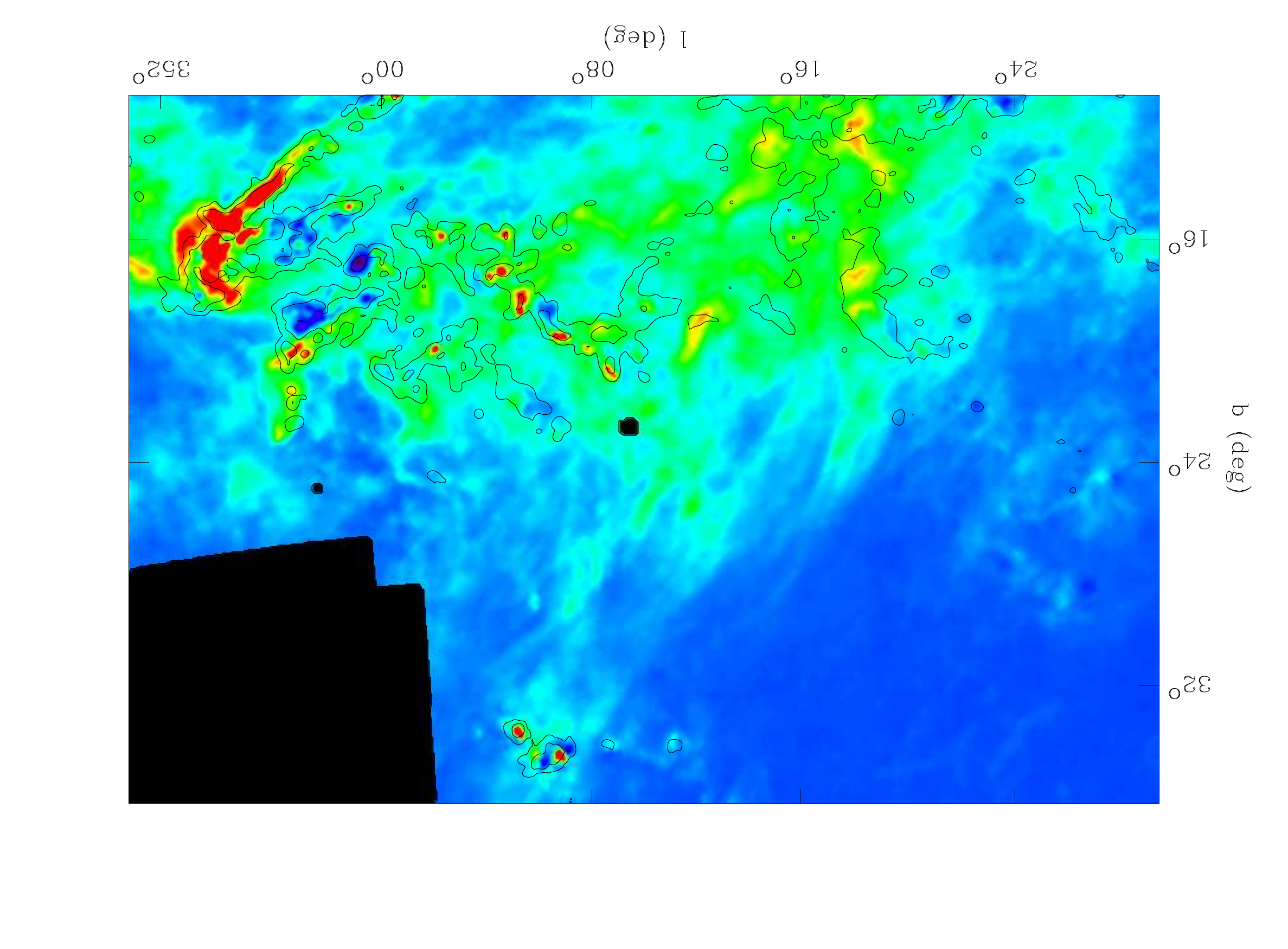}
\includegraphics[width=6cm,angle=180]{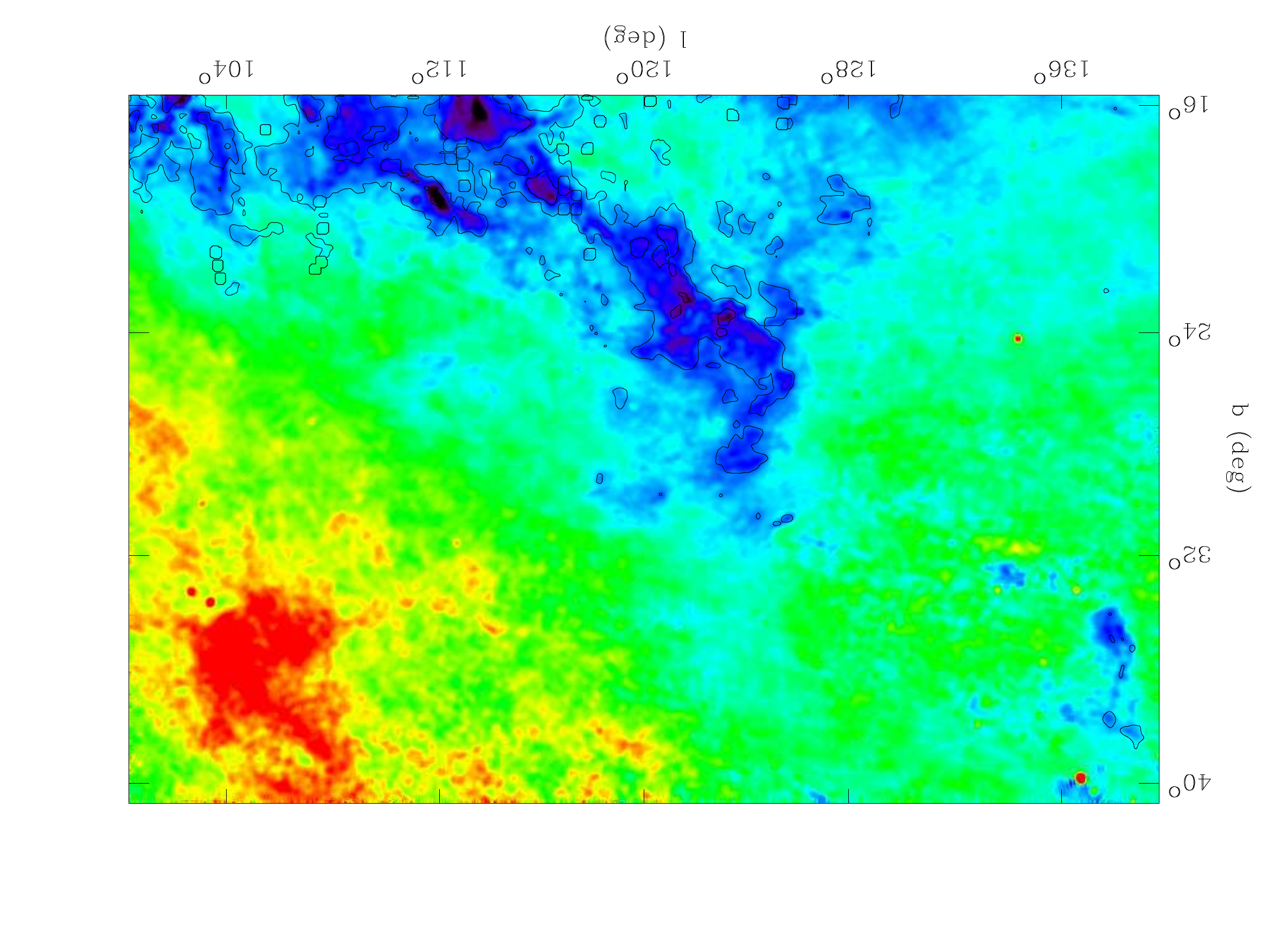}
\includegraphics[width=6cm,angle=180]{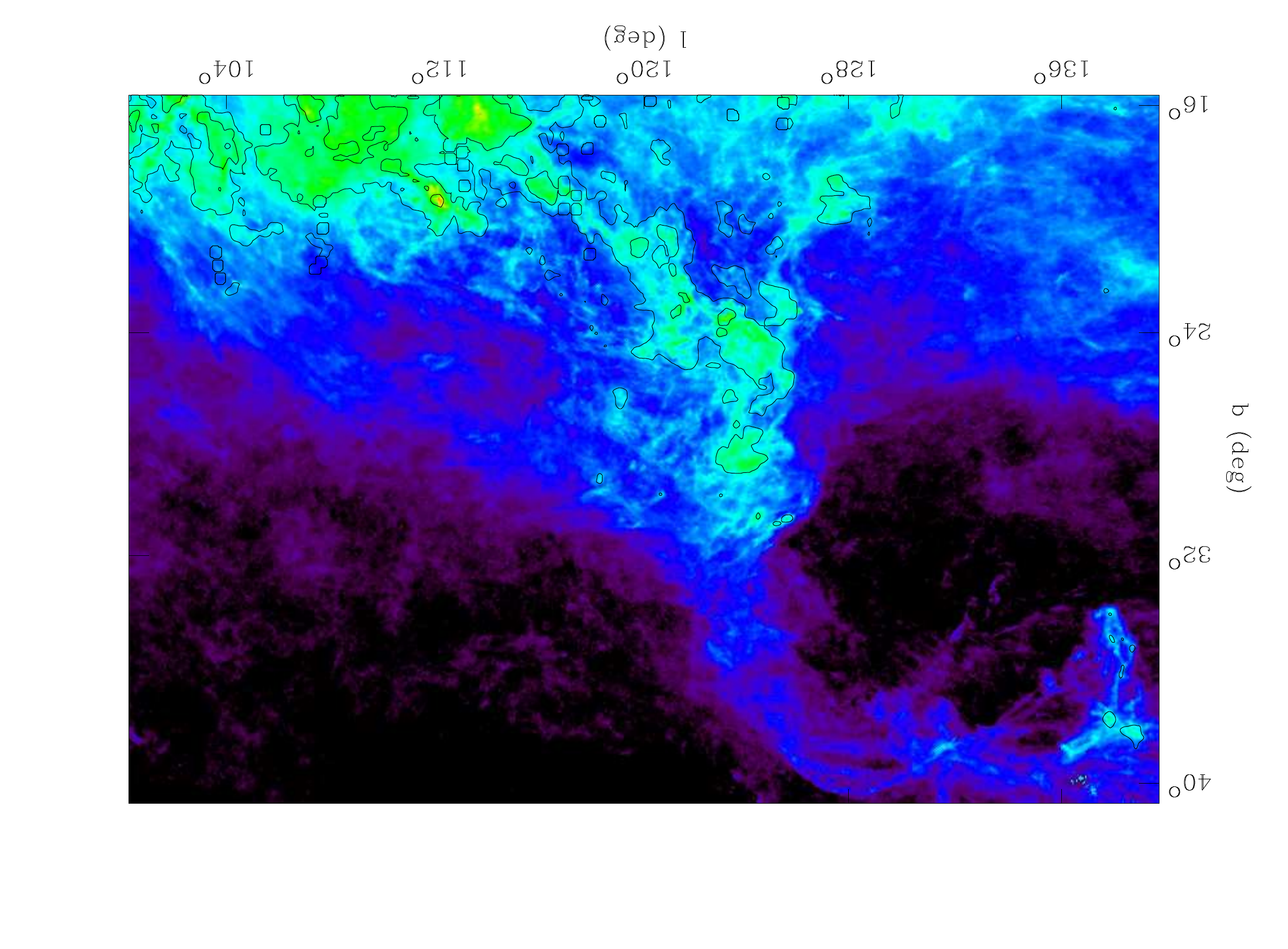}
\includegraphics[width=6cm,angle=180]{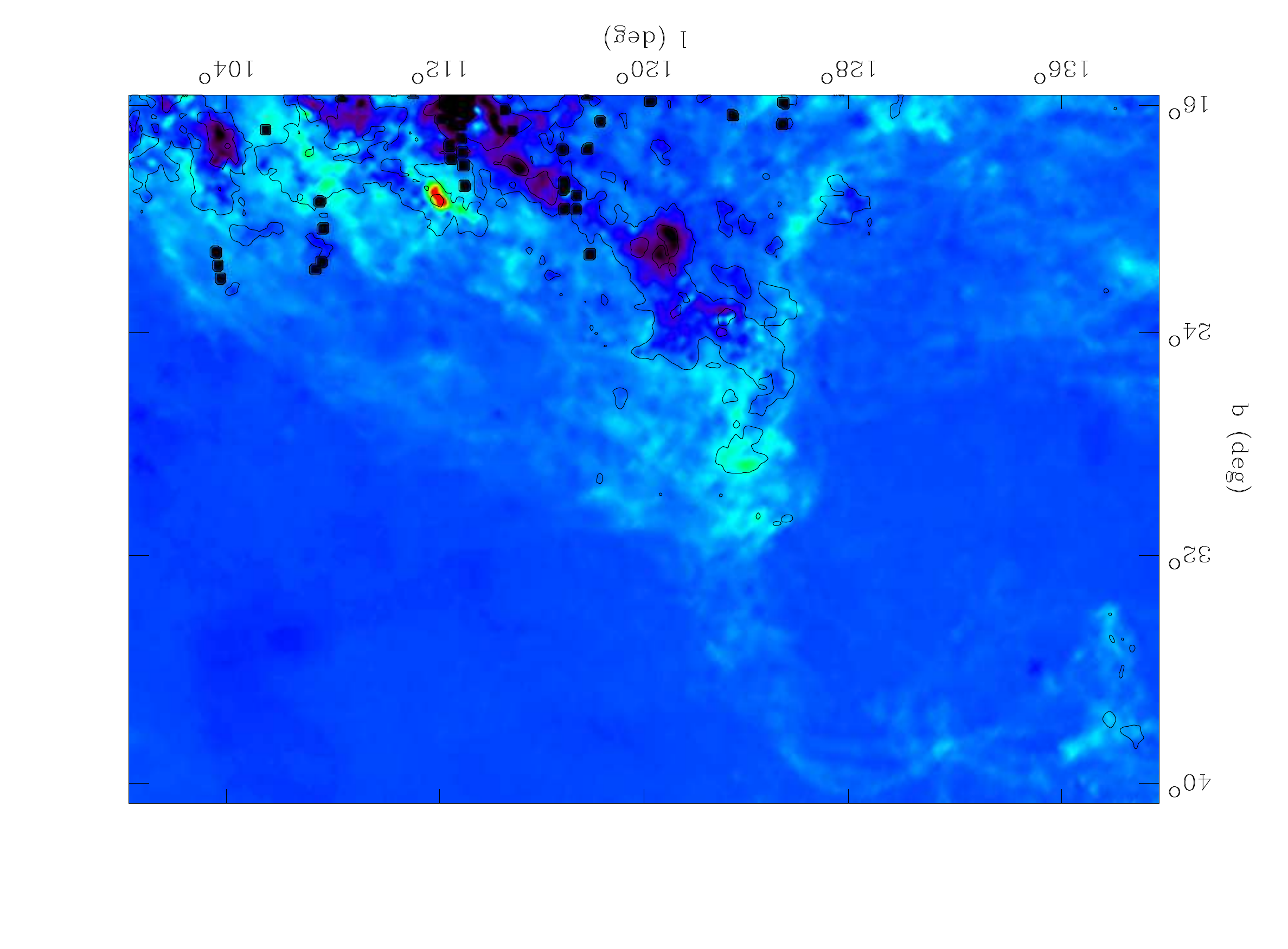}
\includegraphics[width=6cm,angle=180]{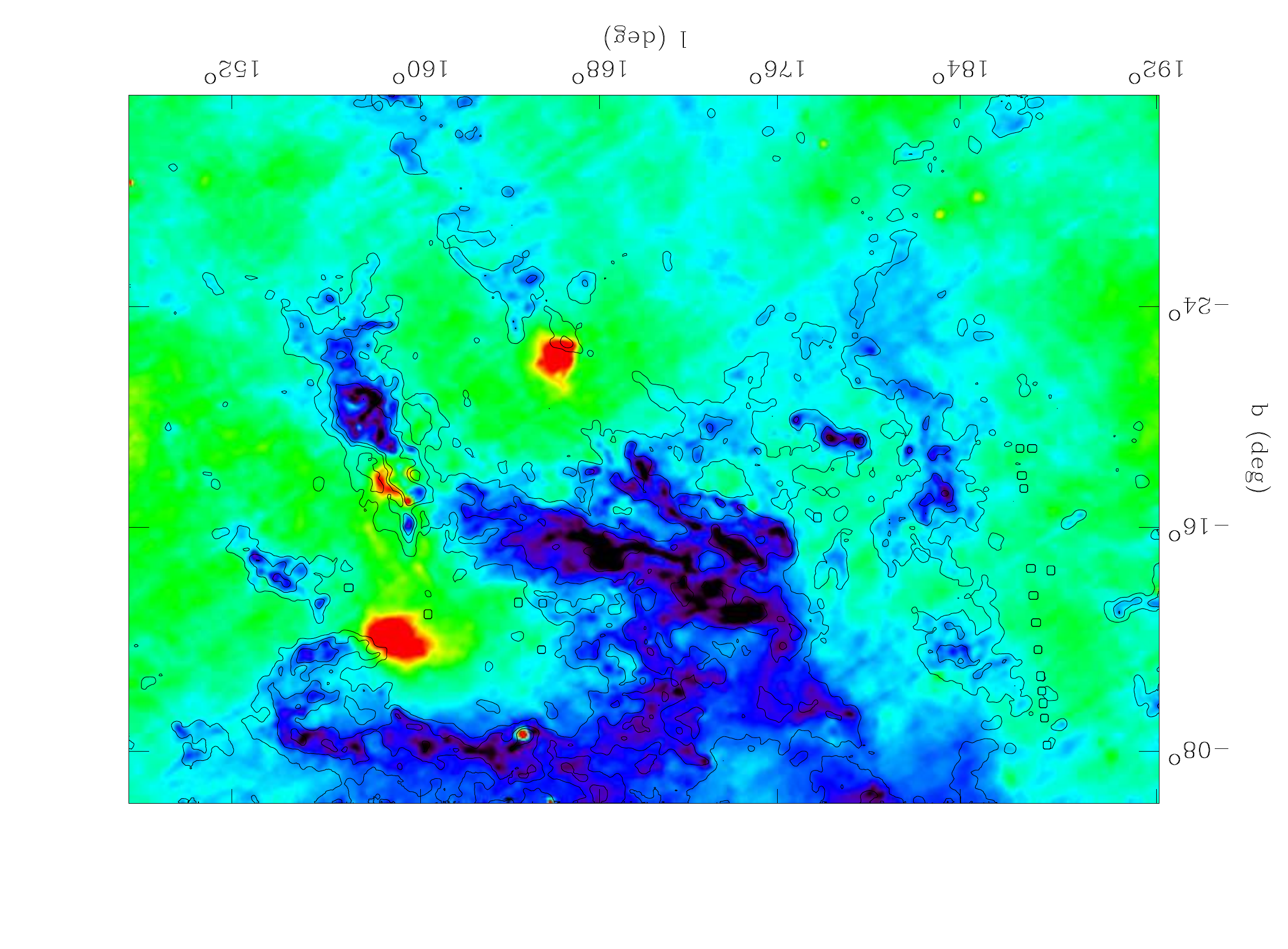}
\includegraphics[width=6cm,angle=180]{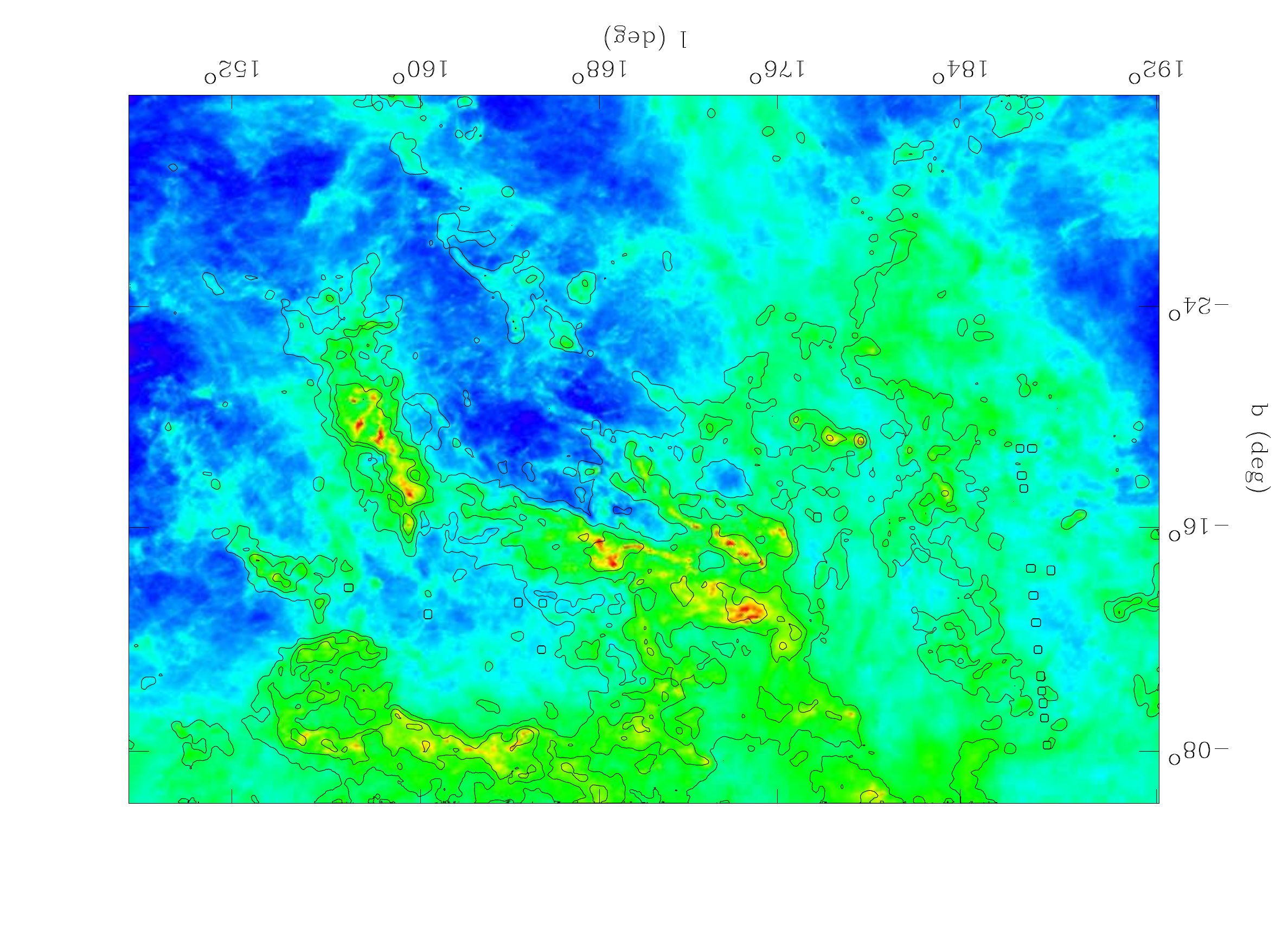}
\includegraphics[width=6cm,angle=180]{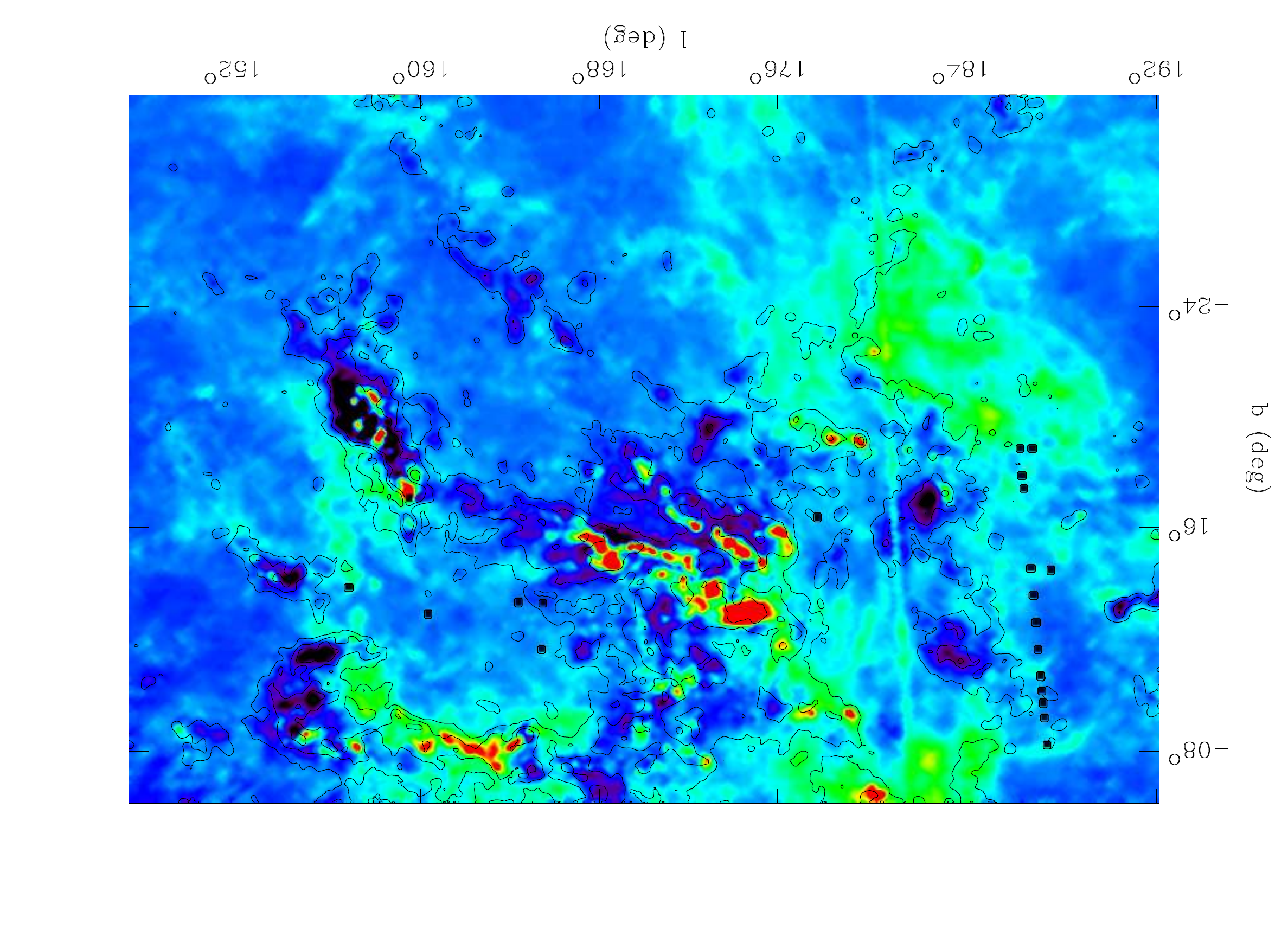}
\caption{
Details of the dust temperature (left column), dust optical depth at
{\HFIonefreq}{\GHz} (central column) and {\DG} column density (right
column) for the Chamaeleon (first line), Aquila-Ophiuchus flare
(second line), Polaris flare (third line) and Taurus (fourth line).
The temperature and optical depth maps are shown in log scale with a
colour scale ranging from 15\,K to 20\,K and $1\times10^{-5}$ to
$3\times10^{-3}$ respectively. The {\DG} column density derived from
the optical depth at {\HFIonefreq}{\GHz} (see Sec.\,\ref{sec_corr})
and is shown in linear scale with a colour scale ranging from -3 to
$7\times10^{21}\,\NHUNIT$ .  The contours show the
$^{12}$CO($J$=1$\rightarrow$0) integrated intensity at 2, 10 and
$20\Kkms$. The double line shows the limit of the CO surveys.
\label{fig:zoomed_maps}}
\end{center}
\end{figure*}

\begin{figure*}[ht!]
\begin{center}
\includegraphics[width=8cm,angle=180]{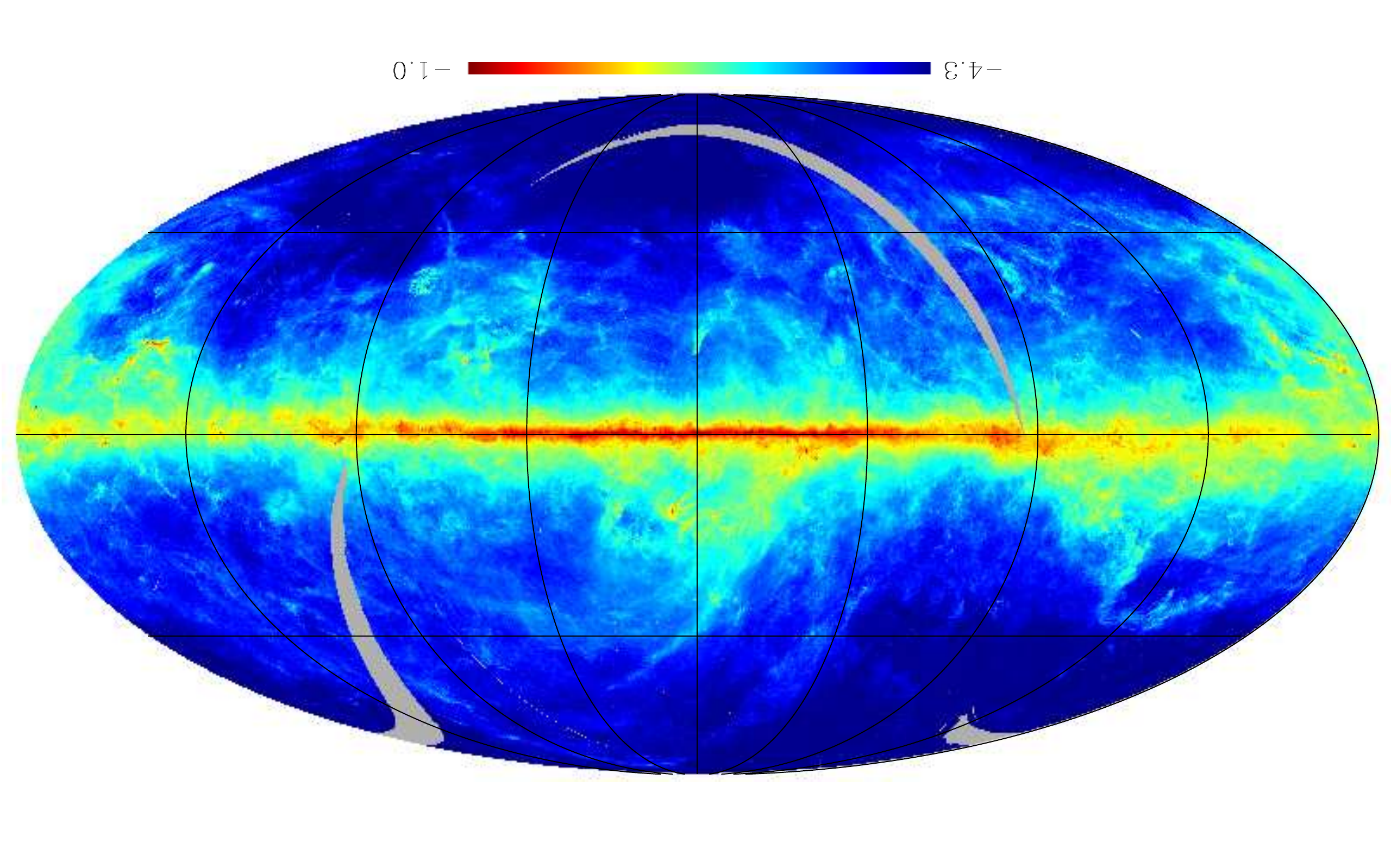}
\includegraphics[width=8cm,angle=180]{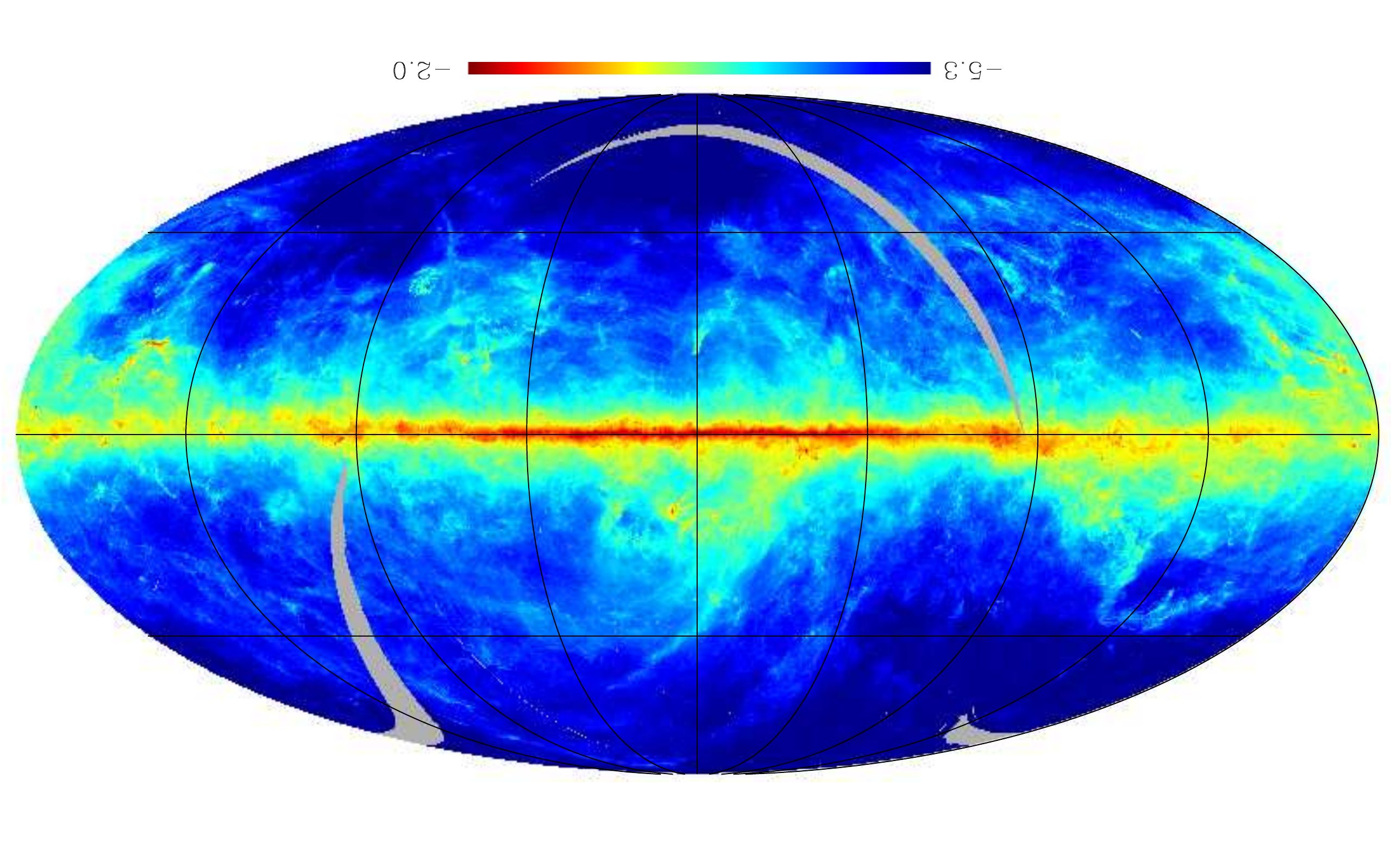}
\includegraphics[width=8cm,angle=180]{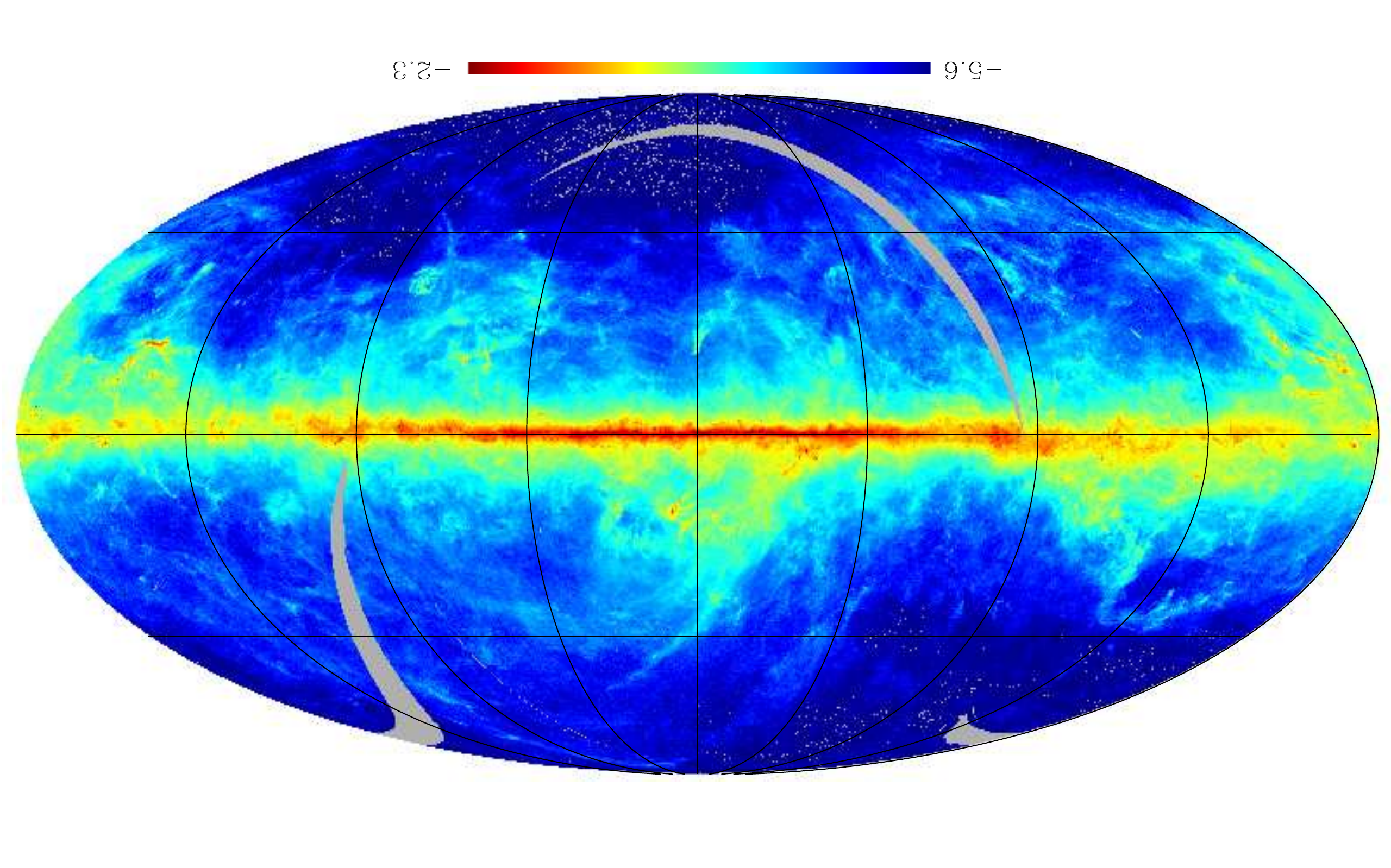}
\includegraphics[width=8cm,angle=180]{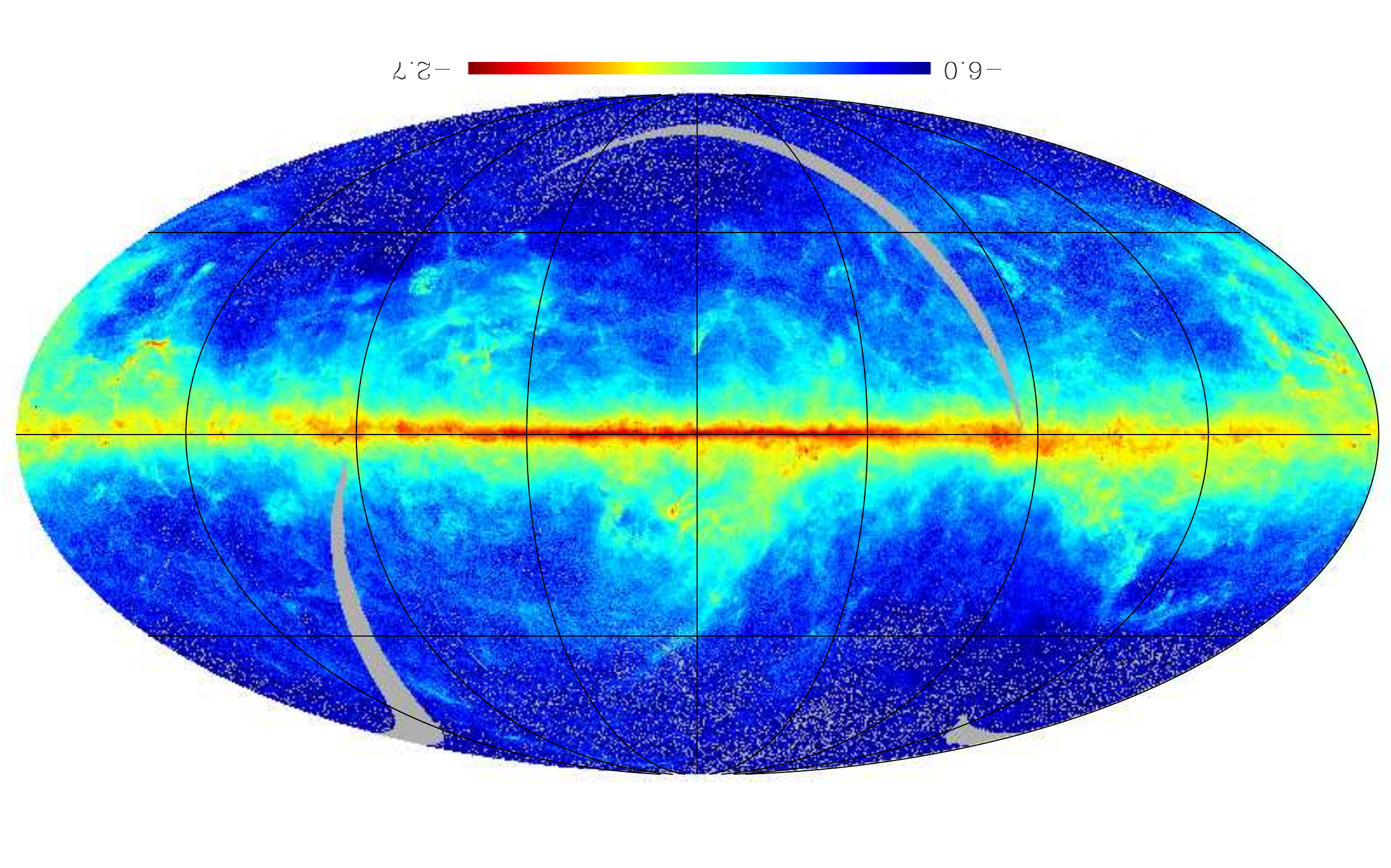}
\includegraphics[width=8cm,angle=180]{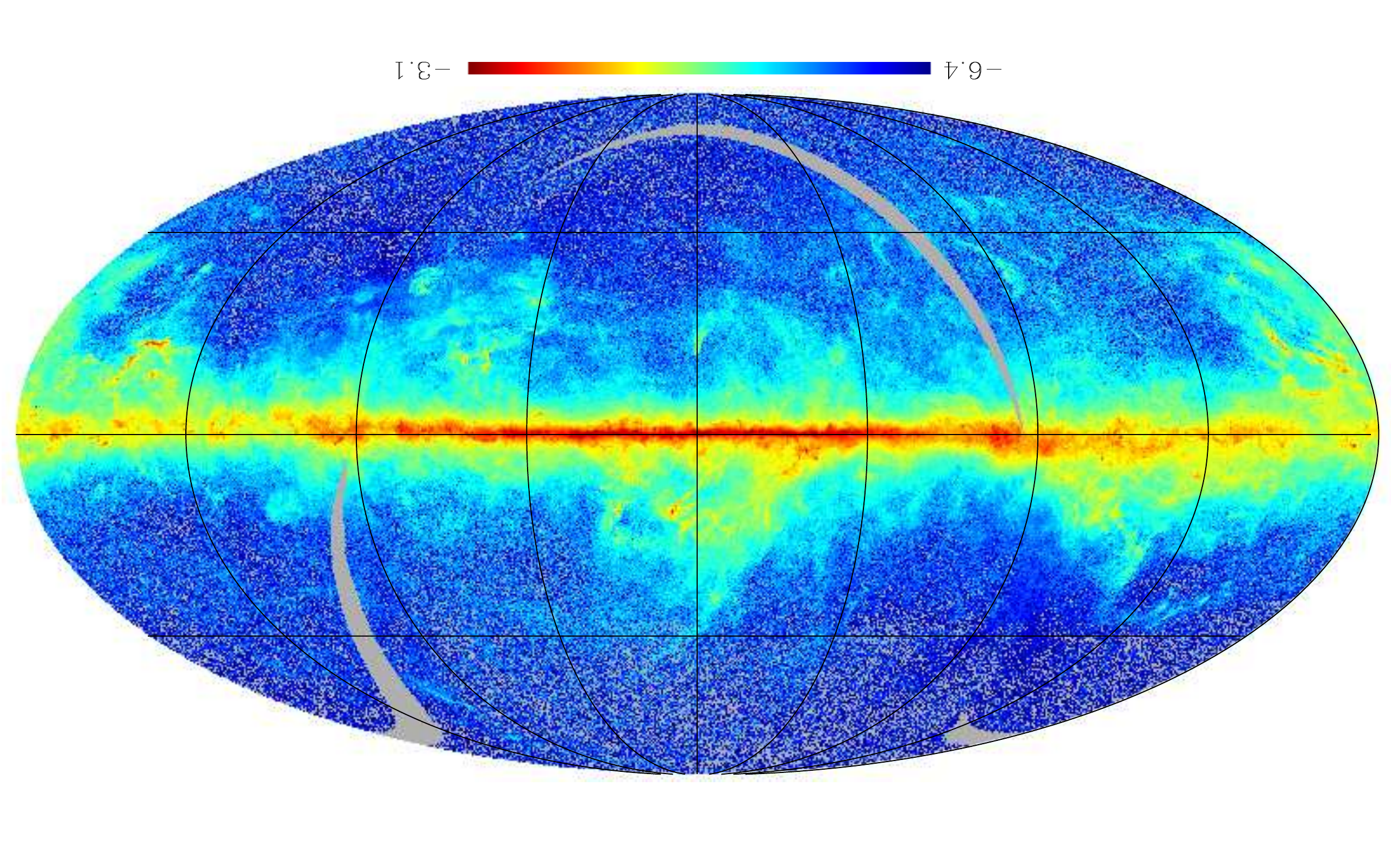}
\includegraphics[width=8cm,angle=180]{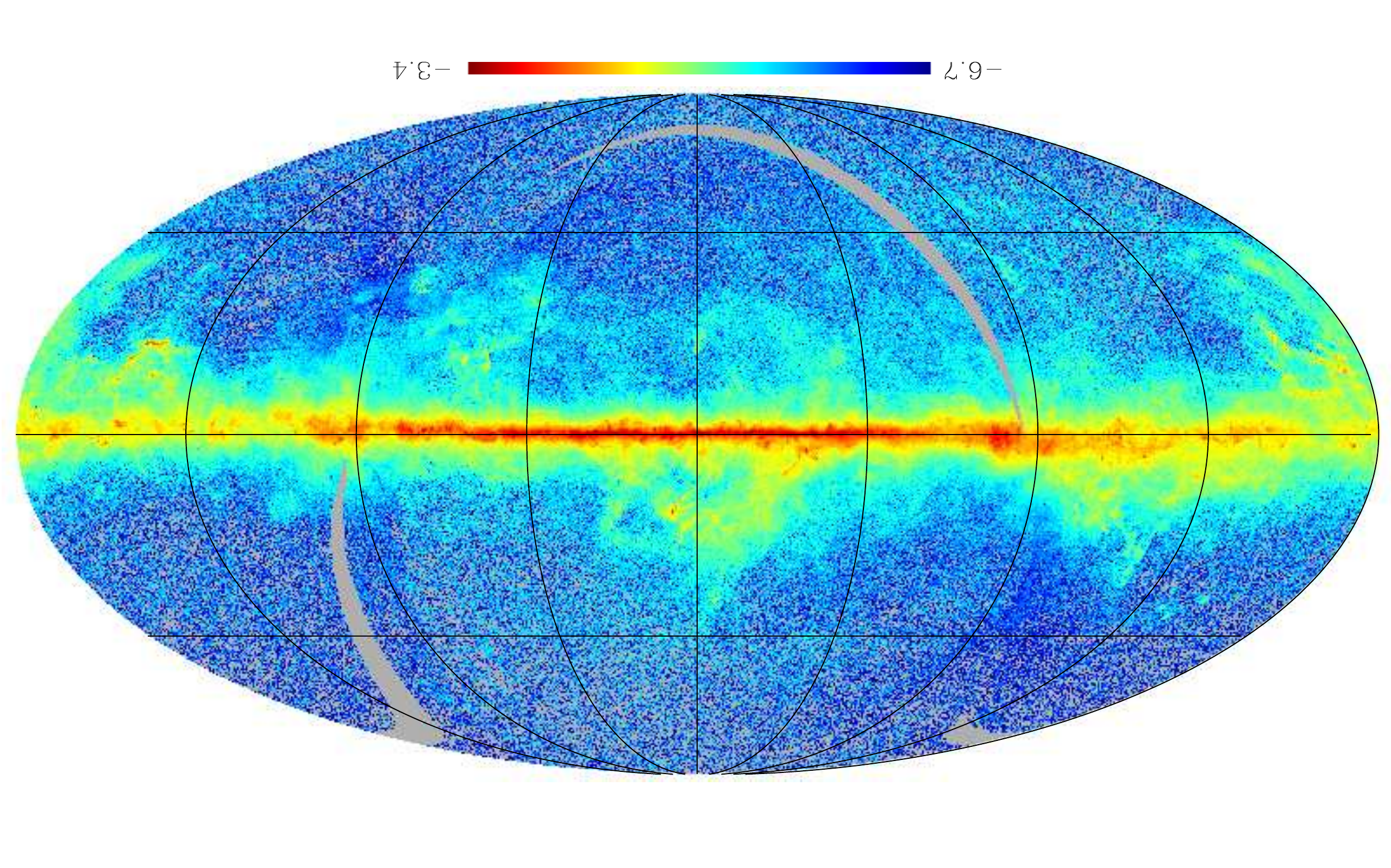}
\includegraphics[width=8cm,angle=180]{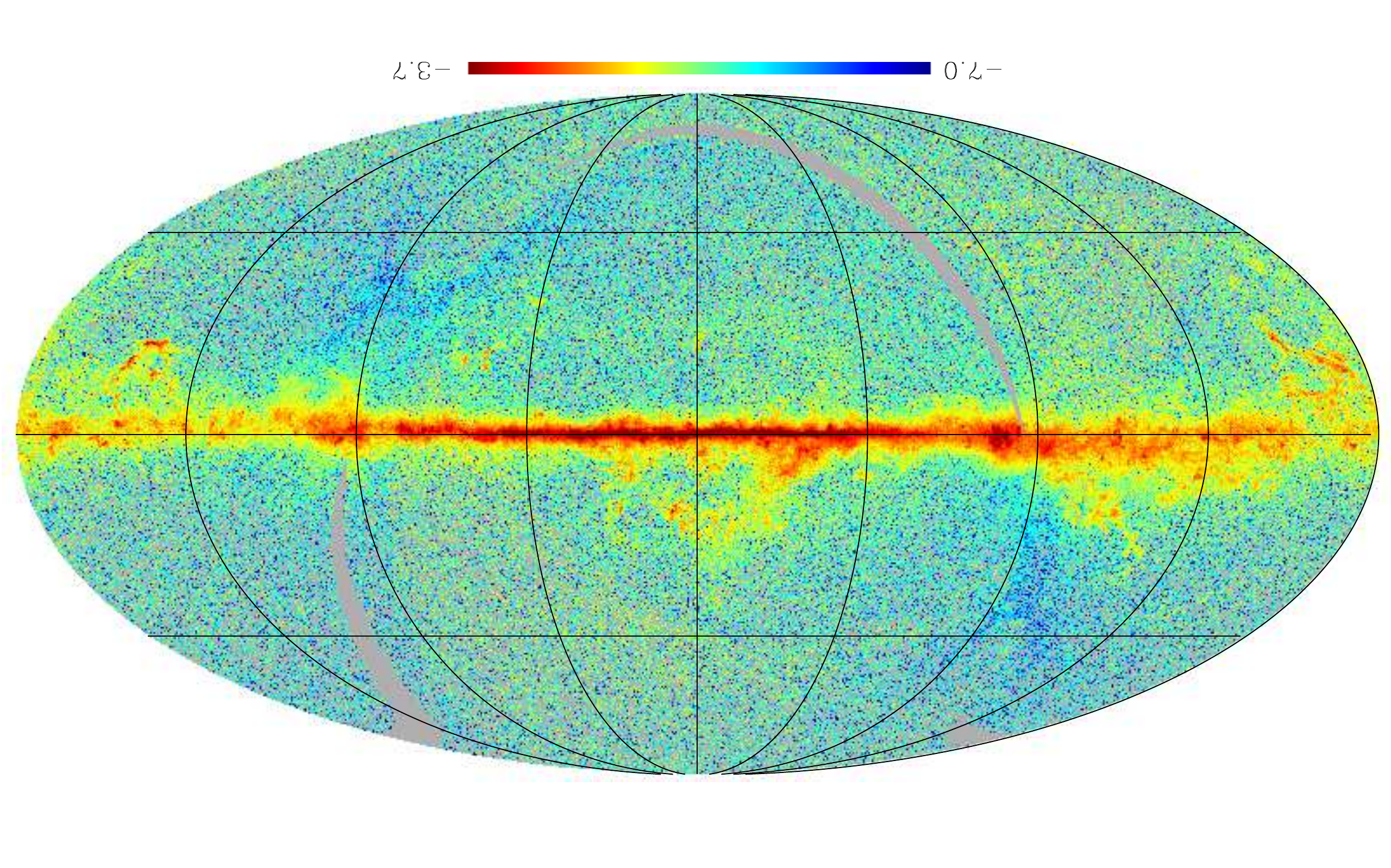}
\caption{
Maps of the dust optical depths on a log scale, in the {\iras}\,$100\mic$
(first row left) and {\Planck}-{\hfi} bands at $\HFIonefreq$ (first
row right), $\HFItwofreq$ (second row left), $\HFIthreefreq$ (second
row right), $\HFIfourfreq$ (third row left), $\HFIfivefreq$ (third row
right) and $\HFIsixfreq${\GHz} (fourth row). All maps are shown in
Galactic coordinates with the Galactic centre at the centre of the
image. The missing data in all images correspond to the {\iras}
gaps. The upper and lower bounds of the colour scale are set to $\rm
\tau_{min}=5\times10^{-5}\times(\lambda/100\mic)^{-1.8}$ and $\rm
\tau_{max}=10^{-2}\times(\lambda/100\mic)^{-1.8}$ respectively.
\label{fig_Tmap}}
\end{center}
\end{figure*}

The all-sky map of the thermal dust temperature computed as described
in Sec.\,\ref{sec:temperature} for $\beta=1.8$ is shown in
Fig.\,\ref{fig_Tmap}. The elongated regions with missing values in the
map correspond to the {\iras} gaps, where the temperature cannot be
determined from the {\Planck}-{\hfi} data alone. The distribution of the
temperature clearly reflects the large scale distribution of the
radiation field intensity.

Along the Galactic plane, a large gradient can be seen from the outer
Galactic regions, with $\rm \Td \simeq 14-15\,K$ to the inner Galactic
regions around the Galactic center regions with $\Td \simeq
19\,K$. This asymmetry was already seen at lower angular resolution in
the \dirbe \citep{Sodroski1994} and the FIRAS \citep{Reach1995}
data. The asymmetry is probably due to the presence of more massive
stars in the inner Milky Way regions, in particular in the molecular
ring. The presence of warmer dust in the inner Galaxy is actually
clearly highlighted by the radial distribution of the dust temperature
derived from Galactic inversion of IR data
\citep[e.g.][]{Sodroski1994,Paladini2007,planck2011-7.3}.  The origin
of the large scale region near (\lII,\bII)=($340\degr$,$-10\degr$)
with $\rm \Td\simeq 20\,K$ is currently unclear, but we note that it
corresponds to a region of enhanced X-ray emission in the Rosat
All-Sky Survey (RASS).

It may therefore correspond to warm dust associated with hot gas
pervading the local bubble around the Sun, or a pocket of hot gas in
Loop I. Similar large regions with enhanced dust temperature, such as
near (\lII,\bII)=($340\degr$,$-30\degr$) or
(\lII,\bII)=($315\degr$,$+30\degr$) may have a similar origin. Loop I
(\lII,\bII)=($30\degr$,$+45\degr$) is seen as a slightly warmer than
average structure at $\rm \Td \simeq 19\,K$.  Running parallel to it is
the Aquila-Ophiuchus flare (\lII,\bII)=($30\degr$,$+20\degr$) with
apparent $\rm \Td \simeq 14\,K$ extending to latitudes as high as
$60\degr$.  The Cepheus and Polaris Flare
(\lII,\bII)=(100--120$\degr$,+10--+20$\degr$) \citep[see][for a
detailed study]{planck2011-7.12} is also clearly visible as a lower
temperature arch extending up to \bII=$30\degr$ into the North
Celestial Pole loop and containing a collection of even colder
condensations ($\rm \Td \simeq 12-13\,K$).

On small angular scales, which are accessible over the whole sky only
with the combination of the \iras and {\Planck}-{\hfi} data at 5\arcm,
the map shows a variety of structures that can all be identified with
local heating by known single stars or \ion{H}{ii} regions for warmer
spots and with molecular clouds for colder regions.
Figure\,\ref{fig:zoomed_maps} illustrates the high resolution spatial
distribution of dust temperature and dust optical depth around some of
these regions.  Warmer regions include the tangent directions to the
spiral Galactic arms in Cygnus (\lII,\bII)=($80\degr$,$0\degr$) and
Carina (\lII,\bII)=($280\degr$,$0\degr$), hosts to many OB
associations, and many \ion{H}{ii} regions along the plane.  At higher
Galactic latitude, dust heated by individual hot stars such in the
Ophiuchi region (\lII,\bII)=($340\degr$,$+20\degr$) with individual
stars $\sigma-Sco$, $\nu-Sco$, $\rho-Oph$, $\zeta-Oph$, in Orion
(\lII,\bII)=($210\degr$,$-20\degr$) with the Trapezium stars or in
Perseus-Taurus (\lII,\bII)=($160\degr$,$-20\degr$) with the California
Nebula (NGC1499) can clearly be identified. Note the Spica HII region
at (\lII,\bII)=($300\degr$,$+50\degr$) where dust temperatures are
$\Td \simeq 20\,K$ due to heating by UV photons from the nearby (80
pc) early-type, giant (B1III) star $\alpha$ Vir.

At intermediate and high latitudes, nearby molecular clouds generally
stand out as cold dust environments with $\rm \Td \simeq 13\,K$. The
most noticeable ones are Taurus (\lII,\bII)=($160\degr$,$-20\degr$)
\citep[see][for a detailed study]{planck2011-7.13}, RCrA
(\lII,\bII)=($0\degr$,$-25\degr$), Chamaeleon
(\lII,\bII)=($300\degr$,$-20\degr$) and Orion
(\lII,\bII)=($200\degr$,$-20\degr$).  Numerous cold small scale
condensations can readily be found when inspecting the temperature
map, which mostly correspond to cold cores similar to those discovered
at higher resolution in the \herschel data
\citep[e.g.][]{Andre2010,Konyves2010,Molinari2010,Juvela2010}
and in the {\Planck} Cold-Core catalog \citep[see][]{planck2011-7.7a,planck2011-7.7b}.

Individual nearby Galaxies are also readily identified, in particular
the Large (\lII,\bII)=($279\degr$,$-34\degr$) and the Small Magellanic
Cloud (\lII,\bII)=($301\degr$,$-44\degr$) \citep[see][for a detailed
study]{planck2011-6.4b}, as well as M31 and M33.

Near the Galactic poles, the temperature determination becomes noisy
at the 5\arcm resolution due to the low signal levels.

\subsection{Optical depth determination}
\label{sec_tau}

Maps of the thermal dust optical depth ($\taudust(\lambda)$) are derived using:
\begin{equation}
\taudust(\lambda)=\frac{I_\nu(\lambda)}{B_\nu(\Td)},
\end{equation}
where $\rm B_\nu$ is the Planck function and $\rm I_\nu(\lambda)$ is the intensity map at frequency $\nu$.
We used resolution--matched maps of $\Td$ and $\rm I_\nu(\lambda)$ and derived $\rm \taudust(\lambda)$ maps at the
various resolutions of the data used here.  The maps of the uncertainty on
$\rm \taudust(\lambda)$ ($\rm \Delta \taudust$) are computed as:
\begin{equation}
\Delta \taudust(\nu)=\taudust \left( \frac{\sigmaII^2}{I_\nu^2} + \left( \frac{\delta B_\nu}{\delta T}(\Td)\right)^2 \frac{\Delta \Td^2}{B_\nu^2(\Td)}\right)^{1/2}.
\end{equation}

\section{Dust/Gas correlation}
\label{sec_corr}

\begin{figure*}[ht]
\begin{center}
\includegraphics[width=16cm,angle=0]{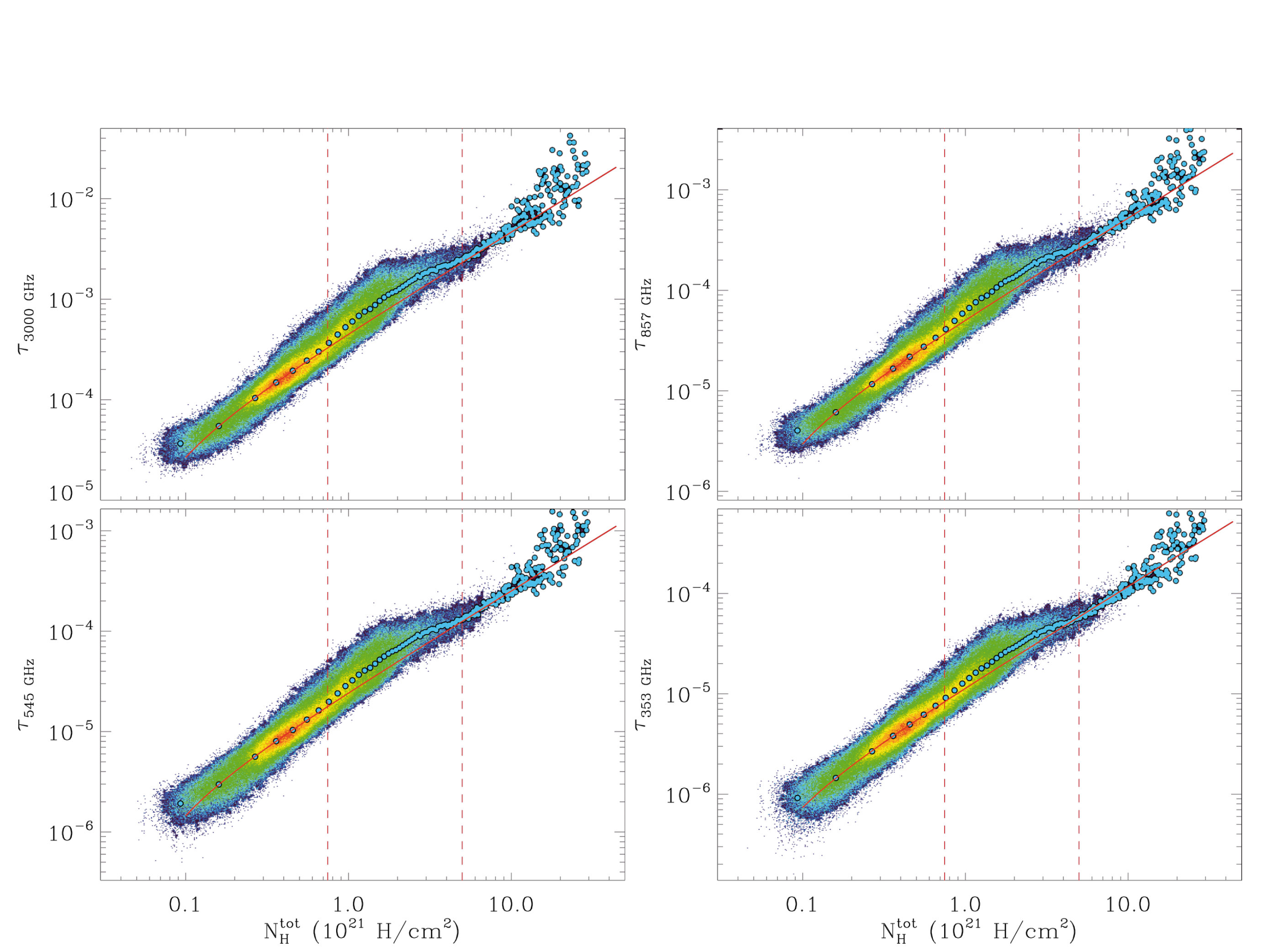}
\caption{
Correlation plots between the dust optical depth at {\iras}\,$100\mic$
(upper left), {\hfi}\,{\HFIonefreq}{\GHz} (upper right),
{\HFItwofreq}{\GHz} (lower left) and {\HFIthreefreq}{\GHz} (lower
right) and the total gas column density $\rm N_H^{obs}$ in the solar
neighbourhood ($\rm |$\bII$|>10\degr$).  The color scale represents the
density of sky pixels on a log scale.  The blue dots show a
$\rm N_H^{obs}$-binned average representation of the correlation. The red
line shows the best linear correlation derived at low $\rm N_H^{obs}$
values ($\rm \tau=\TAUNHREF*N_H^{obs}+cste$).  The vertical lines show
the positions corresponding to $\rm \Av=0.37\,mag$ and $\rm
\Av=2.5\,mag$. These figures are shown for a single
$\XCO=2.3\times10^{20}\XCOUNIT$.
\label{fig_DGcorrel}}
\end{center}
\end{figure*}

We model the dust opacity ($\taumodel$) as
\begin{equation}
\taumodel(\lambda) = \TAUNHREF [ N_{\rm HI} + 2 \XCO \wco],
\end{equation}
where $\TAUNHREF$ is the reference dust emissivity measured in low
$\NH$ regions and $\rm \XCO=N_{\Hdeux}/\wco$ is the traditional
$\Hdeux$/CO conversion factor.  It is implicitly assumed that the dust
opacity per unit gas column density is the same in the atomic and
molecular gas. If this is not the case, this will directly impact our
derived $\XCO$ since only the product of $\XCO$ by the dust emissivity
in the CO phase $\TAUNHCO$ can be derived here.  The fit to derive the
free parameters of the model is performed only in the portion of the
sky covered by all surveys (infrared, \ion{H}{i}, and CO) and where either (1) the
extinction is less than a threshold $\AVHIHtwo$, or (2) the CO is
detected with $\wco>1\Kkms$. Criterion (1) selects the low-column
density regions that are entirely atomic and suffer very small \ion{H}{i}
optical depth effects, so that the dust in this region will be
associated with the \ion{H}{i} emission at 21-cm.  Criterion (2) selects
regions where the CO is significantly detected and the dust is
associated with both the \ion{H}{i} and the $\rm ^{12}CO$ emission lines.  We
fit for the following three free parameters: $\TAUNHREF$, $\XCO$ and
$\AVHIHtwo$.  The threshold $\AVHIHtwo$ measures the extinction (or
equivalently the column density) where the correlation between the
dust optical depth and the {H}{i} column density becomes non-linear.

The correlation between the optical depth for various photometric
channels and the total gas column density ($\rm \NHTOT=N_{HI} + 2 \XCO
\wco$) is shown in Fig.\,\ref{fig_DGcorrel}. The correlations were
computed in the region of the sky where the CO data is available
(about 63\% of the sky) and at Galactic latitudes larger than
\bII$>10\degr$. The $\taudust$ and $\wco$ maps used were smoothed to
the common resolution of the \ion{H}{i} map ($0.6\degr$). For these plots, we
used a fixed value of $\rm \XCO=2.3\times10^{20}\XCOUNIT$. The colours show the
density of points in $\NHTOT$ and $\taudust$ bins. The dots show the
$\NHTOT$ binned average correlation. The larger scatter of these
points at high $\NHTOT$ comes from the limited number of points in the
corresponding bins.  The red line shows the $\taumodel$ model values
derived from the fit (slope=$\TAUNHREF$) to the low $\NHTOT$ part of the data.

It can be seen that the correlation is linear at low $\NHTOT$ values
and then departs from linear at $\rm \NHTOT \simeq \AVGNHHIHtwo\times10^{20}\NHUNIT$
($\AVHIHtwo \simeq \AVGAVHIHtwo\,mag$). Above $\NHTOT \simeq 5\times10^{21}\NHUNIT$
($\AVHtwoCO \simeq 2.5\,mag$), where $\NHTOT$ becomes dominated by the CO
contribution, the dust optical depth again is consistent with the
observed correlation at low $\NHTOT$ for this given choice of the
$\XCO$ value. Between these two limits, the dust optical depth is in
excess of the linear correlation. The same trend is observed in all
photometric channels shown, with a similar value for the threshold.
It is also observed in the {\hfi} bands at lower frequencies, but the
increasing noise at low $\NHTOT$ prevents an accurate determination of
the fit parameters.

\begin{figure}[h!]
\begin{center}
\includegraphics[width=9cm,angle=0]{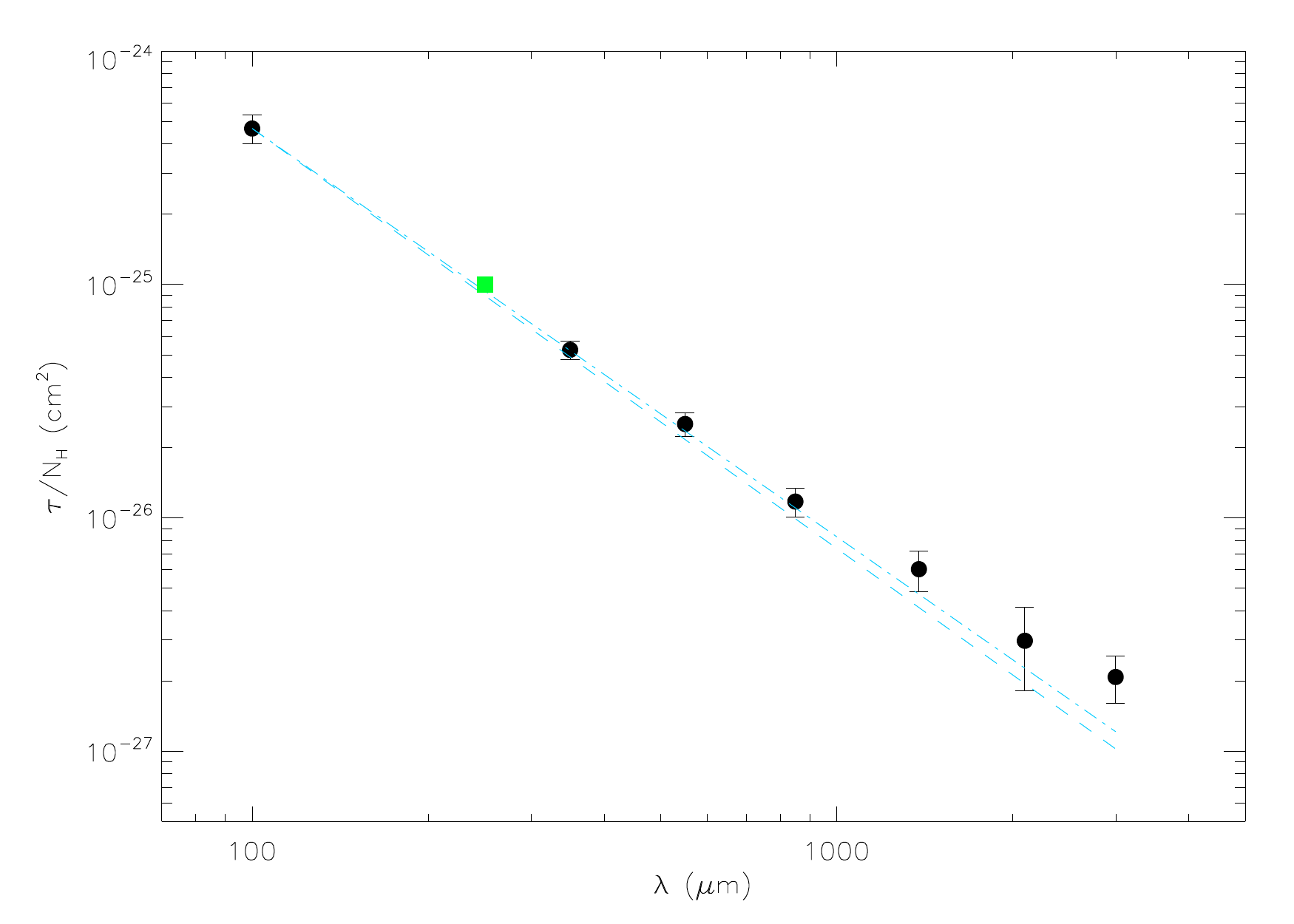}
\caption{
Dust optical depth derived from this study using the {\iras} and
{\Planck}-{\hfi} frequencies. The square symbol shows the emissivity
at $250\mic$ derived by \cite{Boulanger1996}.  The dash and dash-dot
lines show a power law emissivity with $\rm \lambda^{-1.8}$ and $\rm
\lambda^{-1.75}$ respectively, normalized to the data at $100\mic$.
The error bars shown are $\pm$1$\sigma$.
\label{fig:taunh}}
\end{center}
\end{figure}

The best fit parameters for $\TAUNHREF$, $\XCO$ and $\AVHIHtwo$ are
given in Table\,\ref{tab:DGresults}.  They were derived separately for
each frequency. The uncertainty was derived from the analysis of the
fitted $\chi^2$ around the best value.  The $\TAUNHREF$ values
decrease with increasing wavelength, as expected for dust emission.
The resulting dust optical depth SED is shown in
Fig.\,\ref{fig:taunh}. The dust optical depth in low column density
regions is compatible with $\beta=1.8$ at high frequencies.  The best
fit $\beta$ value between the {\iras} $100\mic$ and the
{\hfi}\,{\HFIonefreq}\GHz\_is actually found to be $\beta=1.75$.  The
SED then flattens slightly at intermediate frequencies with a slope of
$\beta=1.57$ around $\lambda=500\mic$ then steepens again to
$\beta=1.75$ above 1\,mm.  The $\XCO$ values derived from the fit are
constant within the error bars, which increase with wavelength.  The
average value, computed using a weight proportional to the inverse
variance is given in Table\,\ref{tab:DGresults} and is found to be
$\XCO=\AVGXCO \pm \AVGXCOERR \times10^{20}\XCOUNIT$.  Similarly, the
$\AVGAVHIHtwo$ parameter does not significantly change over the whole
frequency range and the weighted average value is found to be
$\AVHIHtwo=\AVGAVHIHtwo \pm \AVGAVHIHtwoERR$ mag.

The excess column density is defined using the difference between the
best fit and the observed dust opacity per unit column density using,
\begin{equation}
N_{\rm H}^{\rm x} \equiv (\taudust - \taumodel) / \TAUNHREF.
\end{equation}
The $N_{\rm H}^{\rm x}$ map is used to derive the total excess mass ($M_{\rm H}^{\rm x}$)
assuming a fiducial distance to the gas responsible for the excess.

We also computed the atomic and molecular total gas masses over the
same region of the sky, assuming the same distance.  In the region
covered by the CO survey, the \ion{H}{i} to CO mass ratio derived for
$\rm \XCO=\AVGXCO\times10^{20}\XCOUNIT$ is $M_{\rm HI}/M_{\rm CO}$=4.  Using the
average $\TAUNHREF$ and $\XCO$ values above, the ratio of the {\DG} mass
to the atomic gas mass ($\rm M_H^x/M_H^{HI}$) and to the molecular gas
mass ($\rm M_H^x/M_H^{CO}$) are given in Table\,\ref{tab:DGresults}.
On average, at high Galactic latitudes, the {\DG}
masses are of the order of $\rm \AVGMXMH\%\pm\AVGMXMHERR\%$ of the atomic gas mass and
$\rm \simeq \AVGMXMCO\%\pm\AVGMXMCOERR\%$ of the molecular mass.

\section{Dark-gas spatial distribution}
\label{sec_darkgas}

\begin{figure*}[htb]
\begin{center}
\includegraphics[width=16cm,angle=180]{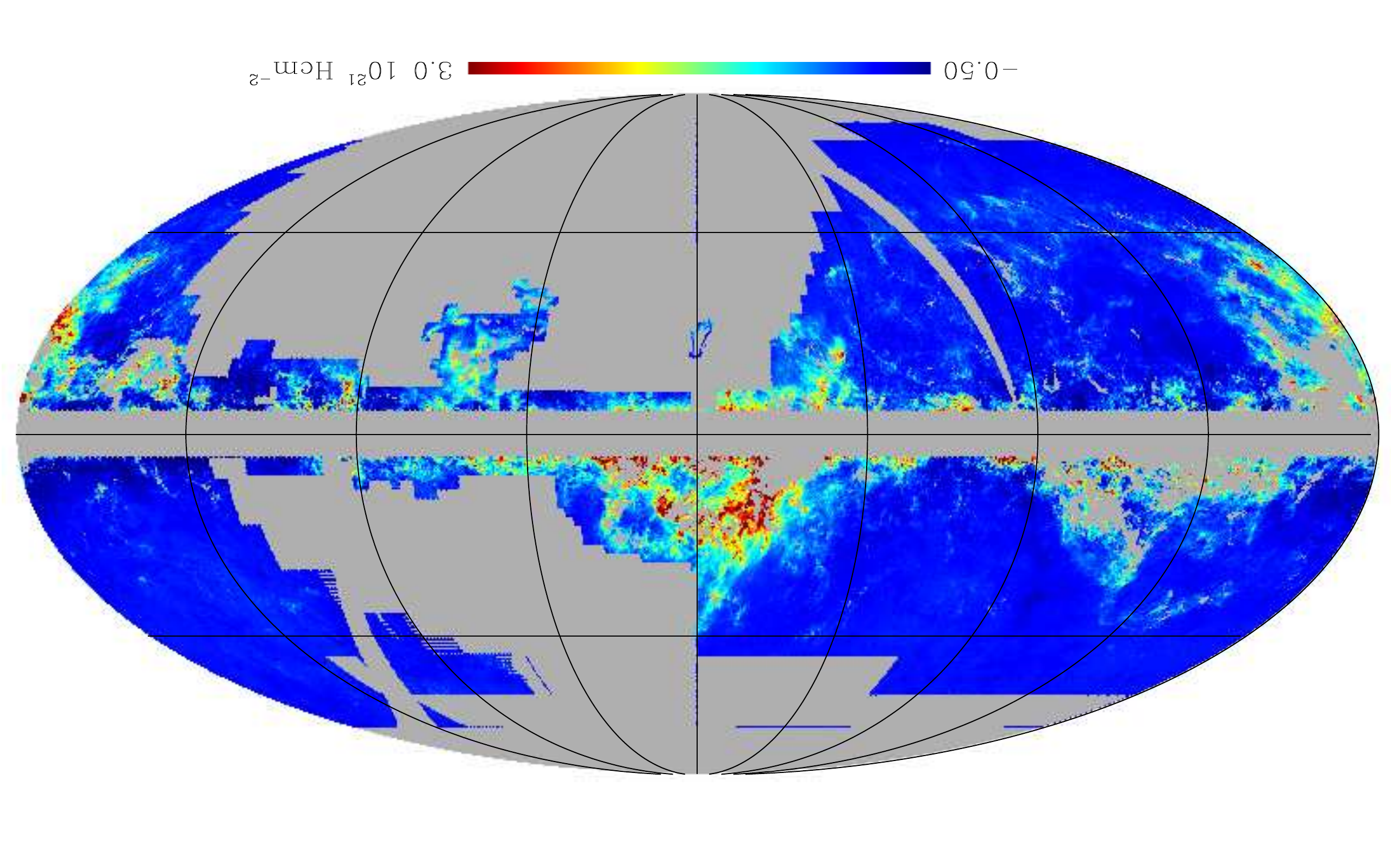}
\caption{
Map of the excess column density derived from the {\HFIonefreq}{\GHz}
data. The map is shown in Galactic coordinates with the Galactic
centre at the centre of the image. The grey regions correspond to
those where no {\iras} data are available, regions with intense CO
emission ($\rm \wco>1\Kkms$) and the Galactic plane ($\rm |$\bII$|<5\degr$).
\label{fig_DGdistrib}}
\end{center}
\end{figure*}

The spatial distribution of the {\DG} as derived from $\taudust$
computed from the {\hfi}\,{\HFIonefreq}\,{\GHz} channel is shown in
Fig.\,\ref{fig_DGdistrib}. It is shown in the region where the CO data
are available and above Galactic latitudes of $\rm |$\bII$|>5\degr$. Regions
where $\rm \wco>1\,Kkm/s$ have also been excluded. The map clearly shows
that the {\DG} is distributed mainly around the best known molecular clouds
such as Taurus, the Cepheus and Polaris flares, Chamaeleon and
Orion. The strongest excess region is in the Aquila-Ophiuchus flare,
which was already evident in \cite{Grenier2005}.

Significant {\DG} is also apparent at high latitudes, south of the
Galactic plane in the anticenter and around known translucent
molecular clouds, such as MBM53 (\lII$=90\degr$, \bII$=-30\degr$). As with
all the molecular clouds, the spatial distribution of the {\DG}
closely follows that of the Gould-Belt \citep{Perrot2003} and
indicates that most of the {\DG} in the solar neighbourhood belongs
to this dynamical structure.

\section{Discussion}
\label{sec:discussion}

\subsection{Dust emissivity in the atomic neutral gas}
\label{sec:diffatomic}

In the solar neighbourhood, \cite{Boulanger1996} measured an emissivity
value in the diffuse medium of $\rm 10^{-25}\,cm^2/H$ at 250 $\mic$
assuming a spectral index $\beta=$ 2 which seemed consistent with
their data. The optical depth of dust derived in our study in the low
$\NHTOT$ regions at $\rm |$\bII$|>10\degr$ is shown in
Fig.\,\ref{fig:taunh}.  The Figure also shows the reference value by
\cite{Boulanger1996} which is in good agreement with the values derived here, interpolated at
$250\mic$  (in fact $10\%$ above when using $\beta=1.8$ and $6\%$ above when
using $\beta=1.75$).  Our study does not allow us to measure the emissivity in
the molecular gas, since we are only sensitive to the product of this
emissivity with the $\XCO$ factor. However, we note that our derived
average $\XCO=\AVGXCO\times10^{20}\XCOUNIT$ is significantly higher than
previously derived values.  Even if we account for the possible
uncertainty in the calibration of the
$^{12}$CO($J$=1$\rightarrow$0)
emission (24\%)
discussed in Sec.\,\ref{sec:codata}, increasing the CO emission by the
corresponding factor would only lower our $\XCO$ estimate to
$\XCO=2.2\,\XCOUNIT$. In comparison, a value of $\rm
(1.8\pm0.3)\times10^{20}\XCOUNIT$ was found at $|$\bII$|>5\degr$ from the
comparison of the \ion{H}{i}, CO, and IRAS $100\mic$ maps
\citep{Dame2001}.  Similarly, values derived from $\gamma$-ray \fermi
data can be as low as $\rm \XCO=0.87\times10^{20}\XCOUNIT$ in Cepheus,
Cassiopea and Polaris \citep{Abdo2010}.  This could be evidence that
the dust emissivity in the high-latitude molecular material could be
larger than in the atomic phase by a factor $\simeq 3$.  Such an
increase in the dust emissivity in molecular regions has been inferred
in previous studies \citep[e.g.][]{Bernard1999,Stepnik2003} and was
attributed to dust aggregation.

\begin{table*}[htmb]
\begingroup
\newdimen\tblskip \tblskip=5pt
\caption{\label{tab:DGresults}Derived parameters for the {\DG}, computed in the
region with available CO data and $\rm |$\bII$|>10\degr$.}
\nointerlineskip
\footnotesize
\vskip -8pt
\setbox\tablebox=\vbox{
\newdimen\digitwidth
\setbox0=\hbox{\rm 0}
\digitwidth=\wd0
\catcode`*=\active
\def*{kern\digitwidth}
\newdimen\signwidth
\setbox0=\hbox{+}
\signwidth=\wd0
\catcode`!=\active
\def!{\kern\signwidth}
\halign{\hbox to 0.7in{#\hfil}\tabskip=2.2em&
	\hfil#\hfil&
	\hfil#\hfil&
	\hfil#\hfil&
	\hfil#\hfil&
	\hfil#\hfil\tabskip 8pt\cr
\noalign{\vskip 3pt\hrule\vskip 1.5pt\hrule\vskip 5pt}
\omit\hfil Frequency\hfil&$\rm \taudust/N_H^{obs}$ & $\XCO$ & $\AVHIHtwo$ & $\rm M_H^X/M_H^{HI}$ & $\rm M_H^X/M_H^{CO}$ \cr
\hfil [{\GHz}] & [$10^{-25} cm^2$] & [$\XCOUNIT$] & [mag] & -- & -- \cr
\noalign{\vskip 4pt\hrule\vskip 6pt}
\IRASfourfreq   & 4.66$\pm$0.65                 & (2.60$\pm$0.18)$\times10^{20}$ & (4.05$\pm$0.39)$\times10^{-1}$ & (2.91$\pm$0.38)$\times10^{-1}$ & 1.27$\pm$0.16\cr
\HFIonefreq     & (5.25$\pm$0.49)$\times10^{-1}$ & (2.52$\pm$0.29)$\times10^{20}$ & (3.92$\pm$0.64)$\times10^{-1}$ & (2.73$\pm$0.66)$\times10^{-1}$ & 1.22$\pm$0.30\cr
\HFItwofreq     & (2.53$\pm$0.29)$\times10^{-1}$ & (2.52$\pm$0.36)$\times10^{20}$ & (3.96$\pm$0.79)$\times10^{-1}$ & (2.77$\pm$0.95)$\times10^{-1}$ & 1.24$\pm$0.43\cr
\HFIthreefreq   & (1.18$\pm$0.17)$\times10^{-1}$ & (2.31$\pm$0.47)$\times10^{20}$ & (4.07$\pm$1.28)$\times10^{-1}$ & (2.26$\pm$0.67)$\times10^{-1}$ & 1.10$\pm$0.33\cr
\HFIfourfreq    & (6.03$\pm$1.19)$\times10^{-2}$ & (2.58$\pm$0.67)$\times10^{20}$ & (4.60$\pm$2.39)$\times10^{-1}$ & (1.71$\pm$4.60)$\times10^{-1}$ & (7.46$\pm$20.09)$\times10^{-1}$\cr
\HFIfivefreq    & (2.98$\pm$1.15)$\times10^{-2}$ & (1.83$\pm$1.00)$\times10^{20}$ & (4.62$\pm$4.81)$\times10^{-1}$ & -- & -- \cr
\HFIsixfreq     & (2.08$\pm$0.48)$\times10^{-2}$ & (4.05$\pm$1.93)$\times10^{20}$ & (6.69$\pm$5.15)$\times10^{-1}$ & -- & -- \cr
\noalign{\vskip 3pt\hrule\vskip 4pt}
 Average        & --                             & (2.54$\pm$0.13)$\times10^{20}$ & (4.03$\pm$0.29)$\times10^{-1}$ & (2.78$\pm$0.28)$\times10^{-1}$ & 1.18$\pm$0.12\cr
\noalign{\vskip 3pt\hrule\vskip 4pt}
}}
\endPlancktable
\endgroup
\end{table*}

\subsection{Dark Molecular gas}
\label{sec:DMG}

The nature of `dark molecular gas' has recently been investigated theoretically
by \cite{Wolfire2010}, who specifically address the HI/$\Hdeux$ and
C/C$^+$ transition at the edges of molecular clouds. The nominal cloud modeled
in their study is relatively large, with total column density
$\rm 1.5\times 10^{22}\,cm^{-2}$, so the applicability of the results to
the more translucent conditions of high-Galactic-latitude clouds is
not guaranteed. The envelope of the cloud has an \ion{H}{i} column density of
$\rm 1.9\times 10^{21}\,cm^{-2}$ which is more typical of the entire
column density measured at high latitudes. \cite{Wolfire2010} define $f_{DG}$
as the fraction of molecular gas that is dark, i.e. not detected by
CO. In the nominal model, the chemical and photodissociation balance
yields a total $\Hdeux$ column density of $\rm 7.0\times 10^{21}\,cm^{-2}$,
while the `dark' $\Hdeux$ in the transition region where CO is
dissociated has a column density of $\rm 1.9\times 10^{21}\,cm^{-2}$. The
fraction of the total gas column density that is molecular,
\begin{equation}
f(\Hdeux) = \frac{2 N(\Hdeux)}{2N(\Hdeux)+N(HI)}
\end{equation}
is 93\% in the nominal model, which suggestss that the line of sight
through such a cloud passes through material which is almost entirely molecular.  To compare the
theoretical model to our observational results, we must put them into
the same units.  We define the {\DG} fraction as the fraction of
the total gas column density that is dark,
\begin{equation}
f_{\rm DARK} = \frac{2 N(\Hdeux^{\rm dark})}{2N(\Hdeux)+N(HI)} = f(\Hdeux) f_{\rm DG}.
\end{equation}
For the nominal \cite{Wolfire2010} model, $f_{\rm DG}=$0.29 so we can
infer $f_{\rm DARK}$=0.27.  The smaller clouds in Figure 11 of their paper
have larger $f_{\rm DG}$, but $f(\Hdeux)$ is also probably smaller (not given
in the paper) so we cannot yet definitively match the model and
observations.  These model calculations are in general agreement with
our observational results, in that a significant fraction of the
molecular gas can be in CO-dissociated `dark' layers.

If we assume that all dark molecular gas in the solar neighbourhood is
evenly distributed to the observed CO clouds, the average $f_{DG}$
measured is in the range $f_{DG}=1.06-1.22$. This is more than three times
larger than predicted by the \cite{Wolfire2010} mass fraction. This
may indicate that molecular clouds less massive than the ones assumed
in the model actually have a {\DG} mass fraction higher by a factor
of about three. This would contradict their conclusion that the dark mass fraction 
does not depend on the total cloud mass.

The location of the \ion{H}{i}-to-$\Hdeux$ transition measured here ($\rm \AVHIHtwo
\simeq \AVGAVHIHtwo\,mag$) is comparable, although slightly higher than that
predicted in the \cite{Wolfire2010} model ($\rm \AVHIHtwo \simeq 0.2\,
mag$). Again, this difference may indicate variations with the cloud
size used, since UV shadowing by the cloud itself is expected to be
less efficient for smaller clouds, leading to a transition deeper into
the cloud.

\subsection{Other possible origins}
\label{sec:DMG}

The observed departure from linearity between $\taudust$ and the
observable gas column density could also in principle be caused by
variations of the dust/gas ratio (D/G). However, such variations with
amplitude of $30\%$ in the solar neighbourhood and a systematic trend
for a higher D/G ratio in denser regions would be difficult to explain
over such a small volume and in the presence of widespread enrichment
by star formation.  However, the fact that the {\DG} is also
seen in the $\gamma$-ray with comparable amplitudes is a strong
indication that it originates from the gas phase.  The {\DG}
column-densities inferred from the $\gamma$-ray observations are also
consistent with a standard D/G ratio \citep{Grenier2005}.

The observed excess optical depth could also in principle be due to
variations of the dust emissivity in the FIR-Submm. We expect such
variations to occur if dust is in the form of aggregates with higher
emissivity \citep[e.g.][]{Stepnik2003} in the {\DG} region.
We note however that such modifications of the optical properties mainly affect
the FIR-submm emissivity and are not expected to modify significantly
the absorption properties in the Visible. Therefore, 
detecting a similar departure from linearity between large-scale extinction
maps and the observable gas would allow us to exclude this possibility.

Sky directions where no CO is detected at the sensitivity of the CO
survey used (0.3-$1.2\Kkms$) may actually host significant CO
emission, which could be responsible for the excess dust optical depth
observed. Evidences for such diffuse and weakly emitting CO gas have
been reported.  For instance, in their study of the large-scale
molecular emission of the Taurus complex, \cite{Goldsmith2008} have
found that half the mass of the complex is in regions of low column
density $\rm \NH < 2\times10^{21}\,cm^{-2}$, seen below $\wco\simeq
1\Kkms$. However, \cite{Barriault2010} reported a poor spatial
correlation between emission by diffuse CO and regions of FIR excess
in two high Galactic latitude regions in the Polaris Flare and Ursa
Major.  The difficulty at finding the CO emission associated to {\DG}
is that the edges of molecular coulds tend to be highly
structured spatially, which could explain why many attempts have been
unsuccessful \citep[see for instance][]{Falgarone1991}.
In our case, it is possible to obtain an upper limit to the
contribution of weak CO emission below the survey detection threshold,
by assuming that pixels with undetected CO emission actually emit with
$\wco=0.5\Kkms$. This is the detection limit of the survey we use
at $|b|>10\degr$ so this should be considered an upper limit to the
contribution of undetected diffuse CO emission. In that case, the {\DG}
mass is reduced by a factor lower than 20\%. This indicates that,
although diffuse weak CO emission could contribute a fraction of the
observed excess emission, it cannot produce the bulk of it.

Finally, we recognize that the optically thin approximation used here
for the \ion{H}{i} emission may not fully account for the whole atomic gas
present, even at high latitude.  \ion{H}{i} emission is subject to self
absorption and $N_{H}$ can be underestimated from applying too high a
spin temperature ($T_s$) while deriving column densities. $T_{s}$ is
likely to vary from place to place depending on the relative abundance
of CNM clumps (with thermodynamical temperatures of 20-100\,K) and WNM
clouds (at several thousand K) in the telescope beam. The
effective spin temperature of 250-400\,K to be applied to correct for
this blending andto  retrieve the total column density from the \ion{H}{i}
spectra does not vary much in the Galaxy \citep{Dickey2003,Dickey2009}.
It indicates that most of the \ion{H}{i} mass is in the
warm phase and that the relative abundance of cold and warm \ion{H}{i} is a
robust fraction across the Galaxy (outside of the inner molecular
ring).  The correlation between the \fermi $\gamma$-ray maps and the
\ion{H}{i} column densities derived for different spin temperatures also
support an average (uniform) effective spin temperature $>250\,K$ on
and off the plane \citep{Ackermann2010}.  In order to test these
effects, we performed the analysis described in this paper using a
very low choice for the \ion{H}{i} spin temperature. We adopted a value of
$\rm T_s=80\,K$ when the observed \ion{H}{i} peak temperature is below 80\,K and
$\rm T_s=1.1\times T_{peak}$ when above. Under this hypothesis, we
obtained {\DG} fractions which are about half of those given in
Table\,\ref{tab:DGresults} under the optically thin approximation.  We
consider this to indicate that significantly less than half of the
detected {\DG} could be dense, cold atomic gas. We further note that,
under the optically thin \ion{H}{i} hypothesis, the {\DG} fraction appears very
constant with Galactic latitude down to $\rm |$\bII$|\simeq3\degr$ (see
Sec.\,\ref{sec:spatialvar}), while it varies more strongly using
$\rm T_s=80\,K$. This does not support the interpretation that the bulk of
the dust excess results from underestimated \ion{H}{i} column densities.

\subsection{Dark-Gas variations with latitude}
\label{sec:spatialvar}

\begin{figure}
\begin{center}
\includegraphics[width=9cm,angle=0]{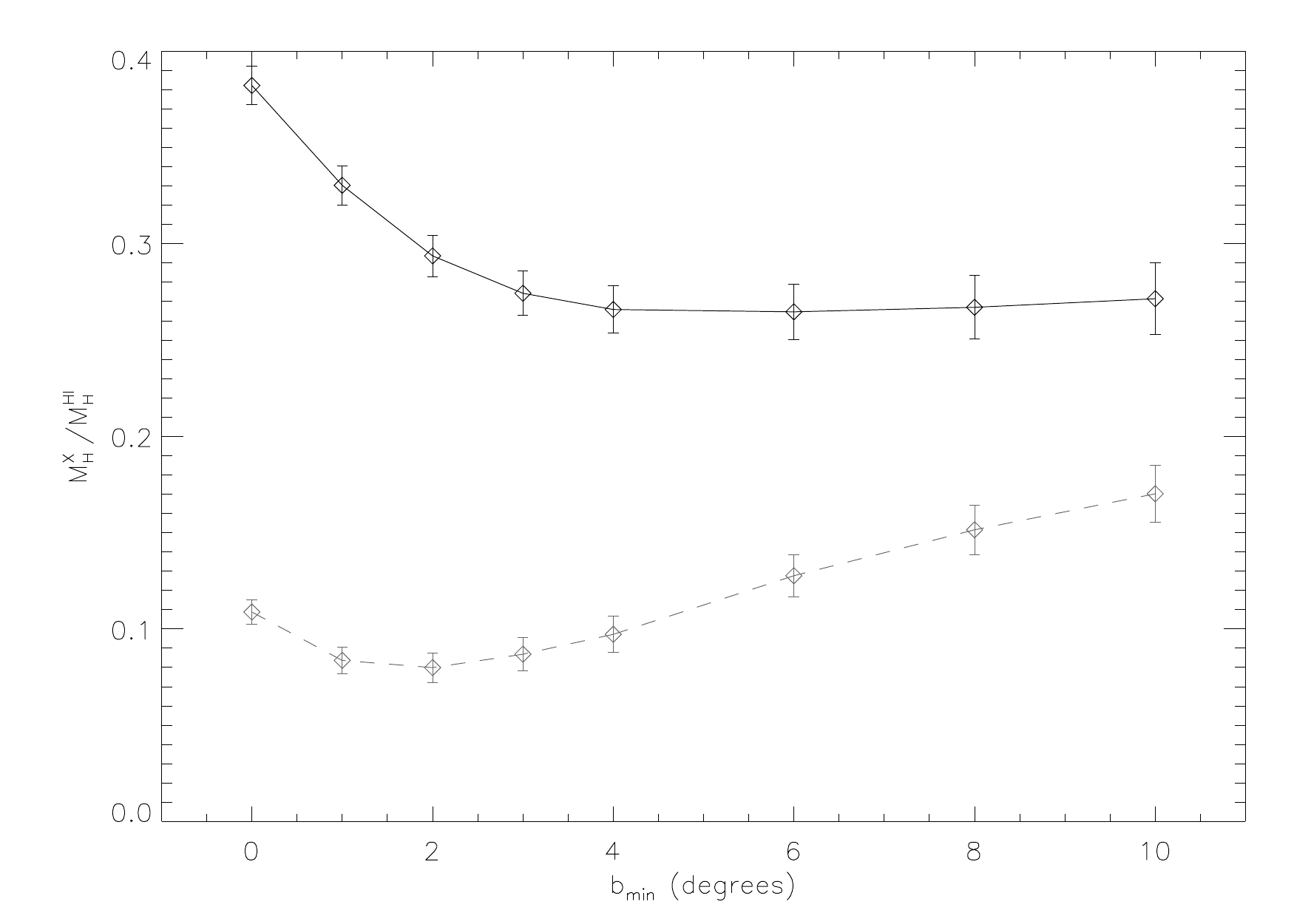}
\caption{
Fractional mass of the {\DG} with respect to the neutral gas mass as a
function of the lower \bII value used in the analysis.  The solid
curve is computed under the assumption of optically thin \ion{H}{i},
the dashed curve is for $\rm N_H^{\ion{H}{i}}$ computed using $\rm
T_s=80\,K$. Error bars are 1$\sigma$.
\label{fig:mass_bmin}}
\end{center}
\end{figure}

We investigate the distribution of the {\DG} as a function of Galactic
latitude.  This is important, since the {\DG} template produced here
for the solar neighbourhood is also used in directions toward the plane
for Galactic inversion purpose in \cite{planck2011-7.3}.  We performed
the calculations described in Sec.\,\ref{sec_corr} for various values
of the Galactic latitude lower cutoff ($\rm b_{min}$) in the range
$\rm b_{min}<|$\bII$|<90\degr$ with $\rm b_{min}$ varying from $\rm
0\degr$ to $\rm 10\degr$. For each value, we used the best fit
parameters derived from $\rm b_{min}=10\degr$ and given in
Table\,\ref{tab:DGresults}.

Figure\,\ref{fig:mass_bmin} shows the evolution of the {\DG} mass
fraction with respect to the atomic gas mass as a function of $\rm
b_{min}$. It can be seen that the ratio changes only mildly (increases
by a factor 1.12 from $b_{min}=10\degr$ to $\rm b_{min}=2\degr$) as we
approach the Galactic plane. This indicates that a fairly constant
fraction of the {\DG} derived from the solar neighbourhood can be
applied to the rest of the Galaxy.

Figure\,\ref{fig:mass_bmin} also shows the same quantity computed
using the \ion{H}{i} column density derived using $\rm T_s=80\,K$. It
can be seen that, in that case, the {\DG} fraction is predicted to
decrease by a factor 2.12 from $\rm b_{min}=10\degr$ to $\rm
b_{min}=2\degr$. This is caused by the much larger inferred \ion{H}{i}
masses toward the plane under this hypothesis. We consider it unlikely
that the {\DG} fraction varies by such a large factor from the solar
neighbourhood to the Galactic plane, and consider it more likely that
the correction applied to $\rm N_H$ by using a spin temperature as low
as $\rm T_s=80\,K$ actually strongly overestimates the \ion{H}{i}
opacity, and therefore the fraction of the {\DG} belonging to atomic
gas.

\section{Conclusions}
\label{sec_conclusions}

We used the {\Planck}-{\hfi} and {\iras} data to determine all sky maps of the
thermal dust temperature and optical depth. The temperature map
traces the spatial variations of the radiation field intensity associated
with star formation in the Galaxy. This type of map is very important
for the detailed analysis of the dust properties and their spatial
variations.

We examined the correlation between the dust optical depth and gas
column density as derived from \ion{H}{i} and CO observations. These two
quantities are linearly correlated below a threshold column density of
$\rm N_H^{obs}<\AVGNHHIHtwo\times10^{20}\NHUNIT$ corresponding to $\rm
\Av<\AVGAVHIHtwo\,mag$.  Below this threshold, we observed dust emissivities
following a power-law with $\beta \simeq 1.8$ below
$\lambda\simeq500\mic$ and flattening at longer wavelengths. Absolute
emissivity values derived in the FIR are consistent with previous
estimates.

This linear correlation also holds at high column densities
($\rm N_H^{obs}>5\times10^{21}\NHUNIT$) corresponding to $\rm \Av=2.5\,mag$ where
the total column density is dominated by the molecular phase for a
given choice of the $\XCO$ factor. Under the assumption that the dust
emissivity is the same in both phases, this leads to an estimate of
the average local CO to $\Hdeux$ factor of $\XCO=\AVGXCO\times10^{20}\XCOUNIT$.
The optical depth in the intermediate column density range shows an
excess in all photometric channels considered in this study.  We
interpret the excess as dust emission associated with {\DG}, probably
in the molecular phase where $\Hdeux$ survives photodissociation, while
the CO molecule does not.

In the solar neighbourhood, the derived mass of the {\DG}, assuming the
same dust emissivity as in the \ion{H}{i} phase is found to correspond
to $\simeq \AVGMXMH\%$ of the atomic mass and $\simeq \AVGMXMCO\%$ of the
molecular gas mass. The comparison of this value with the recent
calculations for dark molecular gas around clouds more massive than
the ones present in the solar neighbourhood indicates a {\DG} fraction
about three times larger in the solar neighbourhood. The threshold for
the onset of the {\DG} transition is found to be $\simeq \AVGAVHIHtwo$\,\mag
and appears compatible to, although slightly larger than, the
thresholds predicted by this model.  Finally, we stress that the
\ion{H}{i} 21 cm line is unlikely to be optically thin and to measure
all the atomic gas. Therefore, the {\DG} detected here could well
represent a mixture of dark molecular and dark atomic gas seen through
its dust emission.  For an average \ion{H}{i} spin temperature of $\rm
80\,K$, the mixture is predicted to be 50\% atomic and 50\% molecular.

\begin{acknowledgements}
A description of the Planck Collaboration and a list of its members can be found at \url{http://www.rssd.esa.int/index.php?project=PLANCK&page=Planck_Collaboration}
\end{acknowledgements}

\bibliographystyle{aa}

\bibliography{Planck_all}

\begin{thebibliography}{68}
\expandafter\ifx\csname natexlab\endcsname\relax\def\natexlab#1{#1}\fi

\bibitem[{{Abdo} {et~al.}(2010){Abdo}, {Ackermann}, {Ajello}, {Baldini},
  {Ballet}, {Barbiellini}, {Bastieri}, {Baughman}, {Bechtol}, {Bellazzini},
  {Berenji}, {Bloom}, {Bonamente}, {Borgland}, {Bregeon}, {Brez}, {Brigida},
  {Bruel}, {Burnett}, {Buson}, {Caliandro}, {Cameron}, {Caraveo}, {Casandjian},
  {Cecchi}, {{\c C}elik}, {Chekhtman}, {Cheung}, {Chiang}, {Ciprini}, {Claus},
  {Cohen-Tanugi}, {Cominsky}, {Conrad}, {Dermer}, {de Palma}, {Digel}, {Silva},
  {Drell}, {Dubois}, {Dumora}, {Farnier}, {Favuzzi}, {Fegan}, {Focke},
  {Fortin}, {Frailis}, {Fukazawa}, {Funk}, {Fusco}, {Gargano}, {Gehrels},
  {Germani}, {Giavitto}, {Giebels}, {Giglietto}, {Giordano}, {Glanzman},
  {Godfrey}, {Grenier}, {Grondin}, {Grove}, {Guillemot}, {Guiriec}, {Harding},
  {Hayashida}, {Horan}, {Hughes}, {Jackson}, {J{\'o}hannesson}, {Johnson},
  {Johnson}, {Kamae}, {Katagiri}, {Kataoka}, {Kawai}, {Kerr}, {Kn{\"o}dlseder},
  {Kuss}, {Lande}, {Latronico}, {Lemoine-Goumard}, {Longo}, {Loparco}, {Lott},
  {Lovellette}, {Lubrano}, {Makeev}, {Mazziotta}, {McEnery}, {Meurer},
  {Michelson}, {Mitthumsiri}, {Mizuno}, {Monte}, {Monzani}, {Morselli},
  {Moskalenko}, {Murgia}, {Nolan}, {Norris}, {Nuss}, {Ohsugi}, {Okumura},
  {Omodei}, {Orlando}, {Ormes}, {Paneque}, {Pelassa}, {Pepe}, {Pesce-Rollins},
  {Piron}, {Porter}, {Rain{\`o}}, {Rando}, {Razzano}, {Reimer}, {Reimer},
  {Reposeur}, {Rodriguez}, {Ryde}, {Sadrozinski}, {Sanchez}, {Sander}, {Saz
  Parkinson}, {Sgr{\`o}}, {Siskind}, {Smith}, {Spandre}, {Spinelli}, {Starck},
  {Strickman}, {Strong}, {Suson}, {Takahashi}, {Tanaka}, {Thayer}, {Thayer},
  {Thompson}, {Tibaldo}, {Torres}, {Tosti}, {Tramacere}, {Uchiyama}, {Usher},
  {Vasileiou}, {Vilchez}, {Vitale}, {Waite}, {Wang}, {Winer}, {Wood}, {Ylinen},
  \& {Ziegler}}]{Abdo2010}
{Abdo}, A.~A., {Ackermann}, M., {Ajello}, M., {et~al.} 2010, \apj, 710, 133

\bibitem[{{Ackermann} {et~al.}(2010){Ackermann}, {Ajello}, {Baldini}, {Ballet},
  {Barbiellini}, {Bastieri}, {Bechtol}, \& {Bellazzini}}]{Ackermann2010}
{Ackermann}, M., {Ajello}, M., {Baldini}, L., {et~al.} 2010, \apj

\bibitem[{{Andr{\'e}} {et~al.}(2010){Andr{\'e}}, {Men'shchikov}, {Bontemps},
  {K{\"o}nyves}, {Motte}, {Schneider}, {Didelon}, {Minier}, {Saraceno},
  {Ward-Thompson}, {di Francesco}, {White}, {Molinari}, {Testi}, {Abergel},
  {Griffin}, {Henning}, {Royer}, {Mer{\'{\i}}n}, {Vavrek}, {Attard},
  {Arzoumanian}, {Wilson}, {Ade}, {Aussel}, {Baluteau}, {Benedettini},
  {Bernard}, {Blommaert}, {Cambr{\'e}sy}, {Cox}, {di Giorgio}, {Hargrave},
  {Hennemann}, {Huang}, {Kirk}, {Krause}, {Launhardt}, {Leeks}, {Le Pennec},
  {Li}, {Martin}, {Maury}, {Olofsson}, {Omont}, {Peretto}, {Pezzuto}, {Prusti},
  {Roussel}, {Russeil}, {Sauvage}, {Sibthorpe}, {Sicilia-Aguilar}, {Spinoglio},
  {Waelkens}, {Woodcraft}, \& {Zavagno}}]{Andre2010}
{Andr{\'e}}, P., {Men'shchikov}, A., {Bontemps}, S., {et~al.} 2010, \aap, 518,
  L102+

\bibitem[{{Arnal} {et~al.}(2000){Arnal}, {Bajaja}, {Larrarte}, {Morras}, \&
  {P{\"o}ppel}}]{Arnal2000}
{Arnal}, E.~M., {Bajaja}, E., {Larrarte}, J.~J., {Morras}, R., \& {P{\"o}ppel},
  W.~G.~L. 2000, \aaps, 142, 35

\bibitem[{{Bajaja} {et~al.}(2005){Bajaja}, {Arnal}, {Larrarte}, {Morras},
  {P{\"o}ppel}, \& {Kalberla}}]{Bajaja2005}
{Bajaja}, E., {Arnal}, E.~M., {Larrarte}, J.~J., {et~al.} 2005, \aap, 440, 767

\bibitem[{{Barriault} {et~al.}(2010){Barriault}, {Joncas}, {Falgarone},
  {Marshall}, {Heyer}, {Boulanger}, {Foster}, {Brunt}, {Miville-Desch{\^e}nes},
  {Blagrave}, {Kothes}, {Landecker}, {Martin}, {Scott}, {Stil}, \&
  {Taylor}}]{Barriault2010}
{Barriault}, L., {Joncas}, G., {Falgarone}, E., {et~al.} 2010, \mnras, 406,
  2713

\bibitem[{{Bernard} {et~al.}(2008){Bernard}, {Reach}, {Paradis}, {Meixner},
  {Paladini}, {Kawamura}, {Onishi}, {Vijh}, {Gordon}, {Indebetouw}, {Hora},
  {Whitney}, {Blum}, {Meade}, {Babler}, {Churchwell}, {Engelbracht}, {For},
  {Misselt}, {Leitherer}, {Cohen}, {Boulanger}, {Frogel}, {Fukui}, {Gallagher},
  {Gorjian}, {Harris}, {Kelly}, {Latter}, {Madden}, {Markwick-Kemper},
  {Mizuno}, {Mizuno}, {Mould}, {Nota}, {Oey}, {Olsen}, {Panagia},
  {Perez-Gonzalez}, {Shibai}, {Sato}, {Smith}, {Staveley-Smith}, {Tielens},
  {Ueta}, {Van Dyk}, {Volk}, {Werner}, \& {Zaritsky}}]{Bernard2008}
{Bernard}, J., {Reach}, W.~T., {Paradis}, D., {et~al.} 2008, \aj, 136, 919

\bibitem[{{Bernard} {et~al.}(1999){Bernard}, {Abergel}, {Ristorcelli}, {Pajot},
  {Torre}, {Boulanger}, {Giard}, {Lagache}, {Serra}, {Lamarre}, {Puget},
  {Lepeintre}, \& {Cambr{\'e}sy}}]{Bernard1999}
{Bernard}, J.~P., {Abergel}, A., {Ristorcelli}, I., {et~al.} 1999, \aap, 347,
  640

\bibitem[{{Blitz} {et~al.}(1990){Blitz}, {Bazell}, \& {Desert}}]{Blitz1990}
{Blitz}, L., {Bazell}, D., \& {Desert}, F.~X. 1990, \apjl, 352, L13

\bibitem[{{Bohlin} {et~al.}(1978){Bohlin}, {Savage}, \& {Drake}}]{Bohlin1978}
{Bohlin}, R.~C., {Savage}, B.~D., \& {Drake}, J.~F. 1978, \apj, 224, 132

\bibitem[{{Boulanger} {et~al.}(1996){Boulanger}, {Abergel}, {Bernard},
  {Burton}, {Desert}, {Hartmann}, {Lagache}, \& {Puget}}]{Boulanger1996}
{Boulanger}, F., {Abergel}, A., {Bernard}, J., {et~al.} 1996, \aap, 312, 256

\bibitem[{{Dame}(2011)}]{Dame2011}
{Dame}, T.~M. 2011, in preparation

\bibitem[{{Dame} {et~al.}(2001){Dame}, {Hartmann}, \& {Thaddeus}}]{Dame2001}
{Dame}, T.~M., {Hartmann}, D., \& {Thaddeus}, P. 2001, \apj, 547, 792

\bibitem[{{de Vries} {et~al.}(1987){de Vries}, {Thaddeus}, \&
  {Heithausen}}]{deVries1987}
{de Vries}, H.~W., {Thaddeus}, P., \& {Heithausen}, A. 1987, \apj, 319, 723

\bibitem[{{Dickey} {et~al.}(2003){Dickey}, {McClure-Griffiths}, {Gaensler}, \&
  {Green}}]{Dickey2003}
{Dickey}, J.~M., {McClure-Griffiths}, N.~M., {Gaensler}, B.~M., \& {Green},
  A.~J. 2003, \apj, 585, 801

\bibitem[{{Dickey} {et~al.}(2009){Dickey}, {Strasser}, {Gaensler}, {Haverkorn},
  {Kavars}, {McClure-Griffiths}, {Stil}, \& {Taylor}}]{Dickey2009}
{Dickey}, J.~M., {Strasser}, S., {Gaensler}, B.~M., {et~al.} 2009, \apj, 693,
  1250

\bibitem[{{Falgarone} {et~al.}(1991){Falgarone}, {Phillips}, \&
  {Walker}}]{Falgarone1991}
{Falgarone}, E., {Phillips}, T.~G., \& {Walker}, C.~K. 1991, \apj, 378, 186

\bibitem[{{Finkbeiner} {et~al.}(1999){Finkbeiner}, {Davis}, \&
  {Schlegel}}]{finkbeiner1999}
{Finkbeiner}, D.~P., {Davis}, M., \& {Schlegel}, D.~J. 1999, \apj, 524, 867

\bibitem[{{Fukui} {et~al.}(1999){Fukui}, {Onishi}, {Abe}, {Kawamura},
  {Tachihara}, {Yamaguchi}, {Mizuno}, \& {Ogawa}}]{fukui1999}
{Fukui}, Y., {Onishi}, T., {Abe}, R., {et~al.} 1999, \pasj, 51, 751

\bibitem[{{Gillmon} \& {Shull}(2006)}]{Gillmon2006}
{Gillmon}, K. \& {Shull}, J.~M. 2006, \apj, 636, 908

\bibitem[{{Glover} {et~al.}(2010){Glover}, {Federrath}, {Mac Low}, \&
  {Klessen}}]{Glover2010}
{Glover}, S.~C.~O., {Federrath}, C., {Mac Low}, M., \& {Klessen}, R.~S. 2010,
  \mnras, 404, 2

\bibitem[{{Goldsmith} {et~al.}(2008){Goldsmith}, {Heyer}, {Narayanan}, {Snell},
  {Li}, \& {Brunt}}]{Goldsmith2008}
{Goldsmith}, P.~F., {Heyer}, M., {Narayanan}, G., {et~al.} 2008, \apj, 680, 428

\bibitem[{{G{\'o}rski} {et~al.}(2005){G{\'o}rski}, {Hivon}, {Banday},
  {Wandelt}, {Hansen}, {Reinecke}, \& {Bartelmann}}]{Gorski2005}
{G{\'o}rski}, K.~M., {Hivon}, E., {Banday}, A.~J., {et~al.} 2005, \apj, 622,
  759

\bibitem[{{Grenier} {et~al.}(2005){Grenier}, {Casandjian}, \&
  {Terrier}}]{Grenier2005}
{Grenier}, I.~A., {Casandjian}, J., \& {Terrier}, R. 2005, Science, 307, 1292

\bibitem[{{Hartmann} \& {Burton}(1997)}]{Hartmann1997}
{Hartmann}, D. \& {Burton}, W.~B. 1997, {Atlas of Galactic Neutral Hydrogen},
  ed. {Hartmann, D.~\& Burton, W.~B.}

\bibitem[{{Hauser} {et~al.}(1998){Hauser}, {Arendt}, {Kelsall}, {Dwek},
  {Odegard}, {Weiland}, {Freudenreich}, {Reach}, {Silverberg}, {Moseley},
  {Pei}, {Lubin}, {Mather}, {Shafer}, {Smoot}, {Weiss}, {Wilkinson}, \&
  {Wright}}]{Hauser1998}
{Hauser}, M.~G., {Arendt}, R.~G., {Kelsall}, T., {et~al.} 1998, \apj, 508, 25

\bibitem[{{Heiles} {et~al.}(1988){Heiles}, {Reach}, \& {Koo}}]{Heiles1988}
{Heiles}, C., {Reach}, W.~T., \& {Koo}, B. 1988, \apj, 332, 313

\bibitem[{{Juvela} {et~al.}(2010){Juvela}, {Ristorcelli}, {Montier},
  {Marshall}, {Pelkonen}, {Malinen}, {Ysard}, {T{\'o}th}, {Harju}, {Bernard},
  {Schneider}, {Vereb{\'e}lyi}, {Anderson}, {Andr{\'e}}, {Giard}, {Krause},
  {Lehtinen}, {Macias-Perez}, {Martin}, {McGehee}, {Meny}, {Motte}, {Pagani},
  {Paladini}, {Reach}, {Valenziano}, {Ward-Thompson}, \&
  {Zavagno}}]{Juvela2010}
{Juvela}, M., {Ristorcelli}, I., {Montier}, L.~A., {et~al.} 2010, \aap, 518,
  L93+

\bibitem[{{Kalberla} {et~al.}(2005){Kalberla}, {Burton}, {Hartmann}, {Arnal},
  {Bajaja}, {Morras}, \& {P{\"o}ppel}}]{Kalberla2005}
{Kalberla}, P.~M.~W., {Burton}, W.~B., {Hartmann}, D., {et~al.} 2005, \aap,
  440, 775

\bibitem[{{Kawamura} {et~al.}(1999){Kawamura}, {Onishi}, {Mizuno}, {Ogawa}, \&
  {Fukui}}]{kawamura1999}
{Kawamura}, A., {Onishi}, T., {Mizuno}, A., {Ogawa}, H., \& {Fukui}, Y. 1999,
  \pasj, 51, 851

\bibitem[{{K{\"o}nyves} {et~al.}(2010){K{\"o}nyves}, {Andr{\'e}},
  {Men'shchikov}, {Schneider}, {Arzoumanian}, {Bontemps}, {Attard}, {Motte},
  {Didelon}, {Maury}, {Abergel}, {Ali}, {Baluteau}, {Bernard}, {Cambr{\'e}sy},
  {Cox}, {di Francesco}, {di Giorgio}, {Griffin}, {Hargrave}, {Huang}, {Kirk},
  {Li}, {Martin}, {Minier}, {Molinari}, {Olofsson}, {Pezzuto}, {Russeil},
  {Roussel}, {Saraceno}, {Sauvage}, {Sibthorpe}, {Spinoglio}, {Testi},
  {Ward-Thompson}, {White}, {Wilson}, {Woodcraft}, \& {Zavagno}}]{Konyves2010}
{K{\"o}nyves}, V., {Andr{\'e}}, P., {Men'shchikov}, A., {et~al.} 2010, \aap,
  518, L106+

\bibitem[{{Kulkarni} \& {Heiles}(1988)}]{Kulkarni1988}
{Kulkarni}, S.~R. \& {Heiles}, C. 1988, {Neutral hydrogen and the diffuse
  interstellar medium}, ed. {Kellermann, K.~I.~\& Verschuur, G.~L.}, 95--153

\bibitem[{{Leroy} {et~al.}(2007){Leroy}, {Bolatto}, {Stanimirovic}, {Mizuno},
  {Israel}, \& {Bot}}]{Leroy2007}
{Leroy}, A., {Bolatto}, A., {Stanimirovic}, S., {et~al.} 2007, \apj, 658, 1027

\bibitem[{{Matsunaga} {et~al.}(2001){Matsunaga}, {Mizuno}, {Moriguchi},
  {Onishi}, {Mizuno}, \& {Fukui}}]{matsunaga2001}
{Matsunaga}, K., {Mizuno}, N., {Moriguchi}, Y., {et~al.} 2001, \pasj, 53, 1003

\bibitem[{{Mennella et al.}(2011)}]{planck2011-1.4}
{Mennella et al.} 2011, {Planck early results 03: First assessment of the Low
  Frequency Instrument in-flight performance} ({Submitted to \aap})

\bibitem[{{Meyerdierks} \& {Heithausen}(1996)}]{Meyerdierks1996}
{Meyerdierks}, H. \& {Heithausen}, A. 1996, \aap, 313, 929

\bibitem[{{Miville-Desch{\^e}nes} \& {Lagache}(2005)}]{Miville2005}
{Miville-Desch{\^e}nes}, M. \& {Lagache}, G. 2005, \apjs, 157, 302

\bibitem[{{Miville-Desch{\^e}nes} {et~al.}(2002){Miville-Desch{\^e}nes},
  {Lagache}, \& {Puget}}]{Miville2002}
{Miville-Desch{\^e}nes}, M., {Lagache}, G., \& {Puget}, J. 2002, \aap, 393, 749

\bibitem[{{Mizuno} \& {Fukui}(2004)}]{mizuno2004}
{Mizuno}, A. \& {Fukui}, Y. 2004, in Astronomical Society of the Pacific
  Conference Series, Vol. 317, Milky Way Surveys: The Structure and Evolution
  of our Galaxy, ed. {D.~Clemens, R.~Shah, \& T.~Brainerd}, 59--+

\bibitem[{{Mizuno} {et~al.}(2001){Mizuno}, {Yamaguchi}, {Tachihara}, {Toyoda},
  {Aoyama}, {Yamamoto}, {Onishi}, \& {Fukui}}]{mizuno2001}
{Mizuno}, A., {Yamaguchi}, R., {Tachihara}, K., {et~al.} 2001, \pasj, 53, 1071

\bibitem[{{Molinari} {et~al.}(2010){Molinari}, {Swinyard}, {Bally}, {Barlow},
  {Bernard}, {Martin}, {Moore}, {Noriega-Crespo}, {Plume}, {Testi}, {Zavagno},
  {Abergel}, {Ali}, {Anderson}, {Andr{\'e}}, {Baluteau}, {Battersby},
  {Beltr{\'a}n}, {Benedettini}, {Billot}, {Blommaert}, {Bontemps}, {Boulanger},
  {Brand}, {Brunt}, {Burton}, {Calzoletti}, {Carey}, {Caselli}, {Cesaroni},
  {Cernicharo}, {Chakrabarti}, {Chrysostomou}, {Cohen}, {Compiegne}, {de
  Bernardis}, {de Gasperis}, {di Giorgio}, {Elia}, {Faustini}, {Flagey},
  {Fukui}, {Fuller}, {Ganga}, {Garcia-Lario}, {Glenn}, {Goldsmith}, {Griffin},
  {Hoare}, {Huang}, {Ikhenaode}, {Joblin}, {Joncas}, {Juvela}, {Kirk},
  {Lagache}, {Li}, {Lim}, {Lord}, {Marengo}, {Marshall}, {Masi}, {Massi},
  {Matsuura}, {Minier}, {Miville-Desch{\^e}nes}, {Montier}, {Morgan}, {Motte},
  {Mottram}, {M{\"u}ller}, {Natoli}, {Neves}, {Olmi}, {Paladini}, {Paradis},
  {Parsons}, {Peretto}, {Pestalozzi}, {Pezzuto}, {Piacentini}, {Piazzo},
  {Polychroni}, {Pomar{\`e}s}, {Popescu}, {Reach}, {Ristorcelli}, {Robitaille},
  {Robitaille}, {Rod{\'o}n}, {Roy}, {Royer}, {Russeil}, {Saraceno}, {Sauvage},
  {Schilke}, {Schisano}, {Schneider}, {Schuller}, {Schulz}, {Sibthorpe},
  {Smith}, {Smith}, {Spinoglio}, {Stamatellos}, {Strafella}, {Stringfellow},
  {Sturm}, {Taylor}, {Thompson}, {Traficante}, {Tuffs}, {Umana}, {Valenziano},
  {Vavrek}, {Veneziani}, {Viti}, {Waelkens}, {Ward-Thompson}, {White},
  {Wilcock}, {Wyrowski}, {Yorke}, \& {Zhang}}]{Molinari2010}
{Molinari}, S., {Swinyard}, B., {Bally}, J., {et~al.} 2010, \aap, 518, L100+

\bibitem[{{Onishi} {et~al.}(1999){Onishi}, {Kawamura}, {Abe}, {Yamaguchi},
  {Saito}, {Moriguchi}, {Mizuno}, {Ogawa}, \& {Fukui}}]{onishi1999}
{Onishi}, T., {Kawamura}, A., {Abe}, R., {et~al.} 1999, \pasj, 51, 871

\bibitem[{{Onishi} {et~al.}(2001){Onishi}, {Yoshikawa}, {Yamamoto}, {Kawamura},
  {Mizuno}, \& {Fukui}}]{Onishi2001}
{Onishi}, T., {Yoshikawa}, N., {Yamamoto}, H., {et~al.} 2001, \pasj, 53, 1017

\bibitem[{{Paladini} {et~al.}(2007){Paladini}, {Montier}, {Giard}, {Bernard},
  {Dame}, {Ito}, \& {Macias-Perez}}]{Paladini2007}
{Paladini}, R., {Montier}, L., {Giard}, M., {et~al.} 2007, \aap, 465, 839

\bibitem[{{Paradis} {et~al.}(2009){Paradis}, {Bernard}, \&
  {M{\'e}ny}}]{Paradis2009b}
{Paradis}, D., {Bernard}, J., \& {M{\'e}ny}, C. 2009, \aap, 506, 745

\bibitem[{{Paradis} \& {et. al.}(2011)}]{Paradis2011}
{Paradis}, D. \& {et. al.} 2011, in preparation

\bibitem[{{Perrot} \& {Grenier}(2003)}]{Perrot2003}
{Perrot}, C.~A. \& {Grenier}, I.~A. 2003, \aap, 404, 519

\bibitem[{{Planck Collaboration}(2011{\natexlab{a}})}]{planck2011-1.1}
{Planck Collaboration}. 2011{\natexlab{a}}, {Planck early results 01: The
  Planck mission} ({Submitted to \aap})

\bibitem[{{Planck Collaboration}(2011{\natexlab{b}})}]{planck2011-5.1a}
{Planck Collaboration}. 2011{\natexlab{b}}, {Planck early results 08: The
  all-sky early Sunyaev-Zeldovich cluster sample} ({Submitted to \aap})

\bibitem[{{Planck Collaboration}(2011{\natexlab{c}})}]{planck2011-6.4b}
{Planck Collaboration}. 2011{\natexlab{c}}, {Planck early results 17: Origin of
  the submillimetre excess dust emission in the Magellanic Clouds} ({Submitted
  to \aap})

\bibitem[{{Planck Collaboration}(2011{\natexlab{d}})}]{planck2011-6.6}
{Planck Collaboration}. 2011{\natexlab{d}}, {Planck early results 18: The power
  spectrum of cosmic infrared background anisotropies} ({Submitted to \aap})

\bibitem[{{Planck Collaboration}(2011{\natexlab{e}})}]{planck2011-7.0}
{Planck Collaboration}. 2011{\natexlab{e}}, {Planck early results 19: All-sky
  temperature and dust optical depth from Planck and IRAS --- constraints on
  the ``dark gas" in our Galaxy} ({Submitted to \aap})

\bibitem[{{Planck Collaboration}(2011{\natexlab{f}})}]{planck2011-7.3}
{Planck Collaboration}. 2011{\natexlab{f}}, {Planck early results 21:
  Properties of the interstellar medium in the Galactic plane} ({Submitted to
  \aap})

\bibitem[{{Planck Collaboration}(2011{\natexlab{g}})}]{planck2011-7.7a}
{Planck Collaboration}. 2011{\natexlab{g}}, {Planck early results 22: The
  submillimetre properties of a sample of Galactic cold clumps} ({Submitted to
  \aap})

\bibitem[{{Planck Collaboration}(2011{\natexlab{h}})}]{planck2011-7.7b}
{Planck Collaboration}. 2011{\natexlab{h}}, {Planck early results 23: The
  Galactic cold core population revealed by the first all-sky survey}
  ({Submitted to \aap})

\bibitem[{{Planck Collaboration}(2011{\natexlab{i}})}]{planck2011-7.12}
{Planck Collaboration}. 2011{\natexlab{i}}, {Planck early results 24: Dust in
  the diffuse interstellar medium and the Galactic halo} ({Submitted to \aap})

\bibitem[{{Planck Collaboration}(2011{\natexlab{j}})}]{planck2011-7.13}
{Planck Collaboration}. 2011{\natexlab{j}}, {Planck early results 25: Thermal
  dust in nearby molecular clouds} ({Submitted to \aap})

\bibitem[{{Planck HFI Core Team}(2011{\natexlab{a}})}]{planck2011-1.5}
{Planck HFI Core Team}. 2011{\natexlab{a}}, {Planck early results 04: First
  assessment of the High Frequency Instrument in-flight performance}
  ({Submitted to \aap})

\bibitem[{{Planck HFI Core Team}(2011{\natexlab{b}})}]{planck2011-1.7}
{Planck HFI Core Team}. 2011{\natexlab{b}}, {Planck early results 06: The High
  Frequency Instrument data processing} ({Submitted to \aap})

\bibitem[{{Reach} {et~al.}(1995){Reach}, {Dwek}, {Fixsen}, {Hewagama},
  {Mather}, {Shafer}, {Banday}, {Bennett}, {Cheng}, {Eplee}, {Leisawitz},
  {Lubin}, {Read}, {Rosen}, {Shuman}, {Smoot}, {Sodroski}, \&
  {Wright}}]{Reach1995}
{Reach}, W.~T., {Dwek}, E., {Fixsen}, D.~J., {et~al.} 1995, \apj, 451, 188

\bibitem[{{Reach} {et~al.}(1994){Reach}, {Koo}, \& {Heiles}}]{Reach1994}
{Reach}, W.~T., {Koo}, B., \& {Heiles}, C. 1994, \apj, 429, 672

\bibitem[{{Reach} {et~al.}(1998){Reach}, {Wall}, \& {Odegard}}]{Reach1998}
{Reach}, W.~T., {Wall}, W.~F., \& {Odegard}, N. 1998, \apj, 507, 507

\bibitem[{{Roman-Duval} {et~al.}(2010){Roman-Duval}, {Israel}, {Bolatto},
  {Hughes}, {Leroy}, {Meixner}, {Gordon}, {Madden}, {Paradis}, {Kawamura},
  {Li}, {Sauvage}, {Wong}, {Bernard}, {Engelbracht}, {Hony}, {Kim}, {Misselt},
  {Okumura}, {Ott}, {Panuzzo}, {Pineda}, {Reach}, \& {Rubio}}]{Roman-Duval2010}
{Roman-Duval}, J., {Israel}, F.~P., {Bolatto}, A., {et~al.} 2010, \aap, 518,
  L74+

\bibitem[{{Savage} {et~al.}(1977){Savage}, {Bohlin}, {Drake}, \&
  {Budich}}]{Savage1977}
{Savage}, B.~D., {Bohlin}, R.~C., {Drake}, J.~F., \& {Budich}, W. 1977, \apj,
  216, 291

\bibitem[{{Sodroski} {et~al.}(1994){Sodroski}, {Bennett}, {Boggess}, {Dwek},
  {Franz}, {Hauser}, {Kelsall}, {Moseley}, {Odegard}, {Silverberg}, \&
  {Weiland}}]{Sodroski1994}
{Sodroski}, T.~J., {Bennett}, C., {Boggess}, N., {et~al.} 1994, \apj, 428, 638

\bibitem[{{Stepnik} {et~al.}(2003){Stepnik}, {Abergel}, {Bernard}, {Boulanger},
  {Cambr{\'e}sy}, {Giard}, {Jones}, {Lagache}, {Lamarre}, {Meny}, {Pajot}, {Le
  Peintre}, {Ristorcelli}, {Serra}, \& {Torre}}]{Stepnik2003}
{Stepnik}, B., {Abergel}, A., {Bernard}, J., {et~al.} 2003, \aap, 398, 551

\bibitem[{{Wakker}(2006)}]{Wakker2006}
{Wakker}, B.~P. 2006, \apjs, 163, 282

\bibitem[{{Wolfire} {et~al.}(2010){Wolfire}, {Hollenbach}, \&
  {McKee}}]{Wolfire2010}
{Wolfire}, M.~G., {Hollenbach}, D., \& {McKee}, C.~F. 2010, \apj, 716, 1191

\end{thebibliography}


\end{document}